\documentclass[preprint,12pt]{elsarticle}
\journal{Computer Physics Communications}

\usepackage[colorlinks, urlcolor=blue,linkcolor=blue, anchorcolor=blue, citecolor=blue]{hyperref}
\hypersetup{bookmarksopen=true, pdftitle={cpu vah}}

\usepackage[title,titletoc]{appendix}
\usepackage{amssymb}
\usepackage{amsmath,amsfonts,amsthm}
\usepackage{commath}
\usepackage{graphicx}
\usepackage{subfigure}
\usepackage{multirow}
\usepackage{makecell}
\usepackage{tabularx}
\usepackage{bm}
\usepackage{xcolor}
\usepackage{appendix}
\usepackage{fancyhdr}
\usepackage[utf8]{inputenc}
\usepackage{caption}
\DeclareCaptionFont{white}{\color{white}}
\DeclareCaptionFormat{listing}{\colorbox{gray}{\parbox{\textwidth}{#1#2#3}}}

\biboptions{sort&compress}

\graphicspath{{fig/}}

\newcounter{bla}

\newcommand{\be}{\begin{equation}}
\newcommand{\ee}{\end{equation}}
\newcommand{\bs}{\begin{subequations}}
\newcommand{\es}{\end{subequations}}
\newcommand{\beal}{\begin{align}}
\newcommand{\munu}{{\mu\nu}}
\newcommand{\ene}{\mathcal{E}}
\newcommand{\Peq}{\mathcal{P}_\text{eq}}
\newcommand{\cpuvah}{{\sc VAH}}
\newcommand{\gpuvh}{{\sc GPU VH}}
\newcommand{\vh}{{\sc VH}}
\newcommand{\beshydro}{{\sc BEShydro}}
\newcommand{\PL}{\mathcal{P}_L}
\newcommand{\Pperp}{\mathcal{P}_\perp}
\newcommand{\piperp}{\pi_\perp^{\munu}}
\newcommand{\Wperp}{W_{\perp z}^\mu}
\newcommand{\LL}{\mathcal{L}}
\newcommand{\W}{\mathcal{W}}
\newcommand{\etas}{\eta / \mathcal{S}}
\newcommand{\zetas}{\zeta / \mathcal{S}}
\newcommand{\half}{\frac{1}{2}}
\newcommand{\ab}{{\alpha\beta}}
\newcommand{\PLhat}{\hat{\mathcal{P}}_L}
\newcommand{\I}{\mathcal{I}}

\newcommand{\trento}{{\sc T}$_\text{\sc R}${\sc{ENTo}}}

\newcommand{\IS}{{\sc iS3D}}

\begin{document}
\begin{frontmatter}

\title{{\bf Anisotropic fluid dynamical simulations of heavy-ion collisions}}

\author[a]{Mike McNelis\corref{corauthor}}
\author[a]{Dennis Bazow}
\author[a,b]{Ulrich Heinz}
\cortext[corauthor] {Corresponding author.\\ Email addresses: mcnelis.9@osu.edu (M. McNelis), 
heinz.9@osu.edu (U. Heinz)}
\address[a]{Department of Physics, The Ohio State University, Columbus, OH 43210-1117, USA}
\address[b]{Institut f\"ur Theoretische Physik, J.~W.~Goethe Universit\"at, Max-von-Laue-Str. 1, D-60438 Frankfurt am Main, Germany}
\date{\today}

\begin{abstract}
    We present \cpuvah{}, a (3+1)--dimensional simulation that evolves the far-from-equilibrium quark-gluon plasma produced in ultrarelativistic heavy-ion collisions with anisotropic fluid dynamics. We solve the hydrodynamic equations on an Eulerian grid using the Kurganov--Tadmor algorithm in combination with a new adaptive Runge--Kutta method. Our numerical scheme allows us to start the simulation soon after the nuclear collision, largely avoiding the need to integrate it with a separate pre-equilibrium dynamics module. We test the code's performance by simulating on the Eulerian grid conformal and non-conformal Bjorken flow as well as conformal Gubser flow, whose (0+1)--dimensional solutions are precisely known. Finally, we compare non-conformal anisotropic hydrodynamics to second-order viscous hydrodynamics in central Pb+Pb collisions and find that the former's longitudinal flow profile responds more consistently to the fluid's gradients along the spacetime rapidity direction.
\end{abstract}


\end{frontmatter}

\newpage

\noindent
{\bf PROGRAM SUMMARY}
\vspace{0.5cm}\\
\begin{small}
\textit{Manuscript Title:} Anisotropic fluid dynamical simulations of heavy-ion collisions\\
\textit{Authors:} Mike McNelis, Dennis Bazow, Ulrich Heinz \\
\textit{Program Title:} \cpuvah\ \\
\textit{Licensing provisions:} GPLv3 \\
\textit{Programming Language:} C++ \\
\textit{Computer:} Laptop, desktop, cluster \\
\textit{Operating System:} GNU/Linux distributions, Mac OS X \\
\textit{Global memory usage:} 1.2 GB (for a $129 \times 129 \times 63$ grid)
\textit{Classification:} 12 Gases and Fluids, 17 Nuclear physics \\
\textit{External routines/libraries:} GNU Scientific Library (GSL) \\
\textit{Nature of problem:\\}
Modeling the far-from-equilibrium dynamics of quark-gluon plasma produced in ultrarelativistic heavy-ion collisions. \\
\textit{Solution method:\\}
Kurganov--Tadmor algorithm, adaptive stepsize method \\
\textit{Running time:\\}
A (3+1)--d non-conformal anisotropic fluid dynamical simulation of a central Pb+Pb collision on a $129 \times 129 \times 63$ grid takes about 530$s$ for an Intel Xeon E5-2680 v4 multi-core processor with OpenMP acceleration (see Sec.~\ref{S5.3}).
\end{small}

\newpage
\tableofcontents
\newpage


\section{Introduction}
\label{S1}

The quark-gluon plasma is one of the most extreme phases of matter in our universe. The incredibly high temperatures and densities required to create it ($T \sim 10^{12}$ K,  $\rho \sim 10^{17}$ kg/m$^3$) are orders of magnitude beyond those of any substance typically encountered in nature. The quark-gluon plasma also has the remarkable property of having the lowest shear viscosity to entropy density ratio of any known fluid; current phenomenological constraints put $\etas \sim 0.1$ at temperatures around the deconfinement temperature $T_c = 154$\,MeV \cite{Everett:2020yty,Everett:2020xug}. This value is very close to the conjectured Kovtun-Son-Starinet bound $\etas{\,\geq\,}1/4\pi$, which follows from the AdS/CFT correspondence in string theory \cite{Kovtun:2003wp,Maldacena:1997re}. Therefore, it is of great interest to the physics community to test whether or not the KSS bound holds for such strongly coupled liquids. Today, we can  produce quark-gluon plasma by colliding beams of heavy ions (e.g. Pb+Pb) at very high energies. The femtoscopically small droplets of plasma formed in these ultrarelativistic nuclear collisions have incredibly short lifetimes ($\tau_f \sim 10^{-23}$\,s), making them difficult to probe directly. To reconstruct the medium's transport properties, one can simulate the various stages of a heavy-ion collision with quantitative precision and predictive power and fit the theoretical model to large sets of soft-momentum ($p_T < 3$\,GeV) hadronic observables \cite{Shen:2014vra, Bernhard:2016tnd, Bernhard:2018hnz, Everett:2020yty, Everett:2020xug, Nijs:2020ors, Nijs:2020roc}.\footnote{%
    High-energy jets, heavy quarks and electromagnetic radiation can also serve as experimental probes but their full integration into hybrid models \cite{Hirano:2012kj, Paquet:2015lta, Shen:2016zpp, Okai:2017ofp, Vujanovic:2019yih, Yao:2020xzw} is more
    involved than for soft-momentum hadrons.}
The hybrid model must be accurate enough that it can make full use of the considerable precision of the available experimental data.

The anisotropically expanding quark-gluon plasma stage is usually modeled with relativistic viscous hydrodynamics \cite{Schenke:2010nt, Schenke:2010rr, Denicol:2012cn, Shen:2014vra, Gale:2013da, Bazow:2016yra}. Conventionally, viscous hydrodynamics only works for fluids that are both near local equilibrium (i.e. inverse Reynolds number $\text{Re}^{-1} \ll 1$) and have small spacetime gradients (i.e. Knudsen number Kn $\ll 1$) \cite{Denicol:2012cn, Rezzolla:2013rehy}. But unlike most fluids found in nature, the quark-gluon plasma is initially far from local-equilibrium ($\text{Re}^{-1} \sim 1$) and sustains moderately large gradients (Kn $\sim 1$) for a good portion of its lifetime. This is primarily due to quantum fluctuations in the initial-state profile \cite{Schenke:2012wb, Loizides:2014vua, Moreland:2014oya} and the rapid longitudinal expansion at early times. Under these extreme conditions, the validity of a fluid dynamical description for the quark-gluon plasma comes into question~\cite{Niemi:2014wta}. Nevertheless, second-order viscous hydrodynamic simulations, in conjunction with other multi-stage modules, have been widely successful at reproducing hadronic observables (e.g. anisotropic flow coefficients $v_n$)~\cite{Gale:2012rq, Shen:2014lye}. State-of-the-art hybrid models now have enough predictive power to quantitatively constrain the shear and bulk viscosities of the quark-gluon plasma using heavy-ion experimental data and Bayesian inference \cite{Bernhard:2016tnd, Bernhard:2018hnz, Everett:2020yty, Everett:2020xug, Nijs:2020ors, Nijs:2020roc}. Hydrodynamics has also been found to be applicable in collisions between light and heavy ions (e.g. He$^3$+Au, p+Pb) and in proton--proton collisions \cite{Shen:2016zpp, Weller:2017tsr}, although the origin of collectivity in these small systems is still being debated \cite{Romatschke:2016hle, Zhao:2020pty, Plumberg:2020jod, Plumberg:2020jux}.

The success of second-order viscous hydrodynamics can be attributed to an underlying resummed hydrodynamic theory that also extends to large gradients \cite{Heller:2015dha, Romatschke:2017ejr, Strickland:2017kux}. As the associated non-hydrodynamic modes decay on microscopic time scales $\tau_r {\,\ll\,} \tau_\text{hydro}$,\footnote{%
    Hydrodynamic simulations often employ relaxation-type equations from relativistic kinetic theory (e.g. DNMR \cite{Denicol:2012cn}), but they likely do not precisely capture the transient dynamics in strongly coupled fluids \cite{Florkowski:2017olj, vanderSchee:2013pia}.}
the system approaches a (generally non-equilibrium) hydrodynamic attractor that is well approximated by viscous hydrodynamics. This happens within the time scale\footnote{%
    The hydrodynamization time $\tau_\text{hydro}$ refers to the time when viscous hydrodynamics becomes applicable, but the hydrodynamic simulation starting time $\tau_0$ can be different from this.}
$\tau_\text{hydro} \sim 1$\,fm/$c$, long before the system thermalizes \cite{Romatschke:2017vte, Romatschke:2017acs, Almaalol:2020rnu}. Whether or not the strongly coupled quark-gluon plasma possesses an attractor at presently experimentally accessible temperatures ($T \sim 0.15 - 0.5$ GeV) is currently unknown. A major obstacle to answering this question are the technical difficulties involved in evaluating its transport properties from first principles, including its shear and bulk viscosities \cite{Bazavov:2019lgz}. Recently, it was demonstrated that resummed hydrodynamics can be obtained by expanding the microscopic Green's function of a linearized kinetic system, an alternative to a direct resummation of the gradient expansion~\cite{McNelis:2020jrn}. It might be possible to extend this concept to the non-equilibrium quark-gluon plasma by perturbing the system around local-equilibrium and expanding the resulting correlation function(s) \cite{Jeon:2015dfa, Keegan:2016cpi, Kurkela:2018wud, Kamata:2020mka}.

Despite the robustness of second-order viscous hydrodynamics, it is still susceptible to breaking down when gradients are very large (Kn $\gg 1$) \cite{Florkowski:2013lya, Martinez:2017ibh, Bazow:2016yra}. The largest gradients in heavy-ion collisions occur at very early times ($\tau{\,<\,}0.2$\,fm/$c$), especially around the edges of the fireball. If one starts the viscous hydrodynamic simulation too early, one usually encounters negative longitudinal pressures $\PL$ due to huge pressure anisotropies~\cite{Bazow:2017ewq}; near the edges of the fireball even the transverse pressure $\Pperp$ can turn negative if the bulk viscosity $\zetas$ peaks strongly near the quark-hadron phase transition. Not only does this cause excessive viscous heating\footnote{%
    Viscous heating refers to internal entropy production by dissipative processes, not due to external thermal sources.}
but large negative pressures also redirect the matter inward, potentially resulting in cavitation. Regulation schemes can be used to tamp down the viscous pressures and prevent the simulation from crashing, but they cloud the physical predictions of the original hydrodynamic theory \cite{Shen:2014vra,Bazow:2016yra}.\footnote{%
    Some regulation schemes are less extreme than others. For example, the hydrodynamic code {\sc MUSIC} only regulates the dissipative currents outside the fireball region \cite{Schenke:2010nt} while {\sc iEBE-VISHNU} regulates the entire grid \cite{Shen:2014vra}.}
Because of the technical issues involved, it is more suitable to use a \textit{pre-equilibrium dynamics} model before transitioning to viscous hydrodynamics at $\tau_0 = \tau_\text{hydro}$ \cite{Schenke:2012wb, Gale:2012rq, Liu:2015nwa, Bernhard:2016tnd, Bernhard:2018hnz, Everett:2020yty, Everett:2020xug, Keegan:2016cpi, Kurkela:2018vqr, Kurkela:2018wud, Berges:2020fwq}. Still, the conformal approximation usually made in the former results in a mismatch to the latter's non-conformal equation of state, producing (in spite of the system's {\em expansion}) artificially {\it positive} bulk viscous pressures that can be as large as $\Pi \sim \Peq$ around the edges of the fireball \cite{NunesdaSilva:2020bfs}. In some situations, one cannot even switch between dynamical models instantaneously. For example, in low-energy collisions ($\sqrt{s_\text{NN}} \sim 10 - 50$ GeV) where the nuclear interpenetration time is comparable to the fireball lifetime, one needs to run viscous hydrodynamics in the background as soon as the participant nucleons start feeding thermal energy and net-baryon number into the newly formed fireball \cite{Shen:2017bsr, Shen:2018pty, Akamatsu:2018olk, Du:2018mpf, Du:2019obx}.

The shortcomings of second-order viscous hydrodynamics have motivated the development of hydrodynamic models that can better handle far-from-equilibrium situations \cite{Florkowski:2010cf, Martinez:2010sc, Martinez:2012tu}. The most promising candidate is anisotropic hydrodynamics, which treats the two largest dissipative effects arising in heavy-ion collisions (the pressure anisotropy $\PL - \Pperp$ and the bulk viscous pressure $\Pi$) non-perturbatively \cite{Bazow:2013ifa, Bazow:2015cha, Tinti:2015xwa, Molnar:2016vvu, McNelis:2018jho}. While current formulations of anisotropic hydrodynamics are partially based on weakly coupled kinetic theory,\footnote{%
    Second-order viscous hydrodynamic simulations also rely on microscopic approaches such as relativistic kinetic theory to compute the relaxation times and other second-order transport coefficients \cite{Denicol:2012cn, Denicol:2014vaa, Ryu:2015vwa}.}
they are able to capture exact kinetic solutions more accurately than standard viscous hydrodynamics \cite{Florkowski:2013lya, Bazow:2013ifa, Bazow:2015cha, Tinti:2015xwa, Molnar:2016gwq, Nopoush:2014qba, Martinez:2017ibh}. In addition, their ability to maintain positive longitudinal and transverse pressures makes them less prone to cavitation in (3+1)--dimensional simulations \cite{Bazow:2017ewq}. This means they can run at early times with little interference from viscous regulations and remain numerically stable.

The application of anisotropic hydrodynamics to heavy-ion phenomenology is still relatively new; Au+Au collisions at RHIC ($\sqrt{s_\text{NN}} = 200$ GeV) and Pb+Pb collisions at the LHC ($\sqrt{s_\text{NN}} = 2.76$ and 5.02\,TeV) have been modeled reasonably well but so far only using smooth initial conditions and a few model parameters to adjust the fit to experimental data \cite{Alqahtani:2017jwl, Alqahtani:2017tnq, Almaalol:2018gjh}. Anisotropic hydrodynamics has yet to undergo the same extensive tests and trials as viscous hydrodynamics, but it can potentially serve as an additional discrete model for the fluid dynamical stage of heavy-ion collisions. This would further increase the flexibility of the Bayesian framework developed by the JETSCAPE collaboration \cite{Everett:2020yty, Everett:2020xug}, which has already incorporated a number of discrete models for the particlization stage \cite{McNelis:2019auj}. The hope is that, by controlling large dissipative flows nonperturbatively, anisotropic hydrodynamics can constrain the transport coefficients of QCD matter more accurately.

In this paper we introduce our (3+1)--dimensional anisotropic fluid dynamical simulation for heavy-ion collisions called \cpuvah{}.\footnote{%
    The code package can be downloaded from the GitHub repository \url{https://github.com/mjmcnelis/cpu_vah}.}
The C++ module is based on the GPU--accelerated viscous hydrodynamic code \gpuvh{} \cite{Bazow:2016yra, Bazow:2017ewq}, except that it implements anisotropic hydrodynamics with the ($\PL$, $\Pperp$)--matching scheme developed in Ref.~\cite{McNelis:2018jho}.\footnote{%
    At this point \cpuvah{} itself has not yet been ported to GPUs.}
We evolve the dynamical equations on an Eulerian grid using the Kurganov--Tadmor algorithm \cite{Kurganov:2000}, a popular method also used in other viscous hydrodynamic codes \cite{Schenke:2010nt, Schenke:2010rr, Bazow:2016yra, Pang:2018zzo}. The main improvement in our code is the ability to automatically adjust the time step $\Delta \tau_n$ after each iteration, as opposed to using a fixed value for the entire simulation. Our adaptive Runge--Kutta scheme initially uses a fine time step to resolve the rapid longitudinal expansion at early times while speeding up the evolution at later times with a coarser time step given by the Courant-Friedrichs-Lewy (CFL) condition. As a result, we can start anisotropic hydrodynamics at a very early time $\tau_0 = 0.05$\,fm/$c$ to model both the far-off-equilibrium dynamics stage\footnote{\label{free_stream}%
    We assume the system is longitudinally free-streaming (i.e. $\PL/\Peq \approx 0$ and $\ene \propto 1/\tau$) in the time interval $0 < \tau \leq \tau_0$ before starting anisotropic hydrodynamics. This closely mirrors the situation found in other pre-hydrodynamic models \cite{Schenke:2012wb, Liu:2015nwa, Kurkela:2018wud}. It has also been argued \cite{Jaiswal:2019cju, Kurkela:2019set} that (at least in weakly coupled systems) for $\tau\to0$ the far-off-equilibrium hydrodynamic attractor approaches the free-streaming attractor in Bjorken flow (which approximates the early evolution stage in heavy-ion collisions).}
with a non-conformal QCD equation of state and smoothly transition to viscous hydrodynamics.\footnote{%
    Anisotropic hydrodynamics reduces to second-order viscous hydrodynamics in the limit of small Knudsen and inverse Reynolds numbers ($\text{Kn}, \text{Re}^{-1} \ll 1$). We do not switch to a separate viscous hydrodynamics model as the Knudsen and inverse Reynolds numbers decrease, but use anisotropic hydrodynamics to evolve both the earliest far-off-equilibrium and subsequent viscous hydrodynamic stages.}

The code package contains other useful features, such as the option to run either anisotropic hydrodynamics or second-order viscous hydrodynamics within the same framework. This provides the user with several discrete models to evolve the fluid dynamical stage for comparison. We further implemented user options to accelerate the simulation on a multi-core processor with OpenMP, and to use automated grid settings that optimize the overall size of the spatial grid for a given set of runtime parameters (e.g. the impact parameter and particlization switching temperature).

This paper is organized as follows: In Sec.~\ref{S2} we review the anisotropic hydrodynamic equations used to evolve the energy-momentum tensor of the quark-gluon plasma. Sec.~\ref{S3} discusses the numerical implementation of these dynamical equations in the code. In Sec.~\ref{S4} we test our anisotropic fluid dynamical simulation for a system subject to (non)conformal Bjorken flow and conformal Gubser flow. Furthermore, we compare anisotropic hydrodynamics to second-order viscous hydrodynamics in (3+1)--dimensions. Finally, we benchmark the typical runtimes of (2+1)--d and (3+1)--d non-conformal hydrodynamic simulations in Sec.~\ref{S5}.

In this work we use Milne spacetime coordinates $x^\mu = (\tau, x, y, \eta_s)$, where $\tau = \sqrt{t^2{-}z^2}$ is the longitudinal proper time and $\eta_s = \tanh^{-1}(z/t)$ is the spacetime rapidity. We adopt the mostly-minus convention for the metric tensor, $g^\munu = \mathrm{diag}(1, -1, -1, -1/\tau^{2})$. In the code, we solve the hydrodynamic equations in natural units $\hbar = c = k_B = 1$; energy, momentum and temperature have units of fm$^{-1}$, energy density and pressure have units of fm$^{-4}$, etc. These units are converted to physical units (e.g. energy densities in GeV/fm$^3$) when outputting and plotting the results. The net-baryon density $n_B$ and baryon diffusion current $V_B^\mu$ are neglected in this version of the code.

\section{Anisotropic fluid dynamics}
\label{S2}
In this section we provide an overview of the anisotropic hydrodynamics equations, including the equation of state and transport coefficients, to be implemented in the code that evolves the energy-momentum tensor $T^\munu(x)$ of the quark-gluon plasma. For more details on the derivation of these hydrodynamic equations we refer the reader to Refs.~\cite{McNelis:2018jho, Molnar:2016vvu}.
\subsection{Energy-momentum tensor}
\label{S2.1}
In anisotropic fluid dynamics, it is convenient to decompose $T^\munu(x)$ in the basis \{$u^\mu(x)$, $z^\mu(x)$, $\Xi^\munu(x)$\}, where the fluid velocity $u^\mu$ represents the temporal direction in the local fluid rest frame (LRF) and the spatial vectors are split into the longitudinal direction $z^\mu$ and the transverse projection tensor $\Xi^\munu = g^\munu - u^\mu u^\nu + z^\mu z^\nu$ \cite{Molnar:2016vvu}. In the Landau frame this results in the decomposition (suppressing the spacetime dependence)
\be
\label{eq:Tmunu}
T^\munu = \ene u^\mu u^\nu + \PL z^\mu z^\nu - \Pperp \Xi^\munu + 2 W^{(\mu}_{\perp z} z^{\nu)} + \piperp \,,
\ee
where round parentheses denote symmetrization, i.e. $W^{(\mu}_{\perp z} z^{\nu)} = \frac{1}{2}(\Wperp z^\nu + W_{\perp z}^\nu z^\mu)$. The major components of $T^\munu$ are the LRF energy density $\ene = u_\mu u_\nu T^\munu$, the longitudinal pressure $\PL =  z_\mu z_\nu T^\munu$ and the transverse pressure $\Pperp = - \frac{1}{2}\Xi_{\mu\nu}T^\munu$. Together, the pressures $\PL$ and $\Pperp$ capture the largest dissipative flows in heavy-ion collisions: the pressure anisotropy $\Delta \mathcal{P} = \PL - \Pperp$ caused by the rapid longitudinal expansion rate at early longitudinal proper times, and the bulk viscous pressure\footnote{%
    Here $\bar{\mathcal{P}} = \frac{1}{3}(\PL + 2 \Pperp)$ is the average (isotropic) pressure and $\Peq = \Peq(\ene)$ is the equilibrium pressure.}
$\Pi = \bar{\mathcal{P}} - \Peq$ due to critical fluctuations near the quark-hadron phase transition. The remaining components of $T^\munu$ are the longitudinal momentum diffusion current $\Wperp = - \Xi^{\mu}_{\alpha}z_\nu T^{\alpha\nu}$ and the transverse shear stress tensor $\piperp = \Xi_{\alpha\beta}^{\mu\nu}T^{\alpha\beta}$, with $\Xi^\munu_\ab = \half(\Xi^\mu_\alpha\Xi^\nu_\beta +\Xi^\nu_\beta\Xi^\mu_\alpha - \Xi^\munu\Xi_{\alpha\beta})$ being the traceless double transverse projector. We refer to these components as residual shear stresses since they are typically smaller than the pressure anisotropy term in the full shear stress tensor
\be
\label{eq:shear_vh}
    \pi^\munu = \frac{1}{3}(\PL {-} \Pperp)(2z^\mu z^\nu {+} \Xi^\munu) + 2 W^{(\mu}_{\perp z} z^{\nu)} + \piperp \,.
\ee
\subsection{Dynamical variables}
\label{S2.2}
The dynamical variables that we propagate in the code are
\be
    \boldsymbol{q} = (T^{\tau\mu}, \PL, \Pperp, \Wperp, \piperp)\,.
\ee
Their evolution equations will be discussed below. Although $\Wperp$ and $\piperp$ each have only two independent components, we evolve all 14 of their components independently to simplify the workflow of the algorithm.\footnote{%
    When propagating these extraneous components numerically, slight violations of the orthogonality and tracelessness conditions \eqref{eq:enforce} can occur. We correct for these errors at each step of the simulation with the regulation scheme described in Sec.~\ref{S3.5}.}
In addition, we propagate the energy density $\ene$ and the fluid velocity's spatial components\footnote{The fluid velocity's temporal component is $u^\tau = \sqrt{1 + (u^x)^2 + (u^y)^2 + (\tau u^\eta)^2}$.} $\boldsymbol{u} =$ ($u^x$, $u^y$, $u^\eta$) since they appear in the hydrodynamic equations. They are inferred from the components $T^{\tau\mu}$:
\be
\label{eq:Ttaumu}
    T^{\tau\mu} = \ene u^\tau u^\mu + \PL z^\tau z^\mu - \Pperp \Xi^{\tau\mu} + 2 W^{(\tau}_{\perp z} z^{\mu)} + \pi_\perp^{\tau\mu} \,.
\ee
To solve these equations, one also needs to know $z^\mu$. From the orthogonality conditions $z_\mu z^\mu = -1$ and $z_\mu u^\mu = 0$, there are only two nonzero components that depend on the fluid velocity:
\be
    z^\mu = \frac{1}{\sqrt{1{+}u_\perp^2}}\left(\tau u^\eta, 0, 0, \frac{u^\tau}{\tau}\right)\,,
\ee
where $u_\perp = \sqrt{(u^x)^2 + (u^y)^2}$ is the transverse velocity. The solution of the inferred variables ($\ene$, $\boldsymbol{u}$) from the algebraic equations~\eqref{eq:Ttaumu} will be discussed in Sec.~\ref{S3.3}.

For (3+1)--dimensionally expanding fluids, we have a total of $20$ dynamical variables and four inferred variables to evolve on an Eulerian grid.\footnote{%
    Dynamical variables are evolved directly using the Kurganov--Tadmor algorithm (see Sec.~\ref{S3.1}). Inferred variables are determined from the dynamical variables algebraically.}$^,$\footnote{%
    For anisotropic fluid dynamics with a QCD equation of state~\cite{McNelis:2018jho}, we further evolve the mean-field $B$ as a dynamical variable and the anisotropic variables ($\Lambda$, $\alpha_L$, $\alpha_\perp$) as additional inferred variables (see Secs.~\ref{S2.4} and~\ref{S2.6.3}).}
If the system is longitudinally boost-invariant, we do not need to propagate the components $T^{\tau\eta}$, $\Wperp$, $\pi_\perp^{\mu\eta}$ and $u^\eta$ since they vanish by symmetry.
\subsection{Conservation laws}
\label{S2.3}
The evolution of the components $T^{\tau\mu}$ is given by the energy-momentum conservation laws
\be
    D_\mu T^\munu = 0 \,,
\ee
where the covariant derivative $D_\mu$ accounts for the curvilinear nature of the Milne coordinates. The conservation equations can be expanded as
\be
\label{eq:conservation}
    \partial_\mu T^\munu + \Gamma^\mu_{\mu\lambda} T^{\lambda\nu} + \Gamma^\nu_{\mu\lambda} T^{\lambda\mu} = 0 \,,
\ee
where $\partial_\mu$ is the partial derivative and $\Gamma^{\mu}_{\nu\lambda}$ are the Christoffel symbols. In Milne spacetime, the only nonzero Christoffel symbols are
\be
\begin{aligned}
    \Gamma^\tau_{\eta\eta} = \tau, \indent & \indent \Gamma^\eta_{\tau\eta} = \Gamma^\eta_{\eta\tau} = \frac{1}{\tau}\,.
\end{aligned}
\ee
Thus, the set of equations~\eqref{eq:conservation} can be rewritten as
\bs
\label{eq:cons_eqs}
\beal
    \partial_\tau T^{\tau\tau} + \partial_i T^{\tau i} &= - \frac{T^{\tau\tau} {+} \tau^2 T^{\eta\eta}}{\tau} \,,
\\
    \partial_\tau T^{\tau x} + \partial_j T^{xj} &= - \frac{T^{\tau x}}{\tau} \,,
\\
    \partial_\tau T^{\tau y} + \partial_j T^{yj} &= - \frac{T^{\tau y}}{\tau} \,,
\\
    \partial_\tau T^{\tau\eta} + \partial_j T^{\eta j} &= - \frac{3T^{\tau\eta}}{\tau}\,,
\end{align}
\es
where the Latin indices $(i,j) \in (x,y,\eta)$ are summed over spatial components. We can eliminate the components $T^{\tau i}$ in (\ref{eq:cons_eqs}a) by using the identity:
\be
\begin{split}
    T^{\tau i} &=  (\ene {+} \Pperp) u^\tau u^i + \LL^{\tau i} + \W^{\tau i} + \pi_\perp^{\tau i}
\\
    &= v^i T^{\tau\tau} + v^i (\Pperp {-} \LL^{\tau\tau} {-} \W^{\tau\tau} {-} \pi_\perp^{\tau\tau}) + \LL^{\tau i} + \W^{\tau i} + \pi_\perp^{\tau i} \,.
\end{split}
\ee
Here we introduced the three-velocity $v^i = u^i / u^\tau$ as well as the tensors $\LL^\munu = \Delta \mathcal{P} z^\mu z^\nu$ and $\W^\munu = 2W^{(\mu}_{\perp z} z^{\nu)}$. Likewise, we can express the components $T^{ij}$ in (\ref{eq:cons_eqs}b-d) in terms of $T^{\tau i}$:
\be
\begin{split}
    T^{ij} =& \,(\ene {+} \Pperp) u^i u^j - \Pperp g^{ij} + \LL^{ij} + \W^{ij} + \pi_\perp^{ij}
\\ 
    =& \, v^j T^{\tau i} - v^j (\LL^{\tau i} {+} \W^{\tau i} {+} \pi_\perp^{\tau i}) - \Pperp g^{ij} + \LL^{ij} + \W^{ij} + \pi_\perp^{ij} \,.
\end{split}
\ee
After some algebra one obtains~\cite{Bazow:2017ewq}
\bs
\allowdisplaybreaks
\label{eq:cons_flux}
\beal
    \partial_\tau T^{\tau\tau} + \partial_i (v^i T^{\tau\tau}) =& - \frac{T^{\tau\tau} {+} \tau^2 T^{\eta\eta}}{\tau} + (\LL^{\tau\tau} {+} \W^{\tau\tau} {+} \pi_\perp^{\tau\tau} {-} \Pperp) \partial_i v^i
\\\nonumber
    & + v^i \partial_i (\LL^{\tau\tau} {+} \W^{\tau\tau} {+} \pi_\perp^{\tau\tau} {-} \Pperp) - \partial_\eta \LL^{\tau\eta} - \partial_i \W^{\tau i} - \partial_i \pi_\perp^{\tau i} \,, \quad
\\\nonumber
\\
    \partial_\tau T^{\tau x} + \partial_i (v^i T^{\tau x}) =& - \frac{T^{\tau x}}{\tau} - \partial_x \Pperp + (\W^{\tau x} {+} \pi_\perp^{\tau x}) \partial_i v^i
\\\nonumber
    &+ v^i \partial_i (\W^{\tau x} {+} \pi_\perp^{\tau x}) - \partial_\eta \W^{x\eta} - \partial_i \pi_\perp^{xi} \,,
\\\nonumber
\\
    \partial_\tau T^{\tau y} + \partial_i (v^i T^{\tau y}) =& - \frac{T^{\tau y}}{\tau} - \partial_y \Pperp + (\W^{\tau y} {+} \pi_\perp^{\tau y}) \partial_i v^i
\\\nonumber
    &+ v^i \partial_i (\W^{\tau y} {+} \pi_\perp^{\tau y}) - \partial_\eta \W^{y\eta} - \partial_i \pi_\perp^{yi} \,,
\\\nonumber
\\
    \partial_\tau T^{\tau\eta} + \partial_i (v^i T^{\tau\eta}) =& - \frac{3T^{\tau\eta}}{\tau} - \frac{\partial_\eta \Pperp}{\tau^2} + (\LL^{\tau\eta} {+} \W^{\tau\eta} {+} \pi_\perp^{\tau\eta}) \partial_i v^i
\\\nonumber
    &+ v^i \partial_i (\LL^{\tau\eta} {+} \W^{\tau\eta} {+} \pi_\perp^{\tau\eta}) - \partial_\eta \LL^{\eta\eta} - \partial_i \W^{\eta i} - \partial_i \pi_\perp^{\eta i}\,.
\end{align}
\es

For the numerical algorithm the hydrodynamic equations must be written in conservative flux form \cite{Kurganov:2000}:\footnote{%
    For a conformal fluid, the energy density's spatial derivatives are needed to evaluate the source term $\partial_i \Pperp = \frac{1}{2}(\partial_i \ene - \partial_i \PL)$.}
\be
\label{eq:KT_eqs}
    \partial_\tau \boldsymbol{q}(x) + \partial_i\boldsymbol{F}^i(x) = \boldsymbol{S}(\tau,\boldsymbol{q}(x),\boldsymbol{u}(x),\ene(x),\partial_m\boldsymbol{q}(x), \partial_\mu\boldsymbol{u}(x))\,.
\ee
Here $\boldsymbol{F}^i = v^i \boldsymbol{q}$ are the currents, $\boldsymbol{S}$ are the source terms and $m \in (x,y,\eta)$ is a spatial index. Naturally, the evolution equations for $T^{\tau\mu}$ already assume this form. In the next subsection we will see that the relaxation equations for the dissipative flows require additional manipulations.
\subsection{Relaxation equations}
\label{S2.4}
The relaxation equations for the dissipative flows $\PL$, $\Pperp$, $\Wperp$ and $\piperp$ are~\cite{McNelis:2018jho, Molnar:2016vvu}
\bs
\allowdisplaybreaks
\label{eq:relax_1}
\beal
    \dot{\mathcal{P}}_L =&\,\frac{\Peq{-}\bar{\mathcal{P}}}{\tau_\Pi} - \frac{\PL{-}\Pperp}{3\tau_\pi / 2} + \bar{\zeta}^L_z \theta_L + \bar{\zeta}^L_\perp \theta_\perp - 2\Wperp \dot{z}_\mu
\\\nonumber
    &+ \bar{\lambda}^L_{Wu} \Wperp D_z u_\mu + \bar{\lambda}^L_{W\perp} \Wperp z_\nu \nabla_{\perp,\mu} u^\nu - \bar{\lambda}^L_{\pi} \piperp \sigma_{\perp,\munu} \,,
\\\nonumber
\\
    \dot{\mathcal{P}}_\perp
    =&\, \frac{\Peq{-}\bar{\mathcal{P}}}{\tau_\Pi} + \frac{\PL{-}\Pperp}{3\tau_\pi} + \bar{\zeta}^\perp_z \theta_L + \bar{\zeta}^\perp_\perp \theta_\perp + \Wperp \dot{z}_\mu
\\\nonumber
      &+ \bar{\lambda}^\perp_{Wu} \Wperp D_z u_\mu - \bar{\lambda}^\perp_{W\perp} \Wperp z_\nu \nabla_{\perp,\mu} u^\nu + \bar{\lambda}^\perp_{\pi} \piperp \sigma_{\perp,\munu}\,,
\\\nonumber
\\
    \dot{W}^{\{\mu\}}_{\perp z}
    =&\, - \frac{\Wperp}{\tau_\pi} + 2\bar{\eta}^W_u \Xi^\munu D_z u_\nu - 2\bar{\eta}^W_\perp z_\nu \nabla_\perp^\mu u^\nu - \big(\bar{\tau}^W_z \Xi^\munu {+} \piperp\big) \dot{z}_\nu
\\\nonumber
    &- \bar{\lambda}^W_{W u} \Wperp  \theta_L + \bar{\delta}^W_W \Wperp \theta_\perp + \bar{\lambda}^W_{W \perp} \sigma_\perp^\munu  W_{\perp z, \nu} + \omega_\perp^\munu W_{\perp z, \nu}
\\\nonumber
    &+ \bar{\lambda}^W_{\pi u} \piperp D_z u_\nu -  \bar{\lambda}^W_{\pi \perp} \piperp z_\lambda \nabla_{\perp,\nu} u^\lambda\,,
\\\nonumber
\\
    \dot{\pi}^{\{\munu\}}_{\perp}
    =&\, - \frac{\piperp}{\tau_\pi} + 2 \bar{\eta}_\perp \sigma_\perp^\munu - 2 W_{\perp z}^{\{\mu} \dot{z}^{\nu\}} + \bar{\lambda}^\pi_\pi \piperp \theta_L - \bar{\delta}^\pi_\pi \piperp \theta_\perp
\\\nonumber
    &- \bar{\tau}^\pi_\pi \pi_\perp^{\lambda \{\mu} \sigma^{\nu\}}_{\perp,\lambda} + 2 \pi_\perp^{\lambda \{\mu} \omega^{\nu\}}_{\perp,\lambda}  - \bar{\lambda}^\pi_{W u} W_{\perp z}^{\{\mu} D_z u^{\nu\}} \,.
\end{align}
\es
Here a dot above any quantity denotes the co-moving time derivative $u^\gamma D_\gamma$, and curly brackets denote either the transverse projection of a vector, $t^{\{\mu\}} = \Xi^\mu_{\alpha} t^{\alpha}$, or the traceless double transverse projection of a rank-2 tensor, $t^{\{\mu\nu\}} = \Xi^\munu_{\alpha\beta} t^{\alpha\beta}$. We also define the longitudinal and transverse expansion rates $\theta_L = z_\mu D_z u^\mu$ and $\theta_\perp = \nabla_{\perp,\mu}u^\mu$, where $D_z = - z^\nu D_\nu$ is the LRF longitudinal derivative and $\nabla_\perp^\mu = \Xi^\munu D_\nu$ is the transverse gradient. The transverse velocity-shear tensor is $\sigma_\perp^\munu = \Xi^\munu_{\alpha\beta} D^{(\alpha} u^{\beta)}$, while the transverse vorticity tensor is given by $\omega_\perp^\munu = \Xi^\mu_\alpha \Xi^\nu_\beta D^{[\alpha} u^{\beta]}$, where the square brackets $D^{[\alpha} u^{\beta]} = \frac{1}{2}(D^\alpha u^\beta - D^\beta u^\alpha)$ denote anti-symmetrization. The shear and bulk viscosity to entropy density ratios $\etas$ and $\zetas$ and their associated relaxation times $\tau_\pi$ and $\tau_\Pi$, along with the anisotropic transport coefficients coupled to the gradient forces, will be discussed in Sec.~\ref{S2.6}.

We recast the relaxation equations in conservative flux form by using the product rule identities
\bs
\beal
    \dot{W}_{\perp z}^{\{\mu\}} &= \Xi^\mu_\alpha u^\gamma D_\gamma W_{\perp z}^\alpha = u^\gamma D_\gamma \Wperp -  W_{\perp z}^\alpha u^\gamma D_\gamma \Xi^\mu_\alpha \,,
\\
    \dot{\pi}_{\perp}^{\{\munu\}} &= \Xi^\munu_{\alpha\beta} u^\gamma D_\gamma \pi_\perp^{\alpha\beta} = u^\gamma D_\gamma \piperp -  \pi_\perp^{\alpha\beta} u^\gamma D_\gamma \Xi^\munu_{\alpha\beta}
\end{align}
\es
to rewrite the l.h.s. of Eqs.~(\ref{eq:relax_1}a-d) as
\bs
\allowdisplaybreaks
\label{eq:relax_3}
\beal
    &\dot{\mathcal{P}}_{L} = u^\gamma \partial_\gamma \mathcal{P}_{L}  \,,
\\
    &\dot{\mathcal{P}}_{\perp} = u^\gamma \partial_\gamma \mathcal{P}_{\perp}  \,,
\\
    &\dot{W}_{\perp z}^{\{\mu\}} = u^\gamma\partial_\gamma W_{\perp z}^\mu + u^\gamma \Gamma^\mu_{\gamma\lambda} W_{\perp z}^\lambda + W_{\perp z}^\alpha (u^\mu a_\alpha {-} z^\mu \dot{z}_\alpha)  \,,
\\
    &\dot{\pi}_\perp^{\{\munu\}} = u^\gamma\partial_\gamma \pi_\perp^{\munu} + u^\gamma \Gamma^\mu_{\gamma\lambda} \pi_\perp^{\nu\lambda} + u^\gamma \Gamma^\nu_{\gamma\lambda} \pi_\perp^{\mu\lambda}
\\\nonumber
    &\qquad\quad + \pi_\perp^{\mu\alpha}(u^\nu a_\alpha {-} z^\nu \dot{z}_\alpha) + \pi_\perp^{\nu\alpha}(u^\mu a_\alpha {-} z^\mu \dot{z}_\alpha)\,,
\end{align}
\es
where $a^\mu = \dot{u}^\mu$ is the fluid acceleration. Finally, we use the product rule identity
\be
    u^\gamma \partial_\gamma \mathcal{P}_{L,\perp} = u^\tau \left[\partial_\tau \mathcal{P}_{L,\perp} + \partial_i (v^i \mathcal{P}_{L,\perp}) - \mathcal{P}_{L,\perp} \partial_i v^i \right]
\ee
to rewrite Eqs.~(\ref{eq:relax_3}a-b) as
\bs
\label{eq:relax_PLPT}
\beal
    \partial_\tau\PL + \partial_i(v^i \PL) &= \PL \partial_i v^i + \frac{1}{u^\tau} \left[ \frac{\Peq{-}\bar{\mathcal{P}}}{\tau_\Pi} - \frac{\PL{-}\Pperp}{3\tau_\pi / 2} + \mathcal{I}_L \right]\,,
\\
    \partial_\tau\Pperp + \partial_i(v^i \Pperp) &= \Pperp \partial_i v^i + \frac{1}{u^\tau} \left[\frac{\Peq{-}\bar{\mathcal{P}}}{\tau_\Pi} + \frac{\PL{-}\Pperp}{3\tau_\pi} + \mathcal{I}_\perp \right] \,;
\end{align}
\es
here
\bs
\beal
    \mathcal{I}_L =&\, \bar{\zeta}^L_z \theta_L + \bar{\zeta}^L_\perp \theta_\perp - 2\Wperp \dot{z}_\mu + \bar{\lambda}^L_{Wu} \Wperp D_z u_\mu + \bar{\lambda}^L_{W\perp} \Wperp z_\nu \nabla_{\perp,\mu} u^\nu
\\\nonumber
  &- \bar{\lambda}^L_{\pi} \piperp \sigma_{\perp,\munu}  \,,
\\\nonumber
\\
    \mathcal{I}_\perp =&\, \bar{\zeta}^\perp_z \theta_L + \bar{\zeta}^\perp_\perp \theta_\perp + \Wperp \dot{z}_\mu + \bar{\lambda}^\perp_{Wu} \Wperp D_z u_\mu - \bar{\lambda}^\perp_{W\perp} \Wperp z_\nu \nabla_{\perp,\mu} u^\nu
\\\nonumber
    &+ \bar{\lambda}^\perp_{\pi} \piperp \sigma_{\perp,\munu}
\end{align}
\es
are the gradient source terms for $\PL$ and $\Pperp$. Similarly Eqs.~(\ref{eq:relax_3}c-d) can be rewritten as
\bs
\allowdisplaybreaks
\label{eq:relax_residual}
\beal
    \partial_\tau\Wperp + \partial_i(v^i \Wperp) &= \Wperp \partial_i v^i + \frac{1}{u^\tau} \left[-\frac{\Wperp}{\tau_\pi} + \mathcal{I}^\mu_W - \mathcal{P}^\mu_W - \mathcal{G}_W^\mu\right]\,,
\\
    \partial_\tau\piperp + \partial_i(v^i \piperp) &= \piperp \partial_i v^i + \frac{1}{u^\tau}\left[-\frac{\piperp}{\tau_\pi} + \mathcal{I}^\munu_\pi - \mathcal{P}^\munu_\pi - \mathcal{G}_\pi^\munu  \right] \,;
\end{align}
\es
here
\bs
\allowdisplaybreaks
\beal
    \mathcal{I}^\mu_W =&\, \Xi^\munu \big(2\bar{\eta}^W_u  D_z u_\nu {-} \bar{\tau}^W_z \dot{z}_\nu\big) - 2\bar{\eta}^W_\perp z_\nu \nabla_\perp^\mu u^\nu - \piperp\dot{z}_\nu  - \bar{\lambda}^W_{W u} \Wperp \theta_L
\\\nonumber
    & + \bar{\delta}^W_W \Wperp \theta_\perp + \bar{\lambda}^W_{W \perp} \sigma_\perp^\munu W_{\perp z, \nu} + \omega_\perp^\munu W_{\perp z, \nu} + \bar{\lambda}^W_{\pi u} \piperp D_z u_\nu
\\\nonumber
    &- \bar{\lambda}^W_{\pi \perp} \piperp z_\lambda \nabla_{\perp,\nu} u^\lambda\,,
\\\nonumber
\\
    \mathcal{I}^\munu_\pi =&\, \Xi^\munu_{\alpha\beta}\big(2 \pi_\perp^{\lambda (\alpha} \omega^{\beta)}_{\perp,\lambda} - \bar{\tau}^\pi_\pi \pi_\perp^{\lambda(\alpha} \sigma^{\beta)}_{\perp,\lambda} - 2 W_{\perp z}^{(\alpha} \dot{z}^{\beta)} - \bar{\lambda}^\pi_{W u} W_{\perp z}^{(\alpha} D_z u^{\beta)}\big)
\\\nonumber
    & + 2 \bar{\eta}_\perp \sigma_\perp^\munu  + \bar{\lambda}^\pi_\pi \piperp \theta_L - \bar{\delta}^\pi_\pi \piperp \theta_\perp
\end{align}
\es
are the gradient source terms,
\bs
\beal
    \mathcal{P}^\mu_W =&\, W_{\perp z}^\alpha (u^\mu a_\alpha {-} z^\mu \dot{z}_\alpha) \,,
\\
    \mathcal{P}^\munu_\pi =&\, \pi_\perp^{\mu\alpha}(u^\nu a_\alpha {-} z^\nu \dot{z}_\alpha) + \pi_\perp^{\nu\alpha}(u^\mu a_\alpha {-} z^\mu \dot{z}_\alpha)
\end{align}
\es
are the transverse projection source terms, and
\bs
\label{eq:geometric_source}
\beal
    \mathcal{G}^\mu_W =&\, u^\gamma \Gamma^\mu_{\gamma\lambda} W_{\perp z}^\lambda \,,
\\
    \mathcal{G}^\munu_\pi =&\, u^\gamma \Gamma^\mu_{\gamma\lambda} \pi_\perp^{\nu\lambda} + u^\gamma \Gamma^\nu_{\gamma\lambda} \pi_\perp^{\mu\lambda}
\end{align}
\es
are the geometric source terms for $\Wperp$ and $\piperp$. The individual components that make up the source terms in the relaxation equations (\ref{eq:relax_PLPT}a-b) and (\ref{eq:relax_residual}a-b) are listed in Appendices \ref{appa} and \ref{appb}.

If we evolve anisotropic fluid dynamics with a QCD equation of state (see Secs.~\ref{S2.5} and \ref{S2.6}) we also propagate a mean field $B$ as a dynamical variable. Its relaxation equation is \cite{Tinti:2016bav,McNelis:2018jho}
\be
\label{eq:relax_B_0}
    \dot{B} = \frac{B_\text{eq} {-} B}{\tau_\Pi} - \frac{\dot{m}}{m}(\ene {-} 2\Pperp {-} \PL {-} 4B)
\ee
or, in conservative flux form,
\be
\label{eq:relax_B}
    \partial_\tau B + \partial_i(v^i B) = B \, \partial_i v^i + \frac{1}{u^\tau} \left[\frac{B_\text{eq} {-} B}{\tau_\Pi} - \frac{\dot{m}}{m}(\ene {-} 2\Pperp {-} \PL {-} 4B)\right]\,,
\ee
where $B_\text{eq}(\ene)$ is the equilibrium mean field and $m(\ene)$ is the quasiparticle mass.

\subsection{Equation of state}
\label{S2.5}
\begin{figure}[t]
\centering
\includegraphics[width=0.9\linewidth]{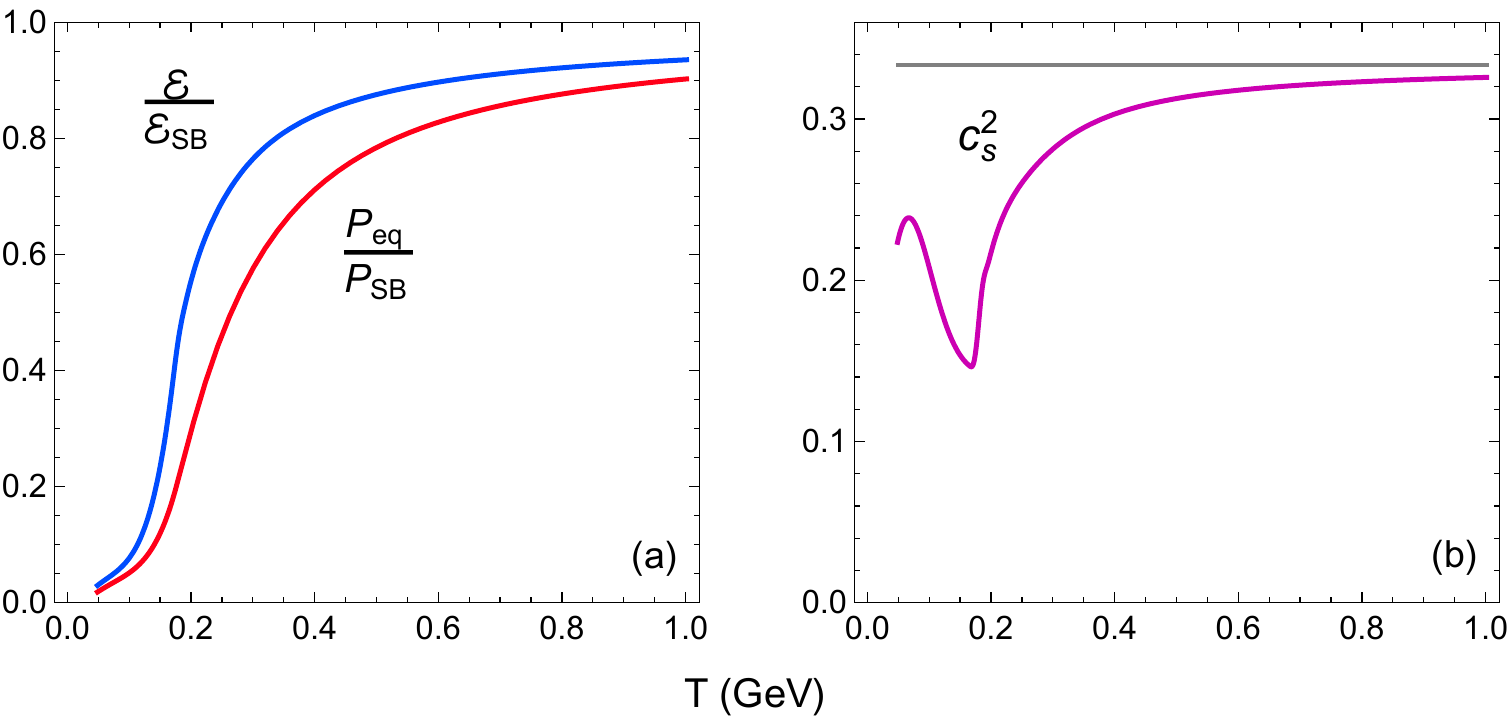}
\caption{(Color online)
\label{eos}
    {\sl Left:} The QCD energy density (blue) and equilibrium pressure  (red), normalized by their Stefan--Boltzmann limits \eqref{eq:stefan}.
    {\sl Right:} The squared speed of sound (purple) as a function of temperature; the gray line indicates the conformal limit $c_s^2 = \frac{1}{3}$.
}
\end{figure}
In the code there are two options for the quark-gluon plasma's equation of state $\Peq(\ene)$: conformal and QCD. The former assumes a non-interacting gas of massless quarks and gluons:
\be
\label{eq:stefan}
    \Peq = \frac{\ene}{3} = \frac{g T^4}{\pi^2} \,,
\ee
where $T$ is the temperature and the degeneracy factor is
\be
\label{eq:degeneracy}
    g = \frac{\pi^4}{90}\left[2\left(N_c^2 {-} 1\right) + \frac{7}{2}N_c N_f \right]\,,
\ee
with $N_c = 3$ colors and $N_f = 3$ massless quark flavors (i.e. up, down and strange). This equation of state is primarily used to test the fluid dynamical simulation subject to either conformal Bjorken expansion or conformal Gubser expansion (see Secs.~\ref{S4.1} and \ref{S4.2}).

The QCD equation of state that we employ for more realistic simulations interpolates between the lattice QCD calculations provided by the HotQCD collaboration~\cite{Bazavov:2014pvz} and a hadron resonance gas composed of the hadrons that can be propagated in the hadronic afterburner code {\sc SMASH} \cite{Weil:2016zrk}.\footnote{%
    In hybrid model simulations of heavy-ion collisions, the equilibrium equation of state used in the hydrodynamic module must be consistent with that in the hadronic afterburner. Otherwise, serious violations of energy and momentum conservation can occur at the hadronization phase.}
Fig.~1 shows the energy density, equilibrium pressure and speed of sound as a function of temperature. The data table in the code covers the temperature range $T \in [0.05, 1.0]$\,GeV.

\subsection{Transport coefficients}
\label{S2.6}
\begin{figure}[t]
\centering
\includegraphics[width=0.9\linewidth]{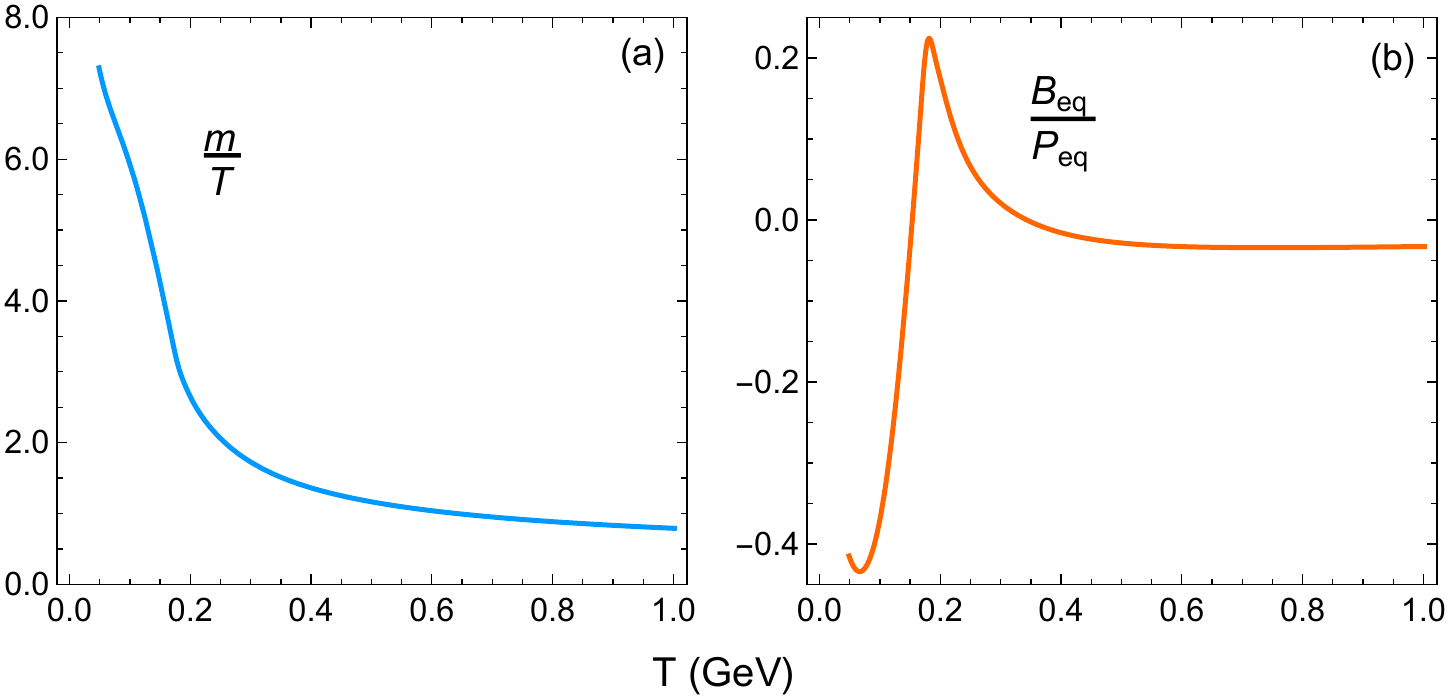}
\caption{(Color online)
\label{quasi}
    The quasiparticle mass to temperature ratio (blue, left) and equilibrium mean field normalized to the QCD equilibrium pressure (orange, right) as functions of temperature (for conformal systems, $m = 0$ and $B_\text{eq} = 0$).
}
\end{figure}
In this section we list the transport coefficients appearing in the relaxation equations in Sec.~\ref{S2.4}, for both QCD and conformal equations of state.

At the present moment, the transport coefficients of QCD matter (especially in the nonperturbative region $T \in [0.15, 0.5]$ GeV) are not explicitly known from first principles. Instead, we parametrize the shear and bulk viscosities $(\etas)(T)$ and $(\zetas)(T)$ as a function of temperature. In this work, we use the best-fit parametrization models from the JETSCAPE collaboration~\cite{Everett:2020yty,Everett:2020xug}. The relaxation times and anisotropic transport coefficients are computed with a quasiparticle kinetic theory model, whose equation of state is fitted to the QCD one. The kinetic model contains quasiparticles with a temperature-dependent mass $m(T)$ and an equilibrium mean-field $B_\text{eq}(T)$; these are shown in Figure~\ref{quasi}.

\subsubsection{Shear and bulk viscosities}
\label{S2.6.1}
\begin{figure}[t]
\centering
\includegraphics[width=0.9\linewidth]{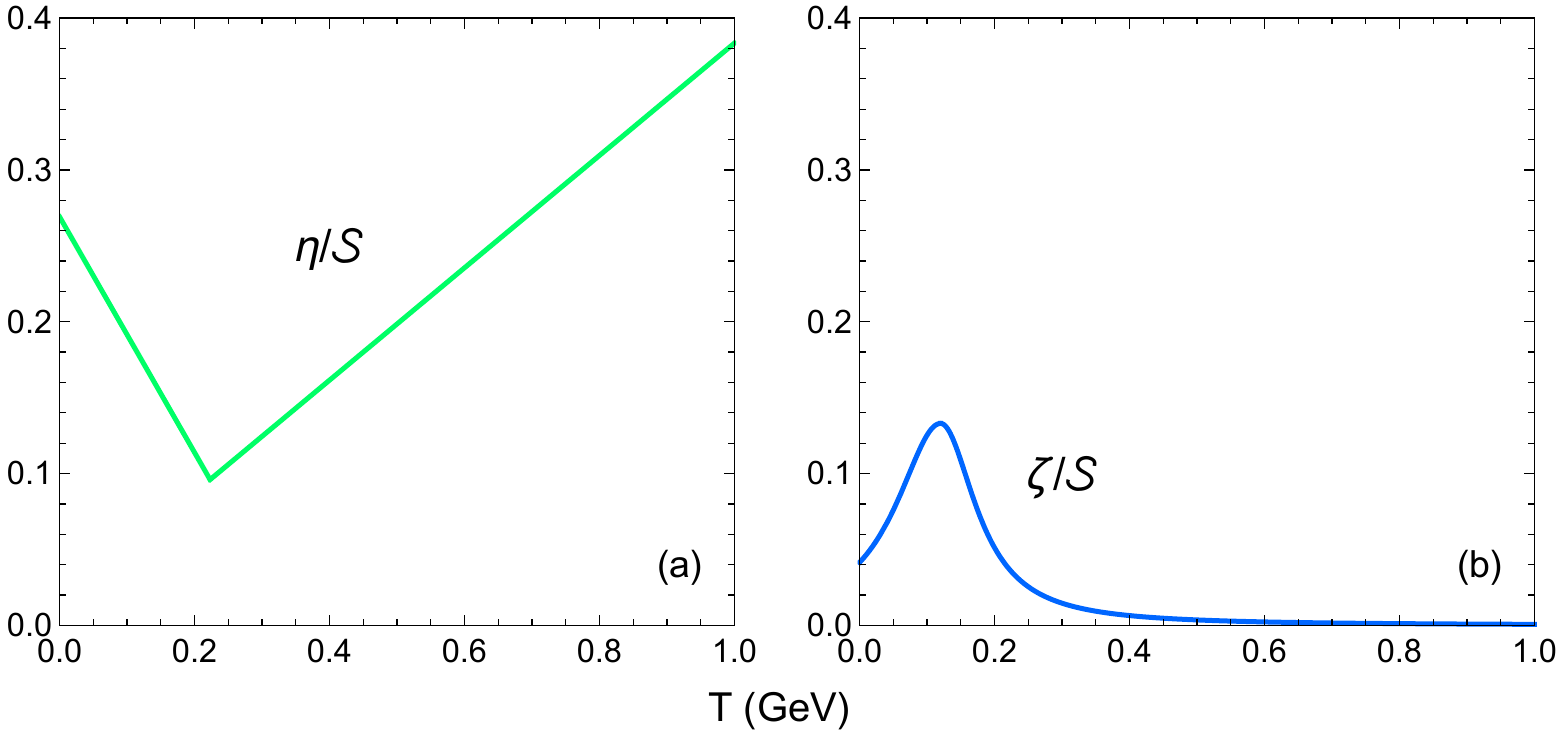}
\caption{(Color online)
\label{viscosities}
The temperature parametrization of $(\eta / \mathcal{S})(T)$ and $(\zeta / \mathcal{S})(T)$ used in this work. (For conformal systems, we set $\eta / \mathcal{S} = 0.2$ and $\zeta / \mathcal{S} = 0$.)
}
\end{figure}
The shear viscosity is modeled as a linear piecewise function with a kink at temperature $T_\eta$:
\be
\label{eq:etas}
    (\eta/\mathcal{S})(T) = (\etas)_\text{kink} + (T {-} T_\eta)\left(a_\text{low} \Theta(T_\eta {-} T) + a_\text{high} \Theta(T {-} T_\eta)\right)\,,
\ee
where $(\etas)_\text{kink}$ is the value of $\eta/\mathcal{S}$ at $T_\eta$, $a_\text{low}$ and $a_\text{high}$ are the left and right slopes, respectively, and $\Theta$ is the Heaviside step function. The bulk viscosity is parametrized as a skewed Cauchy distribution:
\be
\label{eq:zetas}
    (\zeta/\mathcal{S})(T) = \frac{(\zeta/\mathcal{S})_\text{max} \,\Lambda_\zeta(T)^2}{\Lambda_\zeta(T)^2 + (T {-} T_\zeta)^2}\,,
\ee
where $(\zeta/\mathcal{S})_\text{max}$ is the normalization factor, $T_\zeta$ is the peak temperature and
\be
    \Lambda_\zeta(T) = w_\zeta \left(1 + \lambda_\zeta \,\mathrm{sgn}(T{-}T_\zeta)\right)\,,
\ee
with $w_\zeta$ and $\lambda_\zeta$ being the width and skewness parameters, respectively.

The best-fit values for the viscosity parameters\footnote{%
    They correspond to a hybrid model whose particlization phase uses the 14-moment approximation for the $\delta f$ correction in the Cooper-Frye formula \cite{McNelis:2019auj, Everett:2020yty, Everett:2020xug} (the viscosity parameters used for the code validation tests in Sec.~\ref{S4} are slightly different because they were taken from an early draft of Ref.~\cite{Everett:2020xug}).}
are $(\etas)_\text{kink} = 0.096$, $T_\eta = 0.223$ GeV, $a_\text{low} = -0.776$ GeV$^{-1}$, $a_\text{high} = 0.37$ GeV$^{-1}$, $(\zetas)_\text{max} = 0.133$, $T_\zeta = 0.12$ GeV, $w_\zeta = 0.072$ GeV and $\lambda_\zeta = -0.122$ (see Table II in Ref.~\cite{Everett:2020xug}). The resulting temperature dependence of $\eta/\mathcal{S}$ and $\zeta/\mathcal{S}$ is shown in Figure~\ref{viscosities}.

For conformal systems, we fix the shear viscosity to $\eta / \mathcal{S} = 0.2$ and the bulk viscosity to $\zeta / \mathcal{S} = 0$.

\subsubsection{Shear and bulk relaxation times}
\label{S2.6.2}
%
\begin{figure}[t]
\centering
\includegraphics[width=0.9\linewidth]{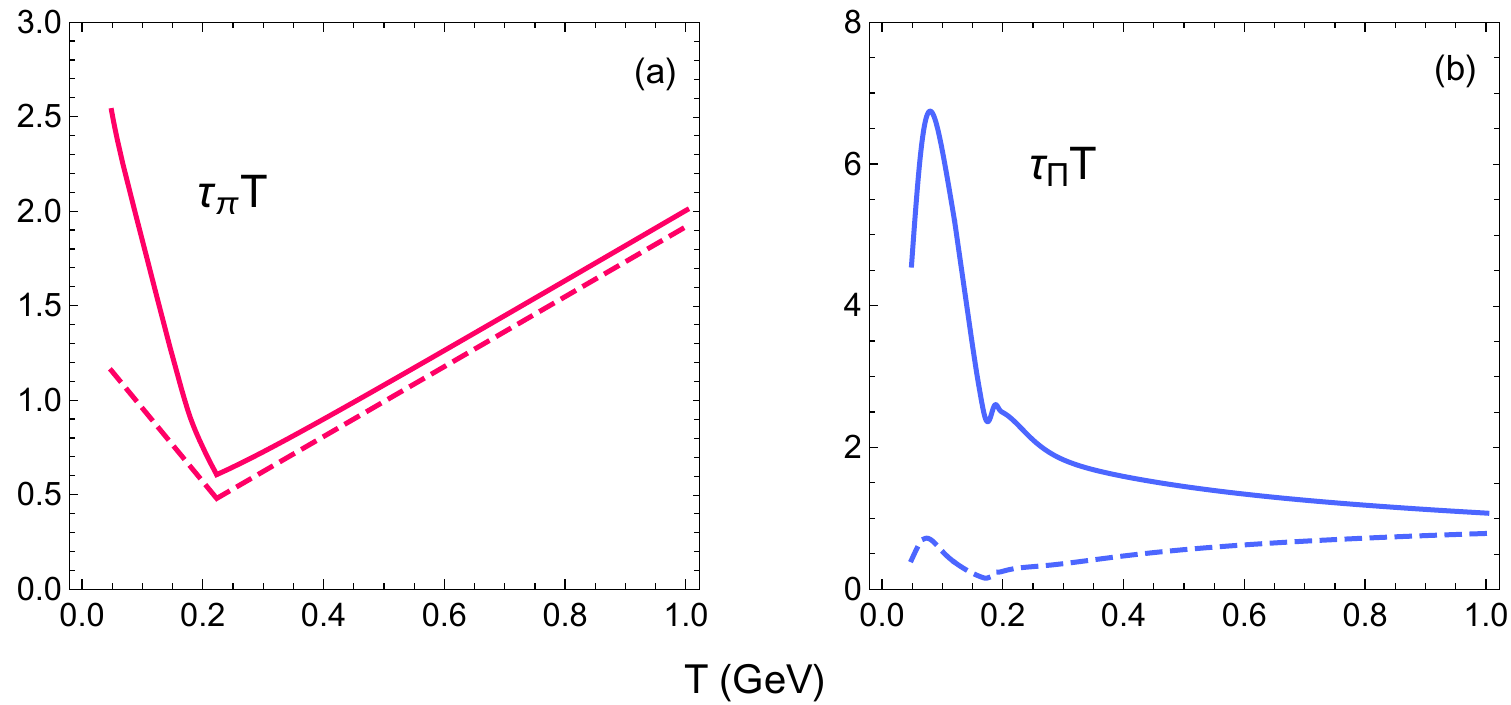}
\caption{(Color online)
\label{relaxation}
The dimensionless shear and bulk relaxation times computed with the quasiparticle kinetic model (solid color) and small-mass approximation (dashed color). (For conformal systems, $\tau_\pi T = 5\etas$ and $\tau_\Pi T = 0$.)
}
\end{figure}
In the quasiparticle kinetic model \cite{Tinti:2016bav, McNelis:2018jho}, the shear and bulk relaxation times are proportional to the shear and bulk viscosity, respectively,
\be
\label{eq:tau_r}
    \tau_\pi = \frac{\eta}{\beta_\pi} \,,\qquad
    \tau_\Pi = \frac{\zeta}{\beta_\Pi}\,,
\ee
where the viscosity to relaxation time ratios $\beta_\pi$ and $\beta_\Pi$ are
\bs
\label{eq:beta_r}
\beal
    \beta_\pi(T) &= \frac{\mathcal{I}_{32}}{T} \,,
\\
    \beta_\Pi(T) &= \frac{5}{3}\beta_\pi + c_s^2
    \big(m \frac{dm}{dT} \mathcal{I}_{11}-(\ene{+}\Peq)\big)\,,
\end{align}
\es
and $c_s^2$ and $\Peq$ are evaluated with the QCD equation of state. Here, we have defined the thermodynamic integrals
\be
\label{eq:Inq_moments}
    \mathcal{I}_{nq} = g\int_P \frac{(u\cdot p)^{n-2q}(-p \cdot \Delta \cdot p)^q f_\text{eq}}{(2q{+}1)!!} \,,
\ee
where $\int_P \dots \equiv \int d^3p / E_p \dots$ indicates integration with the Lorentz-invariant momentum space measure, $p^\mu$ is the quasiparticle momentum, $-p \cdot \Delta \cdot p$ is the square of its spatial LRF momentum, $u \cdot p = \sqrt{m^2(T) - p \cdot \Delta \cdot p}$ is its LRF energy, and $f_\text{eq} = \exp\left[-u\cdot p/T\right]$ is the local-equilibrium distribution function.

The normalized quasiparticle relaxation times $\tau_\pi T$ and $\tau_\Pi T$ are shown in Figure~\ref{relaxation}. These are compared to the relaxation times in standard viscous hydrodynamic models, where the kinetic transport coefficients are computed in the small-mass approximation $\bar{m} = m/T \ll 1$ and $dm/dT = 0$:
\bs
\label{eq:tau_r_small}
\beal
    \tau_\pi T &\approx 5 \etas + O\big(\bar{m}^2\big)\,,
\\
    \tau_\Pi T &\approx \frac{\zeta \,T}{15(\ene {+} \Peq)\big(\frac{1}{3} {-} c_s^2\big)^2} + O\big(\bar{m}^5\big)\,.
\end{align}
\es
One sees that the shear relaxation times are very similar to each other, except at low temperatures $T < 0.2$ GeV. On the other hand, the bulk relaxation times differ by about an order of magnitude for $T < 0.2$ GeV; this is due to the breakdown of the small-mass approximation, even at high temperatures $T \sim 1$ GeV. As a result, the evolution of the bulk viscous pressure $\Pi$ will be more greatly affected by critical slowing down in our anisotropic hydrodynamics model compared to standard viscous hydrodynamics.

For conformal kinetic plasmas ($m = 0$, $dm/dT = 0$), the shear relaxation time is $\tau_\pi = 5\eta / (\mathcal{S}T)$ and the bulk relaxation time is $\tau_\Pi = 0$.
\subsubsection{Anisotropic transport coefficients}
\label{S2.6.3}
Finally, we list the anisotropic transport coefficients~\cite{McNelis:2018jho, Molnar:2016vvu} that are coupled to the gradient source terms in the relaxation equations for the longitudinal pressure $\PL$,
\bs
\allowdisplaybreaks
\label{eq:pl_coeff}
\beal
    \bar{\zeta}^{L}_z & = \I_{2400} - 3 (\PL{+}B) + m\frac{dm}{d\ene}(\ene{+}\PL)\I_{0200} \,,
\\
    \bar{\zeta}^{L}_\perp & = \I_{2210} - \PL - B + m\frac{dm}{d\ene}(\ene{+}\Pperp)\I_{0200} \,,
\\
    \bar{\lambda}^{L}_{Wu} & = \frac{\I_{4410}}{\I_{4210}} + m\frac{dm}{d\ene}\I_{0200} \,,
\\
    \bar{\lambda}^{L}_{W\perp} & =  1 - \bar{\lambda}^{L}_{Wu}\,,
\\
    \bar{\lambda}^{L}_\pi & = \frac{\I_{4220}}{\I_{4020}} + m\frac{dm}{d\ene}\I_{0200} \,,
\end{align}
\es
the transverse pressure $\Pperp$,
\bs
\allowdisplaybreaks
\label{eq:pt_coeff}
\beal
    \bar{\zeta}^{\perp}_z & = \I_{2210} - \Pperp - B + m\frac{dm}{d\ene}(\ene{+}\PL)\I_{0010}\,,
\\
    \bar{\zeta}^{\perp}_\perp & = 2(\I_{2020} {-} \Pperp {-} B) + m\frac{dm}{d\ene}(\ene{+}\Pperp)\I_{0010}\,,
\\
    \bar{\lambda}^{\perp}_{W\perp} & = \frac{2 \,\I_{4220}}{\I_{4210}} + m\frac{dm}{d\ene}\I_{0010}\,,
\\
    \bar{\lambda}^{\perp}_{Wu} & = \bar{\lambda}^{\perp}_{W\perp} - 1 \,,
\\
    \bar{\lambda}^{\perp}_\pi & = 1 - \frac{3\,\I_{4030}}{\I_{4020}} - m\frac{dm}{d\ene}\I_{0010}\,,
\end{align}
\es
the longitudinal momentum diffusion current $\Wperp$,
\bs
\allowdisplaybreaks
\label{eqB3}
\beal
    \bar{\eta}^W_u & = \frac{1}{2}\big(\PL + B  - \I_{2210})\,,
\\
    \bar{\eta}^W_\perp & = \frac{1}{2}(\Pperp + B  - \I_{2210})\,,
\\
    \bar{\tau}^W_z & = \PL  - \Pperp \,,
\\
    \bar{\delta}^W_{W} & = \bar{\lambda}^W_{W\perp} - \frac{1}{2} + m\frac{dm}{d\ene} (\ene{+}\Pperp) \left(\frac{\I_{2210}}{\I_{4210}}\right) \,,
\\
    \bar{\lambda}^W_{Wu} & = 2 - \frac{\I_{4410}}{\I_{4210}} - m\frac{dm}{d\ene} (\ene{+}\PL) \left(\frac{\I_{2210}}{\I_{4210}}\right) \,,
\\
    \bar{\lambda}^W_{W\perp} & = \frac{2 \,\I_{4220}}{\I_{4210}} - 1,
\\
    \bar{\lambda}^W_{\pi u} & = \frac{\I_{4220}}{\I_{4020}} \,,
\\
    \bar{\lambda}^W_{\pi\perp} & = \bar{\lambda}^W_{\pi u} - 1 \,,
\end{align}
\es
and the transverse shear stress tensor $\piperp$:
\bs
\allowdisplaybreaks
\label{eqB4}
\beal
    \bar{\eta}_\perp & = \Pperp^{(k)}  - \I_{2020},
\\
    \bar{\delta}^\pi_\pi & = \frac{3}{4} \bar{\tau}^\pi_\pi + \frac{1}{2} - m\frac{dm}{d\ene} (\ene{+}\Pperp) \left(\frac{\I_{2020}}{\I_{4020}}\right),
\\
    \bar{\tau}^\pi_\pi & =  2 - \frac{4\,\I_{4030}}{\I_{4020}},
\\
    \bar{\lambda}^\pi_{\pi} & = \bar{\lambda}^W_{\pi u} - 1 + m\frac{dm}{d\ene} (\ene{+}\PL) \left(\frac{\I_{2020}}{\I_{4020}}\right),
\\
    \bar{\lambda}^\pi_{Wu} & =  \bar{\lambda}^W_{W\perp} -  1,
\\
    \bar{\lambda}^\pi_{W\perp} & = \bar{\lambda}^\pi_{Wu} + 2.
\end{align}
\es
We define the anisotropic integrals
\be
    \mathcal{I}_{nrqs} = \frac{g}{(2\pi)^3}\int_P \frac{(u \cdot p)^{n-r-2q}(-z\cdot p)^r (-p \cdot \Xi \cdot p)^q E_a^s f_a}{(2q)!!} \,,
\ee
where
\be
    E_a = \sqrt{m^2(T) - (p \cdot \Xi \cdot p) / \alpha_\perp^2 + (-z\cdot p)^2/\alpha_L^2}
\ee
and $f_a = e^{-E_a/\Lambda}$ is the leading-order anisotropic distribution function.

In addition to the effective temperature $\Lambda$ this anisotropic distribution $f_a$ depends (through $E_a$) on two momentum anisotropy parameters $\alpha_L$ and $\alpha_\perp$, which deform the longitudinal and transverse momentum space, respectively. In order to compute the non-conformal anisotropic transport coefficients at each time step of the simulation, we propagate $(\Lambda, \alpha_\perp, \alpha_L)$ as inferred variables. They are obtained from the kinetic part of $\ene$, $\PL$ and $\Pperp$ in the quasiparticle kinetic model:
\bs
\beal
\ene^{(k)} &= \ene - B \,,\\
\PL^{(k)} &= \PL + B \,,\\
\Pperp^{(k)} &= \Pperp + B \,,
\end{align}
\es
where $\ene^{(k)} = \mathcal{I}_{2000}$ is the kinetic contribution to the energy density (i.e. the total energy density minus the mean field contribution \cite{McNelis:2018jho}), $\PL^{(k)} = \mathcal{I}_{2200}$ is the kinetic longitudinal pressure, and $\Pperp^{(k)} = \mathcal{I}_{2010}$ is the kinetic transverse pressure. The numerical method which we use to solve these equations will be discussed in Sec.~\ref{S3.4}.

The anisotropic transport coefficients in the conformal limit are listed in Appendix~\ref{appc}.
\section{Numerical scheme}
\label{S3}
In this section we discuss the numerical implementation of the hydrodynamic equations in the code. The dynamical and inferred variables are evolved on an $(N_x{+}4) \times (N_y{+}4) \times (N_\eta{+}4)$ Eulerian grid, where $N_x$, $N_y$ and $N_\eta$ are the number of physical grid points along each spatial direction.\footnote{For longitudinally boost-invariant systems, the number of spacetime rapidity points is set to $N_\eta = 1$.} A grid point with cell index $(i,j,k)$ that corresponds to the lower left front corner of a fluid cell, has a spatial position
\bs
\beal
x_i &= \big[i - 2 - \frac{1}{2}(N_x {-} 1)\big] \Delta x \,,\\
y_j &= \big[j - 2 - \frac{1}{2}(N_y {-} 1)\big] \Delta y \,,\\
\eta_{s,k} &= \big[k - 2 - \frac{1}{2}(N_\eta {-} 1)\big] \Delta \eta_s \,,
\end{align}
\es
where $\Delta x$, $\Delta y$ and $\Delta \eta_s$ are the lattice spacings. Physical fluid cells have indices $i\in[2,N_x{+}1]$, $j\in[2,N_y{+}1]$ and $k\in[2,N_\eta{+}1]$. The numerical algorithm also requires six sets of ghost cells with depth two, which neighbor the physical grid's faces (for an illustration, see Fig.~3 in Ref.~\cite{Bazow:2016yra}). The boundary conditions that we impose on the ghost cells neighboring the two $(y,\eta_s)$ faces are\footnote{%
    In conformal anisotropic hydrodynamics, the energy density $\ene$ also requires ghost cell boundary conditions.}
\bs
\beal
\boldsymbol{q}_{0,j,k} &= \boldsymbol{q}_{1,j,k} = \boldsymbol{q}_{2,j,k} \,,\\
\boldsymbol{u}_{0,j,k} &= \boldsymbol{u}_{1,j,k} = \boldsymbol{u}_{2,j,k} \,,\\
\boldsymbol{q}_{N_x{+}2,j,k} &= \boldsymbol{q}_{N_x{+}3,j,k} = \boldsymbol{q}_{N_x{+}1 ,j,k} \,,\\
\boldsymbol{u}_{N_x{+}2,j,k} &= \boldsymbol{u}_{N_x{+}3,j,k} = \boldsymbol{u}_{N_x{+}1 ,j,k}\,,
\end{align}
\es
and similarly for $(x,\eta_s)$ and $(x,y)$ faces after permuting the grid indices and replacing $N_x \to N_y$ or $N_\eta$.

The dynamical variables $\boldsymbol{q}$ are updated using a two-stage Runge--Kutta (RK2) scheme, where the time derivatives are evaluated with the Kurganov--Tadmor (KT) algorithm~\cite{Schenke:2010nt, Bazow:2016yra,  Kurganov:2000}. After each intermediate Euler step in the RK2 scheme, we reconstruct the inferred variables from the dynamical variables. In addition, we regulate the residual shear stresses $\Wperp$ and $\piperp$ and the mean-field $B$. The code also provides the user with the option of using an adaptive time step to capture the fluid's longitudinal dynamics at very early times; this will be discussed at the end of the section.

\subsection{Kurganov-Tadmor algorithm}
\label{S3.1}
The time derivative of the dynamical variables $\partial_\tau\boldsymbol{q}$ in the partial differential equations~\eqref{eq:KT_eqs} can be computed on the Eulerian grid at any time $\tau$ using the KT algorithm~\cite{Kurganov:2000}:
\be
\label{eq:KT_algorithm}
\begin{split}
(\partial_\tau \boldsymbol{q})_{ijk} = & -\frac{\boldsymbol{H}^x_{i{+}\half,j,k} {-} \boldsymbol{H}^x_{i{-}\half,j,k}}{\Delta x} - \frac{\boldsymbol{H}^y_{i,j{+}\half,k} {-} \boldsymbol{H}^y_{i,j{-}\half,k}}{\Delta y} - \frac{\boldsymbol{H}^\eta_{i,j,k{+}\half} {-} \boldsymbol{H}^\eta_{i,j,k{-}\half}}{\Delta \eta_s}\\&
+ \boldsymbol{S}_{ijk}\big(\tau, \boldsymbol{q}_{ijk}, \boldsymbol{u}_{ijk},\ene_{ijk}, (\partial_m \boldsymbol{q})_{ijk}, (\partial_\mu \boldsymbol{u})_{ijk}\big)\,,
\end{split}
\ee
where the numerical fluxes evaluated at the left and right faces of a staggered cell centered around the grid point $(i,j,k)$ are
\be
\label{eq:flux}
\boldsymbol{H}^x_{i\pm\half,j,k} = \frac{1}{2}\Big[\boldsymbol{F}^{x+}_{i\pm\half,j,k} + \boldsymbol{F}^{x-}_{i\pm\half,j,k} - s^x_{i\pm\half,j,k}\big(\boldsymbol{q}^+_{i\pm\half,j,k} {-\,} \boldsymbol{q}^-_{i\pm\half,j,k}\big)\Big],
\ee
and similarly for $\boldsymbol{H}^y_{i,j\pm\frac{1}{2},k}$ and $\boldsymbol{H}^\eta_{i,j,k\pm\frac{1}{2}}$ after permuting the $\pm\half$ in the grid indices and the corresponding spatial component (i.e. $x \to y$ or $\eta$).

The first two terms in Eq.~\eqref{eq:flux} take the average of the currents extrapolated to the staggered cell face $(i{+}\half,j,k)$ (or $(i{-}\half,j,k)$)  from the left ($-$) and right ($+$) sides. A first-order expression for the extrapolated currents can be computed using the chain rule:
\bs
\allowdisplaybreaks
\label{eq:Fx_ex}
\beal
\boldsymbol{F}^{x-}_{i+\frac{1}{2},j,k} &= \boldsymbol{F}^x_{ijk} + \frac{\Delta x}{2}\Big[(\partial_x v^x)_{ijk} \boldsymbol{q}_{ijk} + v^x_{ijk} (\partial_x \boldsymbol{q})_{ijk} \Big] \,,\\
\boldsymbol{F}^{x+}_{i+\frac{1}{2},j,k} &= \boldsymbol{F}^x_{i+1,j,k} - \frac{\Delta x}{2}\Big[(\partial_x v^x)_{i+1,j,k} \boldsymbol{q}_{i+1,j,k} + v^x_{i+1,j,k}(\partial_x \boldsymbol{q})_{i+1,j,k}\Big] \,,\\
\boldsymbol{F}^{x-}_{i-\frac{1}{2},j,k} &= \boldsymbol{F}^x_{i-1,j,k} + \frac{\Delta x}{2}\Big[(\partial_x v^x)_{i-1,j,k}\boldsymbol{q}_{i-1,j,k} + v^x_{i-1,j,k}(\partial_x \boldsymbol{q})_{i-1,j,k}\Big]\,,\\
\boldsymbol{F}^{x+}_{i-\frac{1}{2},j,k} &= \boldsymbol{F}^x_{ijk} - \frac{\Delta x}{2}\Big[(\partial_x v^x)_{ijk}\boldsymbol{q}_{ijk} + v^x_{ijk}(\partial_x \boldsymbol{q})_{ijk}\Big] \,,
\end{align}
\es
where $\boldsymbol{F}^x_{ijk} = v^x_{ijk} \boldsymbol{q}_{ijk}$. Similarly, the extrapolated currents $\boldsymbol{F}^{y+}_{i,j\pm\frac{1}{2},k}$, $\boldsymbol{F}^{y-}_{i,j\pm\frac{1}{2},k}$, $\boldsymbol{F}^{\eta+}_{i,j,k\pm\frac{1}{2}}$ and $\boldsymbol{F}^{\eta-}_{i,j,k\pm\frac{1}{2}}$ at the remaining faces of the staggered cell are obtained by permuting the $\pm\half$ (or $\pm 1$) in the grid indices, the spatial components and derivatives, and the lattice spacing.

The final term in Eq.~\eqref{eq:flux} takes into account the wave propagation of the discontinuities $\boldsymbol{q}^+_{i\pm\half,j,k} - \boldsymbol{q}^-_{i\pm\half,j,k}$ at a finite speed~\cite{Kurganov:2000}. We define the local propagation speed component $s^x_{i\pm\half,j,k}$ at the staggered cell faces $(i{\pm}\half,j,k)$ as
\be
s^x_{i\pm\half,j,k} = \max\big(|v^{x-}_{i\pm\half,j,k}|, |v^{x+}_{i\pm\half,j,k}|\big)\,,
\ee
where the extrapolated velocities are
\bs
\label{eq:vx_ex}
\beal
v^{x-}_{i+\frac{1}{2},j,k} &= v^x_{ijk} + \frac{\Delta x}{2}(\partial_x v^x)_{ijk} \,,\\
v^{x+}_{i+\frac{1}{2},j,k} &= v^x_{i+1,j,k} - \frac{\Delta x}{2}(\partial_x v^x)_{i+1,j,k} \,,\\
v^{x-}_{i-\frac{1}{2},j,k} &= v^x_{i-1,j,k} + \frac{\Delta x}{2}(\partial_x v^x)_{i-1,j,k}\,,\\
v^{x+}_{i-\frac{1}{2},j,k} &= v^x_{ijk} - \frac{\Delta x}{2}(\partial_x v^x)_{ijk} \,.
\end{align}
\es
The discontinuities $\boldsymbol{q}^+_{i\pm\half,j,k} - \boldsymbol{q}^-_{i\pm\half,j,k}$ propagating from the staggered cell faces $(i{\pm}\half,j,k)$  depend on the extrapolated dynamical variables
\bs
\label{eq:q_ex}
\beal
\boldsymbol{q}^{-}_{i+\frac{1}{2},j,k} &= \boldsymbol{q}_{ijk} + \frac{\Delta x}{2}(\partial_x\boldsymbol{q})_{ijk} \,,\\
\boldsymbol{q}^{+}_{i+\frac{1}{2},j,k} &= \boldsymbol{q}_{i+1,j,k} - \frac{\Delta x}{2}(\partial_x \boldsymbol{q})_{i+1,j,k} \,,\\
\boldsymbol{q}^{-}_{i-\frac{1}{2},j,k} &= \boldsymbol{q}_{i-1,j,k} + \frac{\Delta x}{2}(\partial_x \boldsymbol{q})_{i-1,j,k}\,,\\
\boldsymbol{q}^{+}_{i-\frac{1}{2},j,k} &= \boldsymbol{q}_{ijk} - \frac{\Delta x}{2}(\partial_x \boldsymbol{q})_{ijk} \,.
\end{align}
\es
The formulae for the local propagation speed components $s^y_{i,j\pm\half,k}$ and $s^\eta_{i,j,k\pm\half}$, extrapolated velocities $v^{y+}_{i,j\pm\half,k}$, $v^{y-}_{i,j\pm\half,k}$, $v^{\eta+}_{i,j,k\pm\half}$ and $v^{\eta-}_{i,j,k\pm\half}$, and extrapolated dynamical variables $\boldsymbol{q}^+_{i,j\pm\half,k}$, $\boldsymbol{q}^-_{i,j\pm\half,k}$, $\boldsymbol{q}^+_{i,j,k\pm\half}$ and $\boldsymbol{q}^-_{i,j,k\pm\half}$ are analogous.

The numerical spatial derivatives appearing in the extrapolated quantities~(\ref{eq:Fx_ex}a-d),~(\ref{eq:vx_ex}a-d) and~(\ref{eq:q_ex}a-d)  are computed with a minmod flux limiter~\cite{Kurganov:2000}:
\bs
\beal
(\partial_x \boldsymbol{q})_{ijk} &= \mathcal{M}\Big(\Theta \, \frac{\boldsymbol{q}_{ijk} - \boldsymbol{q}_{i-1,j,k}}{\Delta x}, \frac{\boldsymbol{q}_{i+1,j,k} - \boldsymbol{q}_{i-1,j,k}}{2\Delta x},\Theta \, \frac{\boldsymbol{q}_{i+1,j,k} - \boldsymbol{q}_{ijk}}{\Delta x}\Big) \,,\\
(\partial_x v^x)_{ijk} &= \mathcal{M}\Big(\Theta \, \frac{v^x_{ijk} - v^x_{i-1,j,k}}{\Delta x}, \frac{v^x_{i+1,j,k} - v^x_{i-1,j,k}}{2\Delta x},\Theta \, \frac{v^x_{i+1,j,k} - v^x_{ijk}}{\Delta x}\Big) \,,
\end{align}
\es
where
\be
\mathcal{M}(a,b,c) = \mathrm{minmod}(a, \mathrm{minmod}(b,c)) \,,
\ee
with
\be
\mathrm{minmod}(a,b) = \frac{\mathrm{sgn}(a) + \mathrm{sgn}(b)}{2} \times \min(|a|, |b|)\,;
\ee
the flux limiter parameter is set to $\Theta = 1.8$~\cite{Marrochio:2013wla}. The flux limiter derivatives $(\partial_y \boldsymbol{q})_{ijk}$, $(\partial_y v^y)_{ijk}$, $(\partial_\eta \boldsymbol{q})_{ijk}$ and $(\partial_\eta v^\eta)_{ijk}$ are analogous.

In contrast, the numerical spatial derivatives appearing in the source terms $\boldsymbol{S}_{ijk}$ are approximated with second-order central differences~\cite{Bazow:2016yra,Pang:2018zzo}:
\bs
\beal
(\partial_x \boldsymbol{q})_{ijk} &= \frac{\boldsymbol{q}_{i+1,j,k} - \boldsymbol{q}_{i-1,j,k}}{2\Delta x} \,,\\
(\partial_x \boldsymbol{u})_{ijk} &= \frac{\boldsymbol{u}_{i+1,j,k} - \boldsymbol{u}_{i-1,j,k}}{2\Delta x} \,,
\end{align}
\es
and similarly for $(\partial_y \boldsymbol{q})_{ijk}$, $(\partial_y \boldsymbol{u})_{ijk}$, $(\partial_\eta \boldsymbol{q})_{ijk}$ and $(\partial_\eta \boldsymbol{u})_{ijk}$. The fluid velocity's time derivative $(\partial_\tau \boldsymbol{u})_{ijk}$ also appears in the source terms; its evaluation will be discussed the next subsection.
\subsection{Two-stage Runge--Kutta scheme}
Because the KT algorithm~\eqref{eq:KT_algorithm} admits a semi-discrete form~\cite{Kurganov:2000} (i.e. the time derivative is continuous while the spatial derivatives are discrete), we can combine it with an RK2 ODE solver to evolve the system in time~\cite{Schenke:2010nt,Bazow:2016yra}. Given the dynamical variables $\boldsymbol{q}_{n,ijk} \equiv \boldsymbol{q}_{ijk}(\tau_n)$ at time $\tau = \tau_n$ (discrete times are labeled with index $n$), we evolve the system one time step $\Delta\tau_n$ with an intermediate Euler step (omitting the spatial indices):
\be
\label{eq:intermediate_step}
\boldsymbol{q}_{\,\text{I},n+1} = \boldsymbol{q}_{n} + \Delta \tau _n \boldsymbol{E}(\tau_n, \boldsymbol{q}_n, \boldsymbol{u}_n,\ene_n; \boldsymbol{u}_{n-1}, \Delta\tau_{n-1}) \,,
\ee
where $\boldsymbol{E} = \partial_\tau \boldsymbol{q}$ is evaluated with r.h.s of Eq.~\eqref{eq:KT_algorithm}. The time derivative $\partial_\tau \boldsymbol{u}$ in the source terms is approximated with a first-order backward difference~\cite{Bazow:2016yra,Schenke:2010rr}:
\be
(\partial_\tau \boldsymbol{u})_n = \frac{ \boldsymbol{u}_n - \boldsymbol{u}_{n-1}}{\Delta\tau_{n-1}}\,,
\ee
where $\boldsymbol{u}_{n-1}$ is the previous fluid velocity and $\Delta\tau_{n-1}$ is the previous time step.\footnote{%
    At the start of the hydrodynamic simulation, $n=0$ or $\tau = \tau_0$, the previous time step $\Delta \tau_{n-1}$ is set to the current time step $\Delta \tau_n$. Unless stated otherwise, we also initialize the previous fluid velocity as $\boldsymbol{u}_{n-1} = \boldsymbol{u}_n$.}
From the intermediate variables \eqref{eq:intermediate_step}, we reconstruct the inferred variables $(\ene_{\text{I},n+1}, \boldsymbol{u}_{\text{I},n+1})$ as well as the anisotropic variables ($\Lambda_{\text{I},n+1}$ $\alpha_{\perp,\text{I},n+1}$, $\alpha_{L,\text{I},n+1}$), as described in Secs.~\ref{S3.3} and \ref{S3.4} below. Afterwards, we regulate the mean field $B$ and residual shear stresses $\Wperp$ and $\piperp$ in $\boldsymbol{q}_{\,\text{I},n+1}$ (see Sec.~\ref{S3.5}) and set the ghost cell boundary conditions for $\boldsymbol{q}_{\,\text{I},n+1}$ and $\boldsymbol{u}_{\text{I},n+1}$.

Next, we evolve the system with a second intermediate Euler step
\be
\label{eq:2nd_intermediate_step}
\boldsymbol{Q}_{n+2} = \boldsymbol{q}_{\,\text{I},n+1} + \Delta \tau _n \boldsymbol{E}(\tau_n {+} \Delta\tau_n, \boldsymbol{q}_{\,\text{I},n+1}, \boldsymbol{u}_{\text{I},n+1},\ene_{\text{I},n+1}; \boldsymbol{u}_n, \Delta\tau_n) \,,
\ee
where the fluid velocity's time derivative is now evaluated as
\be
(\partial_\tau \boldsymbol{u})_{\text{I},n+1} = \frac{ \boldsymbol{u}_{\text{I},n+1} - \boldsymbol{u}_n}{\Delta\tau_n}\,.
\ee
In the RK2 scheme, we average the two intermediate Euler steps in $\boldsymbol{Q}_{n+2}$ to update the dynamical variables at $\tau = \tau_n + \Delta\tau_n$:
\be
\label{eq:RK2}
\boldsymbol{q}_{n+1} = \frac{\boldsymbol{q}_n + \boldsymbol{Q}_{n+2}}{2}\,.
\ee
From this, we update the inferred variables ($\ene_{n+1}$, $\boldsymbol{u}_{n+1}$) and ($\Lambda_{n+1}$, $\alpha_{\perp,n+1}$, $\alpha_{L,n+1}$) and regulate the residual shear stresses and mean field. Finally, we set the ghost cell boundary conditions for $\boldsymbol{q}_{n+1}$ and $\boldsymbol{u}_{n+1}$ and proceed with the next RK2 iteration.

\subsection{Reconstructing the energy density and fluid velocity}
\label{S3.3}
Given the hydrodynamic variables $T^{\tau\mu}$, along with $\PL$, $\Pperp$, $\Wperp$ and $\pi_\perp^{\tau\mu}$, we can reconstruct the energy density and fluid velocity from Eq.~\eqref{eq:Ttaumu}. The solution for the energy density is~\cite{McNelis:2018jho}
\be
\label{eq:energy_recon}
    \ene = M^{\tau} - \mathcal{L}^{\tau\tau} - \frac{(M^x)^2 {+} (M^y)^2}{M^\tau {+} \Pperp {-} \mathcal{L}^{\tau\tau}} - \frac{(\tau M^\eta)^2 (M^\tau {+} \Pperp {-} \mathcal{L}^{\tau\tau})}{(M^\tau {+} \PL)^2}\,,
\ee
where $\mathcal{L}^{\tau\tau} = \Delta \mathcal{P}(z^\tau)^2$ was defined earlier and
\be
    M^\mu = T^{\tau\mu} - 2 W^{(\tau}_{\perp z} z^{\mu)} - \pi_\perp^{\tau\mu}\,.
\ee
The reconstruction formula \eqref{eq:energy_recon} also requires the components $z^\tau$, $z^\eta$, which can be expressed in terms of hydrodynamic variables as follows:\footnote{%
    Writing $F = u^\eta/u^\tau$ implies that the argument $1-(\tau F)^2$ in Eq.~\eqref{eq:z_reconstruct} is always positive.
}
\be
\label{eq:z_reconstruct}
    z^\tau = \frac{\tau F}{\sqrt{1 - (\tau F)^2}} \,,\qquad
    z^\eta = \frac{1}{\tau\sqrt{1 - (\tau F)^2}}\,.
\ee
Here
\be
F = \frac{A - B \sqrt{1 + \tau^2(B^2{-}A^2)}}{1+(\tau B)^2}\,,
\ee
with $A = K^\eta / (K^\tau {+} \PL)$ and $B = W_{\perp z}^\tau / (\tau(K^\tau {+} \PL))$ where $K^\mu = T^{\tau\mu} {-} \pi_\perp^{\tau\mu}$.

In the cold, dilute regions surrounding the fireball (and, occasionally, in cold spots within the fluctuating fireball), the energy density can become much smaller than the freezeout energy density $\ene_\text{sw}$.\footnote{%
    In this work, we construct a particlization hypersurface of constant energy density $\ene_\text{sw} = 0.116$ GeV/fm$^3$, which corresponds to the switching temperature $T_\text{sw} = 0.136$ GeV. The lowest value used for the switching temperature in the JETSCAPE SIMS analysis is $T_\text{sw} = 0.135$ GeV~\cite{Everett:2020yty,Everett:2020xug}.}
While these regions are not phenomenologically important, hydrodynamic simulations are susceptible to crashing there without intervention \cite{Shen:2014vra, Bazow:2016yra, Denicol:2018wdp}. To prevent this, we regulate the energy density with the formula
\be
\label{eq:energy_reg}
    \ene \leftarrow \ene_+ + \ene_\text{min}\, e^{-\ene_+ / \ene_\text{min}}\,,
\ee
where $\ene_\text{min}$ is the minimum energy density allowed in the Eulerian grid and $\ene_+ = \max(0, \ene)$. For $\ene \gg \ene_\text{min}$, the regulation has virtually no effect on the energy density. As $\ene \to 0$, however, the energy density is smoothly regulated to $\ene_\text{min}$. Ideally, $\ene_\text{min}$ should be the lower limit of our QCD equation of state table $\ene_\text{low} = 3.5 \times 10^{-4}$ GeV/fm$^3$. However, we find that the code evolving non-conformal anisotropic hydrodynamics with fluctuating initial conditions encounters fewer technical difficulties if we instead use the larger value $\ene_\text{min} = 0.02$ GeV/fm$^3$, which is still about six times smaller than our choice for $\ene_\text{sw}$.\footnote{%
    The constraint $\ene \geq \ene_\text{min}$ is not imposed for conformal systems.}
We checked that the regulation scheme~\eqref{eq:energy_reg} has little to no impact on the fluid's dynamics in regions where $\ene \sim \ene_\text{sw}$ or larger.

After regulating the energy density, we reconstruct the fluid velocity's spatial components:
\bs
\allowdisplaybreaks
\beal
u^x &= \frac{M^x}{\sqrt{(\ene+\Pperp)(M^\tau + \Pperp - \mathcal{L}^{\tau\tau})}} \,,\\
u^y &= \frac{M^y}{\sqrt{(\ene+\Pperp)(M^\tau + \Pperp - \mathcal{L}^{\tau\tau})}} \,,\\
u^\eta &= F \sqrt{\frac{M^\tau + \Pperp - \mathcal{L}^{\tau\tau}}{\ene+\Pperp}} \,.
\end{align}
\es
Because of the regulation~\eqref{eq:energy_reg}, there will be inconsistencies between the inferred variables ($\ene$, $\boldsymbol{u}$) and the components $T^{\tau\mu}$ but only in the dilute cold regions.
%
%
\subsection{Reconstructing the anisotropic variables}
\label{S3.4}

To compute the non-conformal anisotropic transport coefficients in Sec.~\ref{S2.6.3}, we need to reconstruct the anisotropic variables
\be
  \boldsymbol{X} = \left(
    \begin{array}{c}
  \Lambda \\ \alpha_\perp \\ \alpha_L
  \end{array}
  \right)
\ee
from $\ene$, $\PL$, $\Pperp$, $B$ and $m(\ene)$, by solving the system of equations \cite{McNelis:2018jho}
\be
\label{eq:nonlinear_system}
    \boldsymbol{f}(\boldsymbol{X}) = \boldsymbol{0}\,,
\ee
where
\be
\label{eq:aniso_root_equation}
\boldsymbol{f}(\boldsymbol{X}) = \left(
  \begin{array}{c}
  \I_{2000}(\boldsymbol{X}) - \ene + B \\ \I_{2200}(\boldsymbol{X}) - \PL - B \\ \I_{2010}(\boldsymbol{X})-\Pperp-B
  \end{array}
  \right).
\ee
We solve these nonlinear equations numerically using Newton's method. Taking the anisotropic variables prior to the intermediate Euler step~\eqref{eq:intermediate_step} as our initial guess\footnote{
    At $\tau = \tau_0$ we initialize the anisotropic variables ($\Lambda_0$, $\alpha_{\perp,0}$, $\alpha_{L,0}$) by iterating $\boldsymbol{X}_g = (T, 1, 1)^T$ (where $T$ inside the parentheses is the temperature).}
$\boldsymbol{X}_g = (\Lambda_n,\alpha_{\perp,n}, \alpha_{L,n})^T$, we iterate $\boldsymbol{X}$ along the direction given by the Newton step
\be
    \Delta \boldsymbol{X} = - \boldsymbol{J}^{-1} \boldsymbol{f}\,,
\ee
where $\boldsymbol{J}^{-1}$ is the inverse of the Jacobian
\be
\boldsymbol{J} = \frac{\partial \boldsymbol{f}}{\partial \boldsymbol{X}} =    \left(
  \begin{array}{c c c}
      \dfrac{\I_{2001}(\boldsymbol{X})}{\Lambda^2} \, & \,\dfrac{2\,\I_{401-1}(\boldsymbol{X})}{\Lambda \alpha_\perp^3} \, & \, \dfrac{\I_{420-1}(\boldsymbol{X})}{\Lambda \alpha_L^3} \\ \\
              \dfrac{\I_{2201}(\boldsymbol{X})}{\Lambda^2}\, & \, \dfrac{2\, \I_{421-1}(\boldsymbol{X})}{\Lambda \alpha_\perp^3} \,
                                                                          & \, \dfrac{\I_{440-1}(\boldsymbol{X})}{\Lambda \alpha_L^3} \\ \\
        \dfrac{\I_{2011}(\boldsymbol{X})}{\Lambda^2} \, & \, \dfrac{4\,\I_{402-1}(\boldsymbol{X})}{\Lambda \alpha_\perp^3} \,                                                      & \, \dfrac{\I_{421-1}(\boldsymbol{X})}{\Lambda \alpha_L^3} \\
  \end{array}
  \right).
\ee
Specifically, we iterate the solution as
\be
\boldsymbol{X} \leftarrow \boldsymbol{X} + \lambda \,\Delta\boldsymbol{X}\,,
\ee
where $\lambda \in [0,1]$ is a partial step that is optimized with a line-backtracking algorithm to improve the global convergence of Newton's method~\cite{Press:1992:NRC:148286}. This procedure is repeated until $\boldsymbol{X}$ has converged to the solution of Eq.~\eqref{eq:nonlinear_system}.\footnote{%
    In our simulation tests, we found several instances when Newton's method failed to converge near the edges of the Eulerian grid. Whenever that happens we simply set the anisotropic variables to the initial guess $\boldsymbol{X}_g$.}

For conformal systems ($B = 0$, $m = 0$, $\alpha_\perp = 1$), the anisotropic variables $\Lambda$ and $\alpha_L$ are much easier to solve. The system of equations \eqref{eq:aniso_root_equation} reduces to
\bs
\label{eq:aniso_eqs_conformal}
\beal
    \ene &= \frac{3 g \Lambda^4 \mathcal{R}_{200}(\alpha_L)}{2\pi^2} \,,
\\
    \PL &= \frac{g \Lambda^4 \mathcal{R}_{220}(\alpha_L)}{2\pi^2} \,,
\end{align}
\es
where the functions $\mathcal{R}_{200}$ and $\mathcal{R}_{220}$ are listed in Appendix~\ref{appc}. We numerically invert the longitudinal pressure to energy density ratio for $\alpha_L$:
\be
\label{eq:aL_conformal}
    \frac{\mathcal{R}_{220}(\alpha_L)}{\mathcal{R}_{200}(\alpha_L)} = \frac{\PL}{\ene} \,.
\ee
From this, we can evaluate $\Lambda$ as
\be
\label{eq:Lambda_conformal}
    \Lambda = \bigg(\frac{\pi^2\ene}{3g\mathcal{R}_{200}(\alpha_L)}\bigg)^{1/4}\,.
\ee
Because the solutions~\eqref{eq:aL_conformal} and~\eqref{eq:Lambda_conformal} do not require the previous values for $\Lambda$ and $\alpha_L$ as input, we do not need to store them in memory during runtime.
\subsection{Regulating the residual shear stresses and mean-field}
\label{S3.5}
After reconstructing the inferred variables, we regulate the residual shear stress components $\Wperp$ and $\piperp$ such that they satisfy the orthogonality and traceless conditions
\bs
\allowdisplaybreaks
\label{eq:enforce}
\beal
    u_\mu \Wperp &= 0 \,,
\\
    z_\mu \Wperp &= 0 \,,
\\
    u_\mu \piperp &= 0 \,,
\\
    z_\mu\piperp &= 0 \,,
\\
    \pi^\mu_{\perp,\mu} &= 0\,,
\end{align}
\es
and that their overall magnitude is smaller than the longitudinal and transverse pressures:
\be
\label{eq:regulate_residual}
\sqrt{\pi_{\perp,\munu} \pi_\perp^\munu - 2 W_{\perp z,\mu} \Wperp} \leq \sqrt{\mathcal{P}_L^2 + 2\mathcal{P}_\perp^2}\,.
\ee
The former ensures that $\Wperp$ and $\piperp$ maintain their orthogonal and tracelessness properties within numerical accuracy during the hydrodynamic simulation. The latter prevents the residual shear stresses from overwhelming the anisotropic part of the energy-momentum tensor~\cite{Bazow:2017ewq}.

First, we update these components in the following order:
\bs
\allowdisplaybreaks
\beal
W_{\perp z}^\tau &\leftarrow \frac{u^\tau(W_{\perp z}^x u^x + W_{\perp z}^y u^y)}{1+u_\perp^2}\,,\\
W_{\perp z}^\eta &\leftarrow \frac{W_{\perp z}^\tau u^\eta}{u^\tau}\,,\\
\pi_\perp^{yy} &\leftarrow \frac{2 \pi_\perp^{xy} u^x u^y - \pi_\perp^{xx}\left(1 {+} (u^y)^2\right)}{1 + (u^x)^2} \,,\\
\pi_\perp^{\tau x} &\leftarrow \frac{u^\tau(\pi_\perp^{xx} u^x + \pi_\perp^{xy}u^y)}{1+u_\perp^2}\,,\\
\pi_\perp^{\tau y} &\leftarrow \frac{u^\tau(\pi_\perp^{xy} u^x + \pi_\perp^{yy}u^y)}{1+u_\perp^2}\,,\\
\pi_\perp^{x\eta} &\leftarrow \frac{\pi_\perp^{\tau x}u^\eta}{u^\tau} \,,\\
\pi_\perp^{y\eta} &\leftarrow \frac{\pi_\perp^{\tau y}u^\eta}{u^\tau} \,,\\
\pi_\perp^{\tau\eta} &\leftarrow \frac{u^\tau(\pi_\perp^{x\eta}u^x + \pi_\perp^{y\eta}u^y)}{1+u_\perp^2} \,,\\
\pi_\perp^{\eta\eta} &\leftarrow \frac{\pi_\perp^{\tau \eta}u^\eta}{u^\tau}\,,\\
\pi_\perp^{\tau\tau} &\leftarrow \frac{\pi_\perp^{\tau x} u^x + \pi_\perp^{\tau y}u^y + \tau^2 \pi_\perp^{\tau\eta} u^\eta}{u^\tau} \,,
\end{align}
\es
while leaving the components $W_{\perp z}^x$, $W_{\perp z}^y$, $\pi_{\perp}^{xx}$ and $\pi_{\perp}^{xy}$ unchanged. Next, we rescale all the components by the same factor $\rho_\text{reg} \in [0,1]$:
\bs
\beal
\Wperp &\leftarrow \rho_\text{reg} \Wperp \,,\\
\piperp &\leftarrow \rho_\text{reg} \piperp\,,
\end{align}
\es
where\footnote{%
    For longitudinally boost-invariant systems, we replace the second argument in Eq.~\eqref{eq:rescale_residual} by $\sqrt{2 \mathcal{P}_\perp^2 / \pi_\perp {\cdot\,} \pi_\perp}$.}
\be
\label{eq:rescale_residual}
\rho_\text{reg} = \min\Bigg(1, \sqrt{\frac{\mathcal{P}_L^2 + 2\mathcal{P}_\perp^2}{\pi_\perp {\cdot\,} \pi_\perp - 2 W_{\perp z}{\,\cdot\,} W_{\perp z}}}\Bigg) \,,
\ee
with $\pi_\perp {\cdot\,} \pi_\perp = \pi_{\perp,\munu} \pi_\perp^\munu$. In the validation tests discussed in Sec.~\ref{S4}, we did not encounter a situation where the residual shear stresses are suppressed by Eq.~\eqref{eq:rescale_residual}. This indicates that the residual shear stresses are naturally smaller than the leading-order anisotropic pressures during the simulation. Nevertheless, we keep this procedure in place as a precaution.

For hydrodynamic simulations with two or three spatial dimensions, we find that the relaxation equation \eqref{eq:relax_B_0} for the mean-field $B$ can become unstable in certain spacetime regions ($T \sim 0.15 - 0.16$ GeV), causing the bulk viscous pressure $\Pi$ to grow positive without bound. The origin of this issue is not fully understood, and we will address it in future work. For now, we remove this instability by regulating the nonequilibrium mean-field component $\delta B = B - B_\text{eq}$ as
\be
\delta B \leftarrow \kappa_\text{reg} \delta B,
\ee
where
\be
\label{eq:dB_reg}
\kappa_\text{reg} = \min\Big(1, -\frac{|B_\text{eq}|}{\delta B}\Big) \indent \forall \,\, \delta B < 0\,.
\ee
In practice, we only find it necessary to regulate the mean-field when $\delta B < 0$ (i.e. in regions with $\Pi > 0$).
%
%
%
\subsection{Adaptive time step}
\label{S3.6}
Most relativistic hydrodynamic codes that simulate heavy-ion collisions evolve the system with a fixed time step $\Delta \tau_n = \Delta \tau$~\cite{Schenke:2010nt,Shen:2014vra}. Any choice for $\Delta \tau$ must satisfy the CFL condition so that the hydrodynamic simulation is at least dynamically stable \cite{Kurganov:2000}. However, the time step must also be small enough to resolve the fluid's evolution rate (in particular, the large longitudinal expansion rate $\theta_L \sim 1/\tau$ at early times) but, on the other hand, large enough to finish the simulation within a reasonable runtime. This balancing act places a practical limit on how early the user can start the hydrodynamic simulation (in practice, the smallest value typically used is $\tau_0 \sim 0.2$ fm/$c$ \cite{Gale:2012rq}). This is not much of a concern for second-order viscous hydrodynamics since it is anyhow prone to breaking down for very early initialization times, as discussed in Sec.~\ref{S1}. Anisotropic hydrodynamics, on the other hand, can handle the large pressure anisotropies occurring at early times much better; to realize its full potential it should be initialized at earlier times, but for that we need to move away from a fixed time step. In this section, we introduce a new adaptive stepsize method, which automatically adjusts the successive time step $\Delta \tau_{n+1}$ to be larger or smaller than $\Delta\tau_n$ in such a way that we can push back our fluid dynamical simulation to very early times ($\tau_0 \sim 0.01 - 0.05 \,\text{fm}/c$) without sacrificing numerical accuracy nor computational efficiency.\footnote{%
    This is also useful for constructing the particlization hypersurface in peripheral heavy-ion collisions or small collision systems (e.g. p+p), whose fireball lifetimes are not that much longer than the hydrodynamization time $\tau_\text{hydro}$ (see Sec.~\ref{S1}).}

For a system with no source terms ($\boldsymbol{S} = \boldsymbol{0}$) the KT algorithm is dynamically stable as long as the time step satisfies the CFL condition \cite{Kurganov:2000}
\be
\label{eq:CFL_bound}
    \Delta \tau_n \leq \Delta\tau_\text{CFL} = \frac{1}{8} \min\left(\frac{\Delta x}{s^x_\text{max}(\tau_n)}, \frac{\Delta y}{s^y_\text{max}(\tau_n)}, \frac{\Delta\eta_s}{s^\eta_\text{max}(\tau_n)} \right)\,,
\ee
where $s^i_\text{max}(\tau_n)$ ($i=x,y,\eta$) are the maximum local propagation speed components on the Eulerian grid at time $\tau_n$. In Milne spacetime, the quark-gluon plasma's flow profile is not ultrarelativistic throughout most of its evolution (as long as one uses a QCD equation of state). Thus, the criterium \eqref{eq:CFL_bound} allows one to maintain dynamical stability with an adaptive time step that is generally larger than the fixed time step obtained by taking the limit $s^i_\text{max} \to 1$:
\be
\label{eq:CFL_fixed}
    \Delta \tau_n = \frac{1}{8} \min\left(\Delta x, \Delta y, \Delta \eta_s\right)\,.
\ee
For $\boldsymbol{S} \neq \boldsymbol{0}$, however, the time step $\Delta\tau_\text{CFL}$ from (\ref{eq:CFL_bound}) is too coarse to resolve the fluid's gradients and relaxation rates at early times $\tau < 0.5$ fm/$c$ when the flow profile is very nonrelativistic. Therefore, at early times when the source terms are strongest, the adaptive time step should primarily depend on those source terms. The following implementation ensures that, as the source terms relax and the fluid velocity grows more relativistic over time, the adaptive time step naturally approaches the CFL bound \eqref{eq:CFL_bound}.

First, we consider a homogeneous fluid undergoing Bjorken expansion (i.e. $\ene = T^{\tau\tau}$ and $\boldsymbol{u} = \boldsymbol{0}$). The system of differential equations~\eqref{eq:KT_eqs} for the dynamical variables $\boldsymbol{q}(\tau)$ reduces to
\be
    \partial_\tau \boldsymbol{q}(\tau) = \boldsymbol{S}(\tau, \boldsymbol{q}),
\ee
and we are given the initial values $\boldsymbol{q}_n$ and time step $\Delta \tau_n$ at time $\tau_n$. We evolve the system one time step $\Delta \tau_n$ using the RK2 scheme~\eqref{eq:RK2}:
\be
    \boldsymbol{q}_{n+1} = \boldsymbol{q}_n + \frac{\Delta \tau_n}{2}\left(\boldsymbol{S}(\tau_n, \boldsymbol{q}_n) + \boldsymbol{S}(\tau_n {+} \Delta \tau_n, \boldsymbol{q}_n {+} \Delta \tau_n \boldsymbol{S}(\tau_n, \boldsymbol{q}_n) \right)\,.
\ee
For the next iteration, we determine how much we need to adjust the time step $\Delta \tau_{n+1}$. Adaptive stepsize methods do this by estimating the local truncation error of each time step~\cite{Press:1992:NRC:148286}. In our method, we approximate the local truncation error of the next intermediate Euler step
\be
\label{eq:next_Euler}
    \boldsymbol{q}_{\,\text{I},n+2} = \boldsymbol{q}_{n+1} + \Delta \tau_{n+1} \boldsymbol{S}(\tau_n {+} \Delta \tau_n, \boldsymbol{q}_{n+1})\,,
\ee
which is
\be
    \epsilon_{n+2} = \frac{1}{2}\norm{(\partial_\tau^2\boldsymbol{q})_{n+1}} \Delta \tau_{n+1}^2 + O(\Delta \tau_{n+1}^3)\,,
\ee
where $\norm{...}$ denotes the $\ell^2$-norm. If the second time derivative $(\partial^2_\tau\boldsymbol{q})_{n+1}$ at $\tau = \tau_n + \Delta\tau_n$ is known, we can set the local truncation error to the desired error tolerance (after dropping higher-order terms)
\be
\label{eq:tolerance}
    \frac{1}{2}\norm{(\partial_\tau^2\boldsymbol{q})_{n+1}} \Delta \tau_{n+1}^2 = \delta_0 \times \max(N_q^{1/2}, \,\norm{\boldsymbol{q}_{\,\text{I},n+2}})\,,
\ee
where $\delta_0$ is the error tolerance parameter and $N_q$ is the number of dynamical variables. It is reasonable to use absolute or relative errors when $\norm{\boldsymbol{q}_{\,\text{I},n+2}}$ is small or large, respectively~\cite{Press:1992:NRC:148286}. In this work, we set the error tolerance parameter to $\delta_0 = 0.004$.

To obtain an expression for the second time derivative, we compute the next intermediate Euler step~\eqref{eq:next_Euler} using the old time step (denoted by $\star$):
\be
\label{eq:next_Euler_star}
    \boldsymbol{q}^\star_{\,\text{I},n+2} = \boldsymbol{q}_{n+1} + \Delta \tau_n \boldsymbol{S}(\tau_n {+} \Delta \tau_n, \boldsymbol{q}_{n+1})\,.
\ee
This allows us to approximate $(\partial_\tau^2\boldsymbol{q})_{n+1}$ with central differences:
\be
\label{eq:approx_2nd_derivative}
    (\partial_\tau^2\boldsymbol{q})_{n+1} = \frac{2(\boldsymbol{q}^\star_{\,\text{I},n+2} {-} 2 \boldsymbol{q}_{n+1} {+} \boldsymbol{q}_n)}{\Delta \tau_n^2} + O(\Delta\tau_n)\,.
\ee
Compared to the usual central difference formula there is an additional factor of $2$ that accounts for the local truncation error present in $\boldsymbol{q}_{\,\text{I},n+2}^\star$. Furthermore, the expression \eqref{eq:approx_2nd_derivative} is numerically accurate to $O(\Delta\tau_n)$ rather than $O(\Delta\tau_n^2)$. After substituting Eqs.~\eqref{eq:next_Euler} and \eqref{eq:approx_2nd_derivative} in Eq.~\eqref{eq:tolerance}, one has for the next time step
\be
\label{eq:next_step}
    \Delta \tau_{n+1} = \max\big(\Delta \tau_{n+1}^\text{(abs)}, \Delta \tau_{n+1}^\text{(rel)}\big)\,,
\ee
where
\be
    \Delta\tau_{n+1}^\text{(abs)} = \Delta\tau_n \sqrt{\dfrac{\delta_0\, N_q^{1/2}}{\norm{\boldsymbol{q}^\star_{\,\text{I},n+2} {-} 2 \boldsymbol{q}_{n+1} {+} \boldsymbol{q}_n}}}
\ee
and $\Delta\tau_{n+1}^\text{(rel)}$ is the numerical solution to the algebraic equation
\be
    \frac{N_q^{1/2}\big(\Delta \tau_{n+1}^\text{(rel)}\big)^2}{\big(\Delta \tau_{n+1}^\text{(abs)}\big)^2} = \sqrt{\norm{\boldsymbol{q}_{n+1}}^2 + 2 \boldsymbol{q}_{n+1} {\cdot\,} \boldsymbol{S}_{n+1} \Delta \tau_{n+1}^\text{(rel)} + \norm{\boldsymbol{S}_{n+1}}^2 \big(\Delta \tau_{n+1}^\text{(rel)}\big)^2} \,,
\ee
with $\boldsymbol{S}_{n+1} = \boldsymbol{S}(\tau_n {+} \Delta\tau_n, \boldsymbol{q}_{n+1})$.

For the general case without Bjorken symmetry, $\boldsymbol{q}(x)$ varies across the grid. We then perform the calculation \eqref{eq:next_step} (replacing $\boldsymbol{S}$ by $\boldsymbol{E}$) for all spatial grid points and take the minimum value. Afterwards, we place safety bounds to prevent the time step from changing too rapidly:
\be
    \left(1 {-} \alpha\right) \Delta\tau_n \leq \Delta \tau_{n+1} \leq (1 {+} \alpha)\Delta\tau_n \,,
\ee
where we set the control parameter to $\alpha = 0.5$. Finally, we impose the CFL bound~\eqref{eq:CFL_bound} on $\Delta \tau_{n+1}$. With the new time step at hand, we resume computing the next RK2 iteration. Notice that there are no additional numerical evaluations of the flux and source terms in the adaptive RK2 scheme since we can recompute the next intermediate Euler step $\boldsymbol{q}_{\,\text{I},n+2}$ simply by adjusting the time step in $\boldsymbol{q}_{\,\text{I},n+2}^\star$. This allows our adaptive time step algorithm to be readily integrated into our numerical scheme.

When we start the hydrodynamic simulation, we initialize the time step $\Delta \tau_0$ to be 20 times smaller than the initial time $\tau_0$.\footnote{\label{adaptive_floor}%
    We do not allow the adaptive time step to become any smaller than this (i.e. we impose $\Delta\tau_n \geq 0.05\tau_0$).}
At first, the adaptive time step $\Delta \tau_n$ tends to increase with time since the fluid's longitudinal expansion rate decreases. As the transverse flow builds up, $\Delta \tau_n$ eventually becomes bounded by the CFL condition~\eqref{eq:CFL_bound}.
\subsection{Program summary}
\begin{figure}[thbp]
\includegraphics[width=0.9\linewidth]{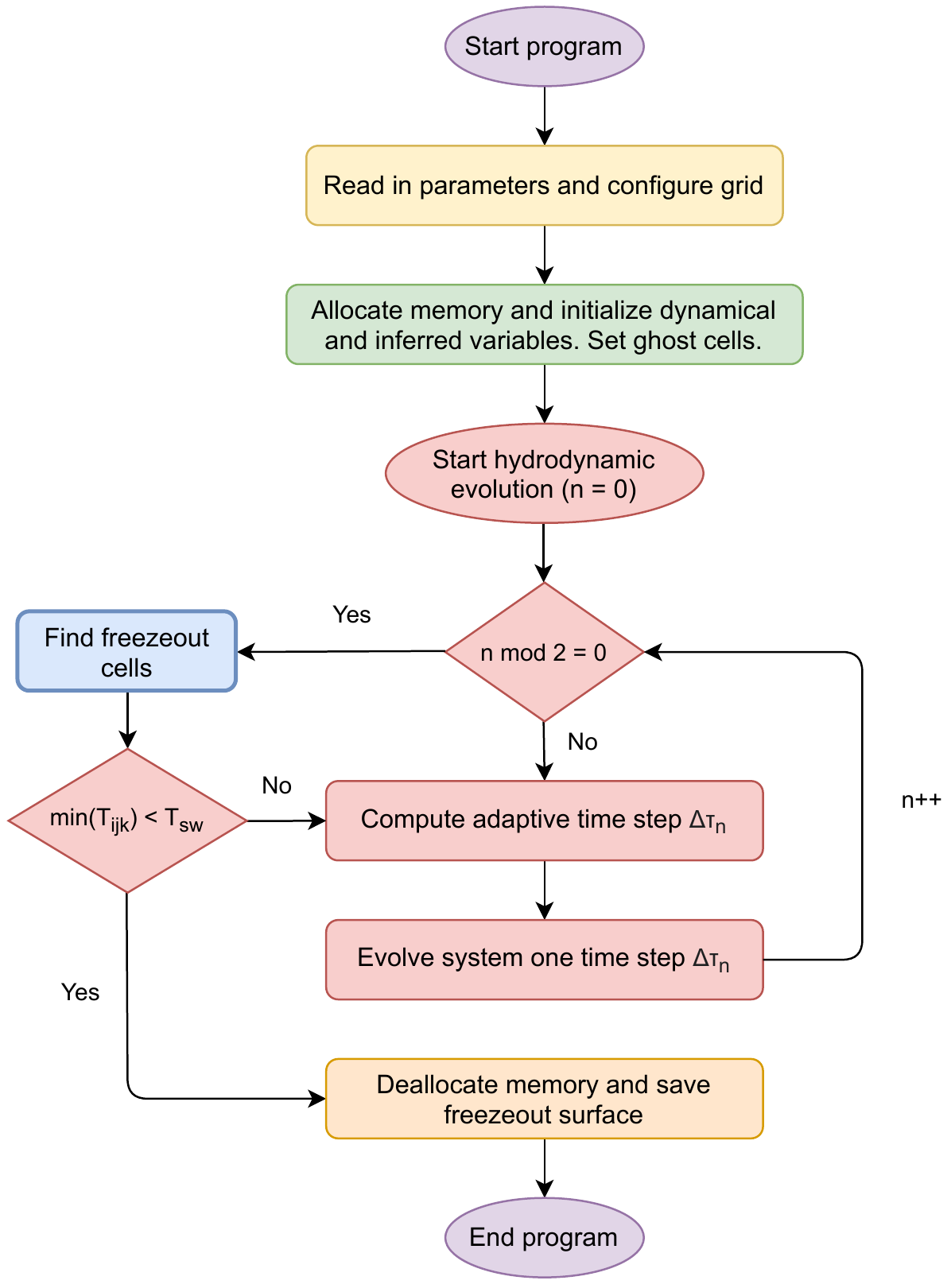}
\centering
\caption{(Color online)
\label{flowchart}
Program flowchart of \cpuvah.
}
\end{figure}

We close this section by summarizing the workflow of the anisotropic fluid dynamical simulation with a QCD equation of state. Figure~\ref{flowchart} shows the flowchart of the code's primary mode,\footnote{%
    The secondary (test) mode outputs the hydrodynamic evolution from the simulation and their corresponding semi-analytic solutions, if they exist.}
which constructs a hypersurface of constant energy density $\ene_\text{sw}$ for a particlization module:

\begin{enumerate}
    \item We read in the runtime parameters and configure the Eulerian grid.\\
    \item We allocate memory to store the dynamical and inferred variables in the Eulerian grid at a given time $\tau_n$. Altogether, there are three dynamical variables ($\texttt{q}$, $\texttt{qI}$, $\texttt{Q}$), two fluid velocity variables ($\texttt{u}$, $\texttt{up}$), one energy density variable $\texttt{e}$ and three anisotropic variables ($\texttt{lambda}$, $\texttt{aT}$, $\texttt{aL}$). They store one of the following variables during the RK2 iteration:
    \begin{enumerate}
        \item \texttt{q} holds the current or updated dynamical variables $\boldsymbol{q}_n$ or $\boldsymbol{q}_{n+1}$.
        \item \texttt{qI} holds the intermediate dynamical variables $\boldsymbol{q}_{\,\text{I},n+1}$ (or $\boldsymbol{q}^\star_{\,\text{I},n+1}$).
        \item \texttt{Q} holds the previous, updated or current dynamical variables $\boldsymbol{q}_{n-1}$, $\boldsymbol{q}_{n+1}$ or $\boldsymbol{q}_n$.
        \item \texttt{u} holds the current, intermediate or updated fluid velocity $\boldsymbol{u}_n$, $\boldsymbol{u}_{\text{I},n+1}$ or $\boldsymbol{u}_{n+1}$.
        \item \texttt{up} holds the previous or current fluid velocity $\boldsymbol{u}_{n-1}$ or $\boldsymbol{u}_n$.
        \item \texttt{e} holds the current, intermediate or updated energy density $\ene_n$, $\ene_{\text{I},n+1}$ or $\ene_{n+1}$.
        \item \texttt{lambda}, \texttt{aT} and \texttt{aL} hold the current,  intermediate or updated anisotropic variables ($\Lambda_n$, $\alpha_{\perp,n}$, $\alpha_{L,n}$), ($\Lambda_{\text{I},n+1}$, $\alpha_{\perp,\text{I},n+1}$, $\alpha_{L,\text{I},n+1}$) or ($\Lambda_{n+1}$, $\alpha_{\perp,n+1}$, $\alpha_{L,n+1}$)\\
    \end{enumerate}

    \item We initialize the variables \texttt{q}, \texttt{u}, \texttt{up}, \texttt{e}, \texttt{lambda}, \texttt{aT} and \texttt{aL} as follows:\footnote{%
        If the code is run with Gubser initial conditions, the hydrodynamic variables are initialized differently (see Sec.~\ref{S4.2}).}
    first, we compute (or read in) the energy density profile given by an initial-state model (e.g. \trento{} \cite{Moreland:2014oya}). Then we initialize the longitudinal and transverse pressures as
    \bs
    \beal
    \PL &= \frac{3 R\,\Peq}{2+R} \,,\\
    \Pperp &= \frac{3\Peq}{2+R} \,,
    \end{align}
    \es
    where $R \in [0,1]$ is a pressure anisotropy ratio parameter set by the user (typically a small value $R \leq 0.3$). This model assumes that only the pressure anisotropy $\PL-\Pperp$ enters in the initial $\mathcal{P}_L$ and $\mathcal{P}_\perp$ while the initial bulk viscous pressure is $\Pi = 0$.
    The initial fluid velocity is static (i.e. $\texttt{u} = \texttt{up} = \boldsymbol{0}$) and the residual shear stresses $\Wperp$ and $\piperp$ are set to zero. From this, we compute the components $T^{\tau\mu}$ with Eq.~\eqref{eq:Ttaumu}. Finally, we set the mean-field to $B = B_\text{eq}$ and initialize the anisotropic variables by solving Eq.~\eqref{eq:nonlinear_system}.\footnote{%
        This initialization scheme is similar to how conformal free-streaming modules are initialized at $\tau_0 \to 0$~\cite{Liu:2015nwa, Bernhard:2016tnd, Bernhard:2018hnz, Everett:2020yty, Everett:2020xug}.
        }\\

    \item We set the ghost cell boundary conditions for \texttt{q}, \texttt{e} and \texttt{u}.\footnote{%
        The ghost cells of \texttt{e} are only used in conformal anisotropic hydrodynamics to approximate the source terms $\propto \partial_i \ene$ on the grid's faces.}
        We also configure the freezeout finder and set the initial time step to $\Delta\tau_0 = 0.05\, \tau_0$.\\
    \item We start the hydrodynamic evolution at $\tau = \tau_0$ (or $n = 0$):
    \begin{enumerate}
        \item We call the freezeout finder every two time steps\footnote{The user can adjust the number of time steps between freezeout finder calls.} (i.e. $n \mod 2  = 0$) to search for freezeout cells between the Eulerian grids from the current and previous calls (if $n = 0$, we only load the initial grid to the freezeout finder). The hypersurface volume elements $d^3\sigma_\mu$ and their spacetime positions are constructed with the code {\sc CORNELIUS}~\cite{Huovinen:2012is}. The hydrodynamic variables at the hypersurface elements' centroid are approximated with a 4d linear interpolation.
        \item We compute the intermediate Euler step~\eqref{eq:intermediate_step} and store the results $\boldsymbol{q}_{\,\text{I},n+1}$ in $\texttt{qI}$ (if $n>0$, we compute $\boldsymbol{q}^\star_{\,\text{I},n+1}$ and the adaptive time step $\Delta\tau_n$ using the algorithm in Sec.~\ref{S3.6} before evaluating $\boldsymbol{q}_{\,\text{I},n+1}$). Afterwards, we swap the variables $\texttt{u} \leftrightarrow \texttt{up}$ so that $\texttt{up} \leftarrow \boldsymbol{u}_n$.
        \item We reconstruct the intermediate inferred variables $\ene_{\text{I},n+1}$, $\boldsymbol{u}_{\text{I},n+1}$, $\Lambda_{\text{I},n+1}$, $\alpha_{\perp,\text{I},n+1}$ and $\alpha_{L,\text{I},n+1}$ from \texttt{qI} and store the results in \texttt{e}, \texttt{u}, \texttt{lambda}, \texttt{aT} and \texttt{aL}, respectively. We regulate the residual shear stresses and mean-field in \texttt{qI} and set the ghost cell boundary conditions for \texttt{qI}, \texttt{e} and \texttt{u}.
        \item We compute the second intermediate Euler step~\eqref{eq:2nd_intermediate_step} and update the dynamical variables $\boldsymbol{q}_{n+1}$ via Eq.~\eqref{eq:RK2}, which is stored in \texttt{Q}. Afterwards, we swap the variables $\texttt{q} \leftrightarrow \texttt{Q}$ so that $\texttt{q} \leftarrow \boldsymbol{q}_{n+1}$ and $\texttt{Q} \leftarrow \boldsymbol{q}_n$.
        \item We update the inferred variables $\ene_{n+1}$, $\boldsymbol{u}_{n+1}$, $\Lambda_{n+1}$, $\alpha_{\perp,n+1}$ and $\alpha_{L,n+1}$ from \texttt{q} and store the results in \texttt{e}, \texttt{u}, \texttt{lambda}, \texttt{aT} and \texttt{aL}, respectively. We regulate the residual shear stresses and mean-field in \texttt{q} and set the ghost cell boundary conditions for \texttt{q}, \texttt{e} and \texttt{u}.
        \item Steps (a) -- (e) are repeated until the temperature of all fluid cells in the grid are below the switching temperature (i.e. $\min(T_{ijk}) < T_\text{sw}$).\\
    \end{enumerate}
    \item After the hydrodynamic evolution, we deallocate the hydrodynamic variables and store the freezeout surface in memory, which can either be passed to another program or written to file.
\end{enumerate}

\section{Validation tests and hydrodynamic model comparisons}
\label{S4}

In this section we test the validity of our anisotropic fluid dynamical simulation for various initial-state configurations, using either the conformal or QCD equation of state. First, we run anisotropic hydrodynamics with conformal Bjorken and Gubser initial conditions \cite{Bjorken:1982qr, Gubser:2010ui, Gubser:2010ze}, whose semi-analytic solutions can be computed accurately using a fourth-order Runge--Kutta ODE solver \cite{Molnar:2016gwq, Martinez:2017ibh}. After these two validation tests, we compare (3+1)--dimensional conformal anisotropic hydrodynamics and second-order viscous hydrodynamics\footnote{\label{VH12}%
    The \cpuvah{} code can also run second-order viscous hydrodynamics, where the kinetic transport coefficients are computed with either quasiparticle masses $m(T)$ (see Fig.~\ref{quasi}a) \cite{Tinti:2016bav, McNelis:2018jho} or light masses (i.e. $m/T \ll 1$)~\cite{Denicol:2014vaa,Bazow:2016yra} (we use the same shear and bulk viscosities as in Fig.~\ref{viscosities}). In this work, we label the former model as \textit{quasiparticle viscous hydrodynamics} (or \vh{}) and the latter as \textit{standard viscous hydrodynamics} (or \vh{}2); for conformal systems, these two models are equivalent. For details on the numerical implementation of second-order viscous hydrodynamics, see Appendix \ref{appe}.}
for a central Pb+Pb collision with a smooth \trento{} initial condition \cite{Moreland:2014oya}.

Next, we run non-conformal anisotropic hydrodynamics and viscous hydrodynamics with Bjorken initial conditions and compare them to their semi-analytic solutions \cite{McNelis:2018jho}. Finally, we study the differences between (3+1)--dimensional non-conformal anisotropic hydrodynamics and viscous hydrodynamics in central Pb+Pb collisions with smooth or fluctuating \trento{} initial conditions, with the goal of identifying situations where \cpuvah{} offers definitive advantages in reliability over standard viscous hydrodynamics.

\subsection{Conformal Bjorken flow test}
\label{S4.1}
For the first test, we evolve a conformal plasma undergoing Bjorken expansion~\cite{Bjorken:1982qr}. In Bjorken flow, the system is longitudinally boost-invariant and homogeneous in the transverse plane. As a result, the fluid velocity $u^\mu = (1,0,0,0)$ is static and the residual shear stresses $\Wperp$ and $\piperp$ vanish. The only two degrees of freedom are the energy density $\ene = T^{\tau\tau}$ and longitudinal pressure $\PL$ (the transverse pressure is $\Pperp = \frac{1}{2}(\ene {-} \PL)$). The anisotropic hydrodynamic equations simplify to \cite{Molnar:2016gwq}
\bs
\label{eq:semi_Bjorken}
\beal
    \partial_\tau \ene &= - \frac{\ene + \PL}{\tau} \,,
\\
    \partial_\tau \PL &= \frac{\ene {-} 3\PL}{3\tau_\pi} + \frac{\bar\zeta_z^L}{\tau}\,,
\end{align}
\es
where the conformal transport coefficient $\bar\zeta_z^L$ is given in Eq.~(\ref{eq:pl_coeff}a).
%
\begin{figure}[t]
\includegraphics[width=\linewidth]{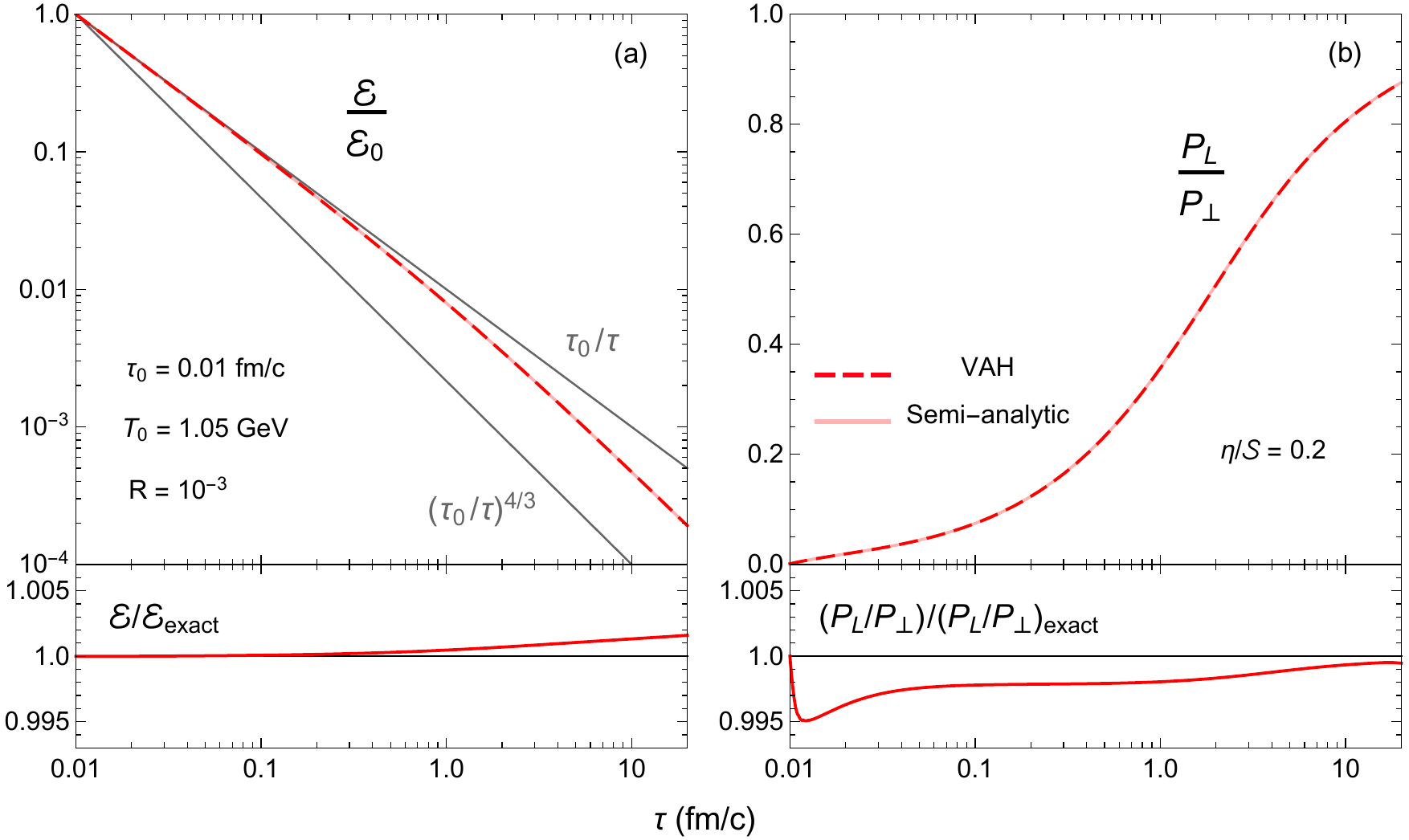}
\centering
\caption{(Color online)
\label{conformal_bjorken}
    Conformal Bjorken evolution of the normalized energy density $\ene/\ene_0$ (a) and pressure ratio $\PL/\Pperp$ (b) from the anisotropic hydrodynamic simulation (dashed red) and semi-analytic solution (transparent red). The gray lines in (a) show two different power laws for comparison. The subpanels at the bottom show the ratio between the numerical simulation and the semi-analytic solutions (solid red).
}
\end{figure}
%
%
\begin{figure}[t]
\includegraphics[width=0.6\linewidth]{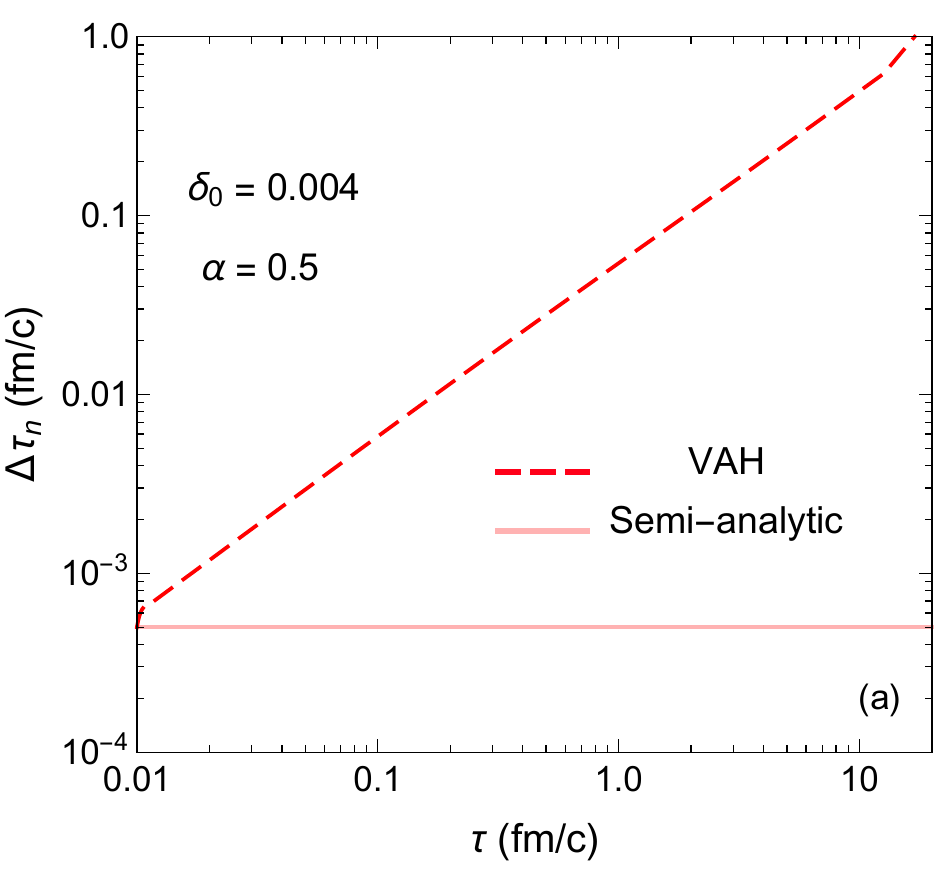}
\centering
\caption{(Color online)
\label{bjorken_adaptive}
The evolution of the adaptive time step $\Delta\tau_n$ (dashed red) in the conformal Bjorken simulation. The semi-analytic method (transparent red) uses a fixed time step of $\Delta\tau_n = 5 \times 10^{-4}$\,fm/$c$.
}
\end{figure}
%
We start the simulation\footnote{%
    Bjorken flow can be simulated on either one fluid cell (i.e. $N_x = N_y = N_\eta = 1)$ or a larger grid with homogeneous initial conditions.}
at $\tau_0 = 0.01$\,fm/$c$ with an initial temperature $T_0 = 1.05$\,GeV, pressure ratio $R = 10^{-3}$ and shear viscosity $\etas = 0.2$. We evolve the system with an adaptive time step, initially set to $\Delta \tau_0 = 5 \times 10^{-4}$\,fm/$c$, until the temperature drops below $T_\text{sw} = 0.136$\,GeV at $\tau_f \sim 20$ fm/$c$. Figure~\ref{conformal_bjorken} shows the evolution of the energy density normalized to its initial value $\ene_0$ and the pressure ratio $\PL / \Pperp$. At very early times, the system is approximately free-streaming due to the rapid longitudinal expansion, resulting in a large Knudsen number and causing the energy density to decrease like $\ene \approx \ene_0 \tau_0/ \tau$. Over time, the longitudinal expansion rate $\theta_L = 1/\tau$ decreases and the system approaches local equilibrium, i.e. $\ene\propto\tau^{-4/3}$ and $\PL / \Pperp \to 1$.

We compare the (0+1)--d \cpuvah{} simulation to the semi-analytic solution of \eqref{eq:semi_Bjorken}, which uses a fixed time step $\Delta \tau = 5 \times 10^{-4}$\,fm/$c$. The simulation is in excellent agreement with the semi-analytic solution, with numerical errors staying below $0.5\%$. We also checked the convergence of the simulation curves as we decrease the error tolerance parameter $\delta_0$ in the adaptive time step algorithm. Figure~\ref{bjorken_adaptive} shows the evolution of the adaptive time step $\Delta \tau_n$ for the parameters $\delta_0 = 0.004$ and $\alpha = 0.5$. One sees that the time step increases linearly with time and is not bounded by the CFL condition \eqref{eq:CFL_bound} since the flow is stationary (i.e. $\Delta\tau_\text{CFL} \to \infty)$. As a result, the adaptive RK2 scheme quickly reaches the switching temperature in 145 time steps, compared to about 40000 steps for the semi-analytic method.


\subsection{Conformal Gubser flow test}
\label{S4.2}
In the next test, we evolve a conformal fluid subject to Gubser flow \cite{Gubser:2010ze, Gubser:2010ui}, which is longitudinally boost-invariant and azimuthally symmetric. The fluid velocity profile takes the following analytic form:
\be
\label{eq:gubser_velocity}
    u^\tau = \cosh\kappa \,,\quad
    u^x = \sinh\kappa \cos\phi \,,\quad
    u^y = \sinh\kappa \sin\phi \,,\quad
    u^\eta = 0\,,
\ee
where
\be
    \kappa(\tau, r) = {\tanh^{-1}}\bigg[\frac{2 q^2 \tau r}{1 + q^2(\tau^2 {+} r^2)}\bigg] \,,
\ee
with $r{\,=\,}\sqrt{x^2{+}y^2}$ being the transverse radius, $\phi = {\tan^{-1}}(y/x)$ the azimuthal angle, and $q$ an inverse length scale parameter that determines the fireball's transverse size $R_\perp \sim 1/q$ at $\tau = 0^+$.

One can transform the polar Milne coordinates $\tilde{x}^\mu = (\tau, r, \phi, \eta_s)$ to the de Sitter coordinates $\hat{x}^\mu = (\rho, \theta, \phi, \eta_s)$ where
\bs
\allowdisplaybreaks
\beal
    \rho(\tau, r) &= -{\sinh^{-1}}\bigg[\frac{1 - q^2\left(\tau^2 {-} r^2\right)}{2q\tau}\bigg] \,,
\\
    \theta(\tau, r) &= {\tan^{-1}}\bigg[\frac{2qr}{1 + q^2\left(\tau^2 {-} r^2\right)}\bigg] \,.
\end{align}
\es
The line element in this de Sitter space possesses $SO(3)_q \otimes SO(1, 1) \otimes Z_2$ symmetry \cite{Gubser:2010ze, Gubser:2010ui}:
\be
    d\hat{s}^2 = - d\rho^2 + {\cosh^2}\rho\,\bigl(d\theta^2 {+} {\sin^2}\theta d\phi^2\bigr) + d\eta^2_s\,,
\ee
which makes the fluid velocity $\hat{u}^\mu = (1, 0, 0, 0)$ stationary.\footnote{%
    A hat denotes a quantity in the de Sitter space $\hat{x}^\mu = (\rho,\theta,\phi,\eta_s)$. All hatted quantities are made unitless by multiplying with appropriate powers of $\tau$.}
Hence, the hydrodynamic variables only depend on the de Sitter time $\rho$, and the anisotropic hydrodynamic equations reduce to \cite{Martinez:2017ibh}
\bs
\label{eq:semi_Gubser}
\allowdisplaybreaks
\beal
    \partial_\rho \hat\ene &= (\PLhat - 3\hat\ene) \tanh\rho \,,
\\
    \partial_\rho \PLhat &= \frac{\hat\ene {-} 3\PLhat}{3\hat\tau_\pi} - \big(4 \PLhat {+} \hat{\bar\zeta}_z^L\big)\tanh\rho \,.
\end{align}
\es
The residual shear stresses $\hat{W}_{\perp z}^\mu$ and  $\hat{\pi}_\perp^\munu$ vanish under Gubser symmetry \cite{Martinez:2017ibh}. In the semi-analytic solution \eqref{eq:semi_Gubser}, we start the evolution at $\rho_0 = -9.2$, which corresponds to a corner in a $14$\,fm\,$\times 14$\,fm transverse grid ($r = 7 \sqrt{2}$\,fm) at the initial time $\tau_0 = 0.01$\,fm/$c$, with an initial temperature $\hat{T}_0 = 0.0017$ and pressure ratio $\hat{R} = 10^{-3}$. We set the inverse fireball size to $q = 1.0$\,fm$^{-1}$ and the shear viscosity to $\eta/\mathcal{S} = 0.2$. We evolve the system with a fixed stepsize $\Delta \rho = 10^{-4}$ until $\rho_f = 1.1$, which corresponds to the center of our Eulerian grid ($r = 0$\,fm) at $\tau_f = 3.01$\,fm/$c$.

Next, we map the semi-analytic solution for the de Sitter energy density $\hat{\ene}(\rho)$ and longitudinal pressure $\PLhat(\rho)$ back to Milne coordinates \cite{Denicol:2014tha}:
\be
\label{eq:semi_Gubser_map}
    \ene(\tau, r) = \frac{\hat\ene\big(\rho(\tau,r)\big)}{\tau^4} \,,\qquad
    \PL(\tau, r) = \frac{\PLhat\big(\rho(\tau,r)\big)}{\tau^4}\,.
\ee
\begin{figure}[thbp]
 \makebox[\textwidth][c]{\includegraphics[width=0.83\linewidth]{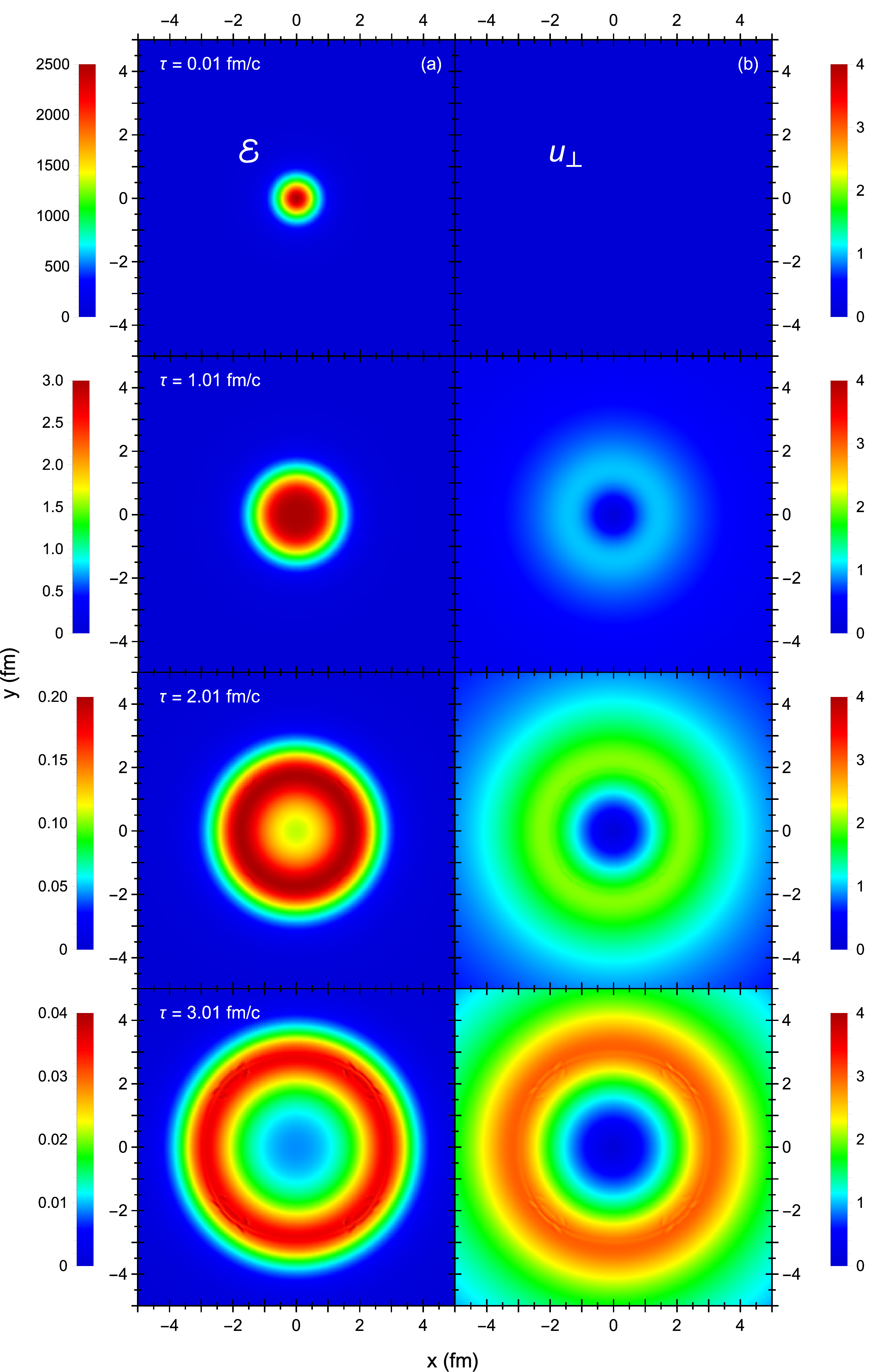}}
\caption{(Color online)
\label{gubser_2d}
    Conformal Gubser evolution of the energy density $\ene$ (GeV/fm$^3$) (left column) and transverse fluid rapidity $u_\perp$ (right column) in the anisotropic hydrodynamic simulation.
}
\end{figure}
\begin{figure}[!t]
\includegraphics[width=\linewidth]{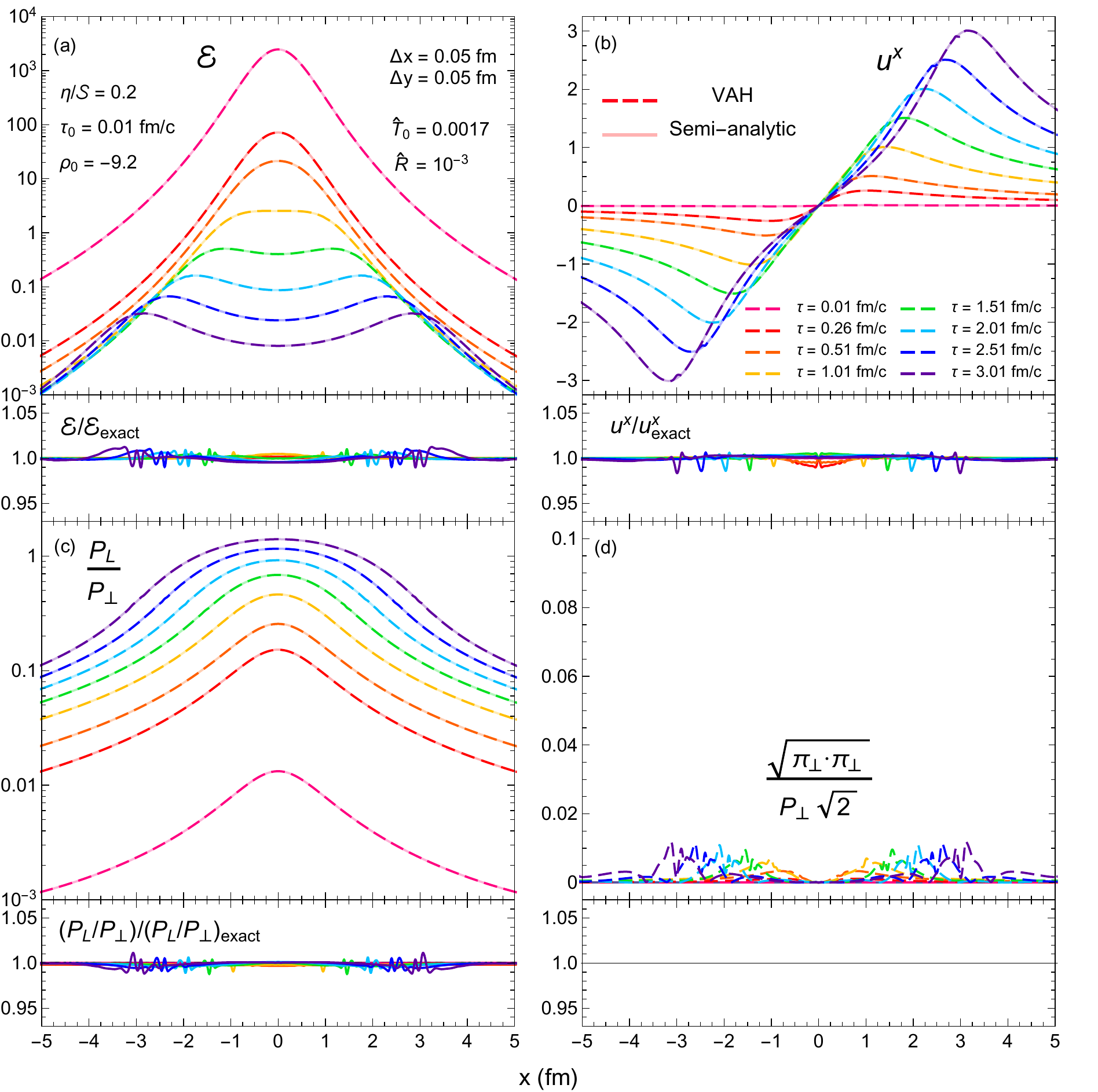}
\centering
\caption{(Color online)
\label{gubser_test}
    Conformal Gubser evolution of (a) $\ene$ (GeV/fm$^3$), (b) $u^x$, (c) $\PL/\Pperp$, and (d) $\sqrt{\pi_\perp {\cdot\,} \pi_\perp }/(\Pperp\sqrt{2})$ along the $x$--axis ($y=0$), given by the anisotropic hydrodynamic simulation (dashed color) and semi-analytic solution (transparent color). The subpanels below panels (a-c) show the ratio between numerical simulation and semi-analytic solution (solid color).
}
\end{figure}
We use this map to set the initial energy density and longitudinal pressure profiles $\ene(\tau_0, x, y)$ and $\PL(\tau_0, x, y)$; the initial temperature at the center of the fireball is $T_{0,\text{center}} = 1.05$\,GeV. We set the initial transverse shear stress to $\piperp = 0$. Finally, we use Eq.~\eqref{eq:gubser_velocity} to initialize the current fluid velocity $\texttt{u} \leftarrow u^\mu(\tau_0,x,y)$ and previous fluid velocity $\texttt{up} \leftarrow u^\mu(\tau_0 - \Delta\tau_0,x,y)$, where the initial time step is $\Delta\tau_0 = 5 \times 10^{-4}$\,fm/$c$. We evolve the Gubser profile on a $14$\,fm\,$\times 14$\,fm transverse grid with a lattice spacing of $\Delta x = \Delta y = 0.05$\,fm.

\begin{figure}[!t]
\includegraphics[width=0.6\linewidth]{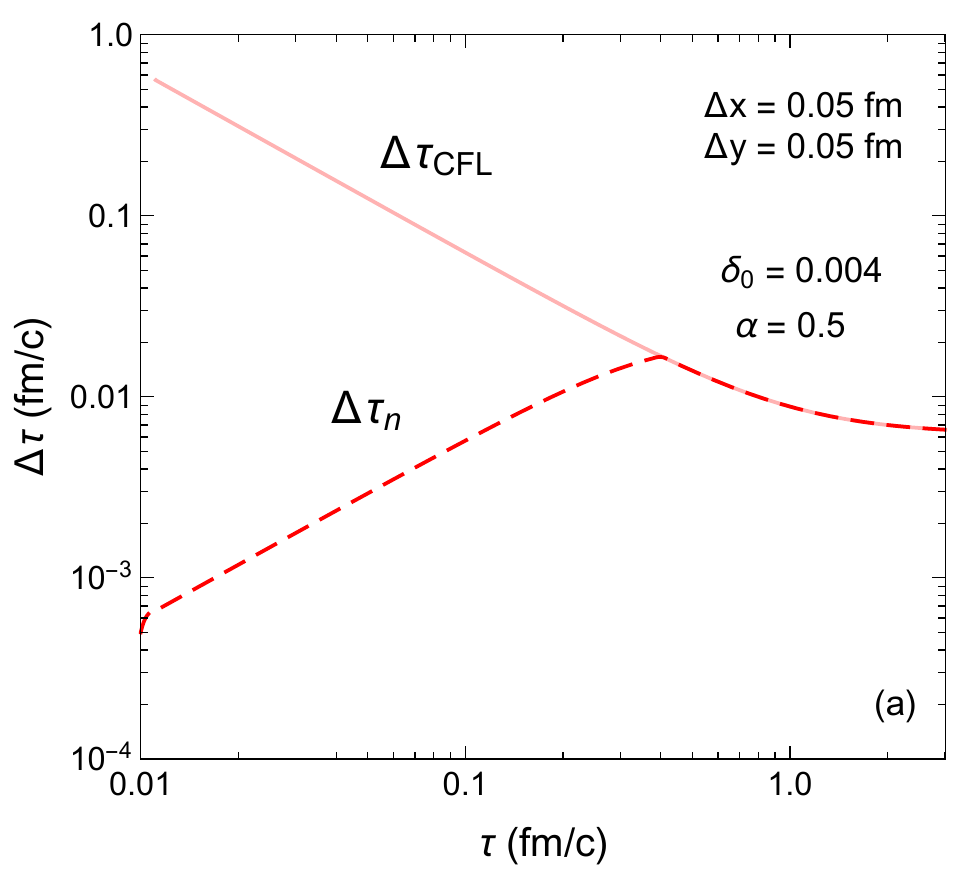}
\centering
\caption{(Color online)
\label{gubser_timestep}
The evolution of the adaptive time step $\Delta \tau_n$ (dashed red) and CFL bound $\Delta \tau_\text{CFL}$ (transparent red) in the conformal Gubser simulation.
}
\end{figure}

Figure~\ref{gubser_2d} shows the evolution of the energy density and transverse fluid velocity $u_\perp = \sqrt{(u^x)^2 + (u^y)^2}$ in the transverse plane at various time frames.\footnote{%
    In order to output the hydrodynamic quantities at specific times, we readjust the adaptive time step whenever we are close to these output times (this is not shown in Fig.~\ref{gubser_timestep}).}
At early times, the very hot and compact fireball expands rapidly along the longitudinal direction and quickly cools down without much change to its transverse shape. Over time, the transverse expansion overtakes the longitudinal expansion, pushing the medium radially outward as a \textit{ring of fire}. One sees that the energy density and transverse fluid velocity maintain their azimuthal symmetry throughout the evolution. However, there are some numerical fluctuations around the lines $y = \pm \,x$ at later times, especially near the peak of $u_\perp$. This is a consequence of using a Cartesian grid with a finite spatial resolution.

Figure~\ref{gubser_test} shows the evolution of the energy density $\ene$, fluid velocity component $u^x$, pressure anisotropy ratio $\PL/\Pperp$ and transverse shear inverse Reynolds number $\sqrt{\pi_\perp {\cdot\,} \pi_\perp }$\,/\,$(\Pperp\sqrt{2})$ along the $x$--axis at multiple time frames. Overall, there is very good agreement between the simulation and semi-analytic solution, except near the local extrema of $u^x$, where the numerical errors are about $1{-}2\%$; these errors can be reduced by using a finer lattice spacing. The transverse shear stress $\piperp$ should remain zero throughout the simulation. However, the transverse shear inverse Reynolds number from the \cpuvah{} simulation is nonzero (albeit small, $\lesssim 2\%$) since the Eulerian grid does not perfectly preserve Gubser symmetry.

Figure~\ref{gubser_timestep} shows the evolution of the adaptive time step in the Gubser simulation. In contrast to Fig.~\ref{bjorken_adaptive}, here $\Delta\tau_n$ becomes bounded by the CFL condition~\eqref{eq:CFL_bound} at $\tau \sim 0.4$ fm/$c$ due to the increasing transverse expansion rate. As a result, the Gubser simulation finishes in 407 time steps, compared to 480 steps if we had used the fixed time step $\Delta \tau = \Delta x / 8$. Although we only gain a slight $1.15\times$ speedup, the adaptive time step is able to resolve the fluid's early-time dynamics more accurately than the fixed time step~\eqref{eq:CFL_fixed} because it is initially independent of the lattice spacing. This property is especially useful when simulating more realistic nuclear collisions, where the lattice spacing required to resolve the participant nucleons' transverse energy deposition is several times coarser than the one used in this test (e.g. $\Delta x = \Delta y \sim 0.1 - 0.3$ fm).
%

\subsection{(3+1)--d conformal hydrodynamic models in central Pb+Pb collisions}
\label{S4.3}
Here we compare (3+1)--dimensional conformal anisotropic hydrodynamics \cpuvah{} to conformal second-order viscous hydrodynamics \vh{} (see Appendix~\ref{appe}), for a central Pb+Pb collision at zero impact parameter ($b = 0$ fm) with static (in Milne coordinates) and smooth initial conditions. To generate the latter we use an azimuthally symmetric \trento{} energy density profile averaged over 2000 fluctuating events \cite{Moreland:2014oya} and extended along the $\eta_s$--direction with a smooth rapidity plateau~\cite{Pang:2018zzo} (see Appendix~\ref{appd} for more information on the \trento{} energy deposition model for high-energy nuclear collisions). The initial temperature at the center of the fireball is $T_{0,\text{center}} = 1.05$ GeV at a starting time of $\tau_0 = 0.01\,\text{fm}/c$ which is the same for both types of hydrodynamic simulations. The fluid is evolved from an initial pressure ratio $R = 10^{-3}$ with a constant specific shear viscosity $\etas = 0.2$; the runs stop once all the fluid cells are below $T_\text{sw} = 0.136$\,GeV, which happens after $\tau_f \sim 7-8$\,fm/$c$.

\begin{figure}[!t]
\includegraphics[width=\linewidth]{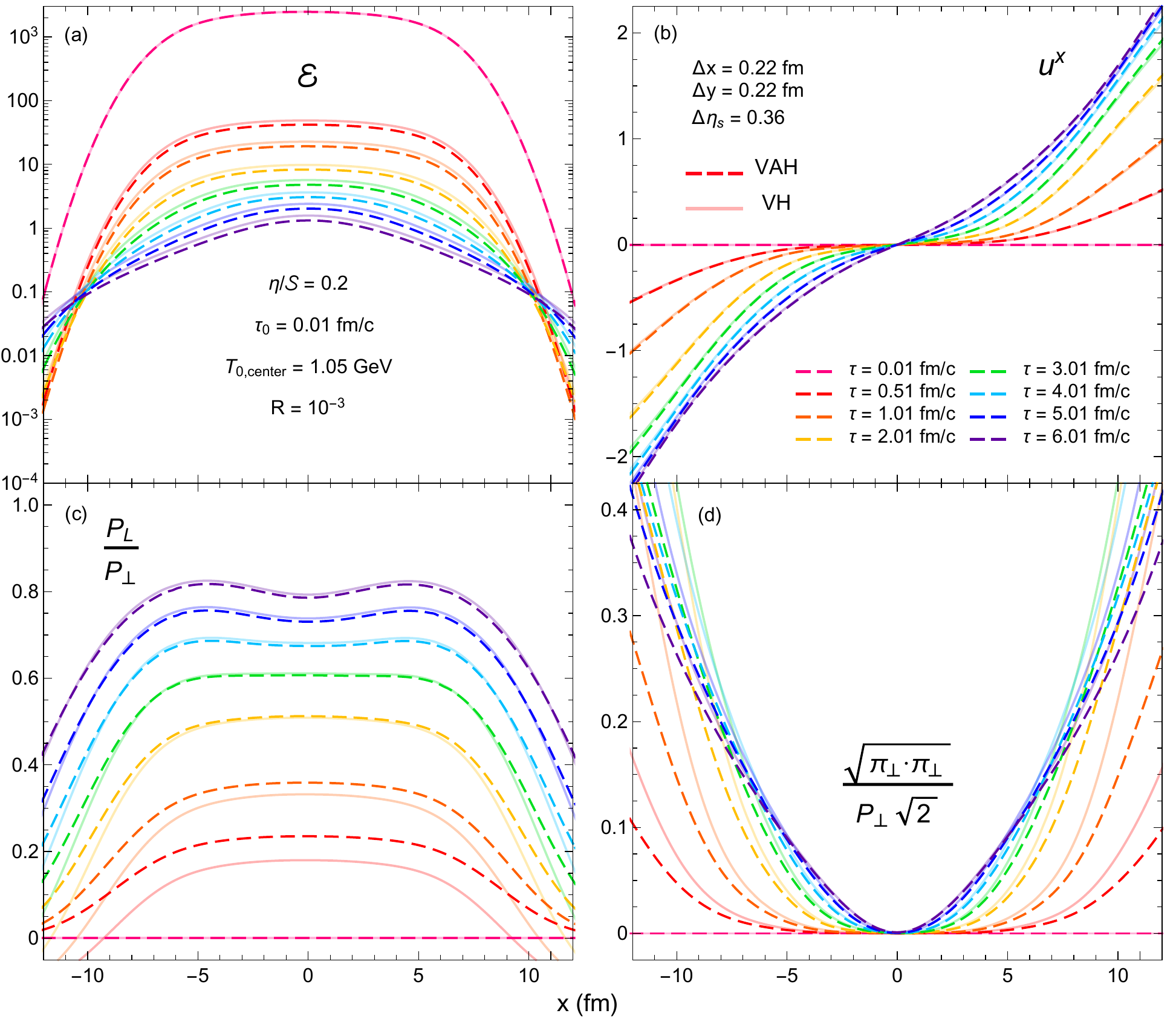}
\centering
\caption{(Color online)
\label{conformal_trento_x}
The evolution of (a) $\ene$ (GeV/fm$^3$), (b) $u^x$, (c) $\PL/\Pperp$ and (d) $\sqrt{\pi_\perp {\cdot\,} \pi_\perp }/(\Pperp\sqrt{2})$ along the $x$--axis ($y{\,=\,}\eta_s{\,=\,}0$), given by conformal anisotropic hydrodynamics (\cpuvah{}, dashed color) and second-order viscous hydrodynamics (\vh{}, transparent color), for the smooth (3+1)--dimensional \trento{} initial condition described in the text.
}
\end{figure}

Figure~\ref{conformal_trento_x} shows the evolution of $\ene$, $u^x$, $\PL/\Pperp$ and $\sqrt{\pi_\perp {\cdot\,} \pi_\perp}$\,/\, $(\Pperp\sqrt{2})$ along the $x$--axis ($y=\eta_s=0)$. Initially, the pressure anisotropy $\PL{-}\Pperp$ in second-order viscous hydrodynamics is so large that the longitudinal pressure turns negative, especially near the grid's edges. In comparison, the pressure ratio $\PL/\Pperp$ in anisotropic hydrodynamics is, at early times, only moderately larger in the central fireball region but quite dramatically different near its transverse edge, staying positive everywhere. The resulting differences in early-time viscous heating create a disparity between the two models for the normalization of the energy density which persists to late times even after the $\PL/\Pperp$ ratios have converged. On the other hand, the transverse flows $u^x$ predicted by the two models are nearly identical even at very early times since they are primarily driven by the transverse pressure gradients $\partial_x \Pperp$ which, in the smooth \trento{} profile, are smaller at early times than the longitudinal gradients $\sim 1/\tau$. The transverse shear stress $\piperp$, which tends to counteract the fluid's transverse acceleration, is larger in viscous hydrodynamics than in anisotropic hydrodynamics, especially along the edges of the fireball.\footnote{The transverse shear stress $\piperp$ is generally nonzero even for azimuthally symmetric flow profiles as long as they do not possess Gubser symmetry.} Overall, however, the hydrodynamic variables are not substantially different along the transverse directions.

\begin{figure}[!t]
\includegraphics[width=\linewidth]{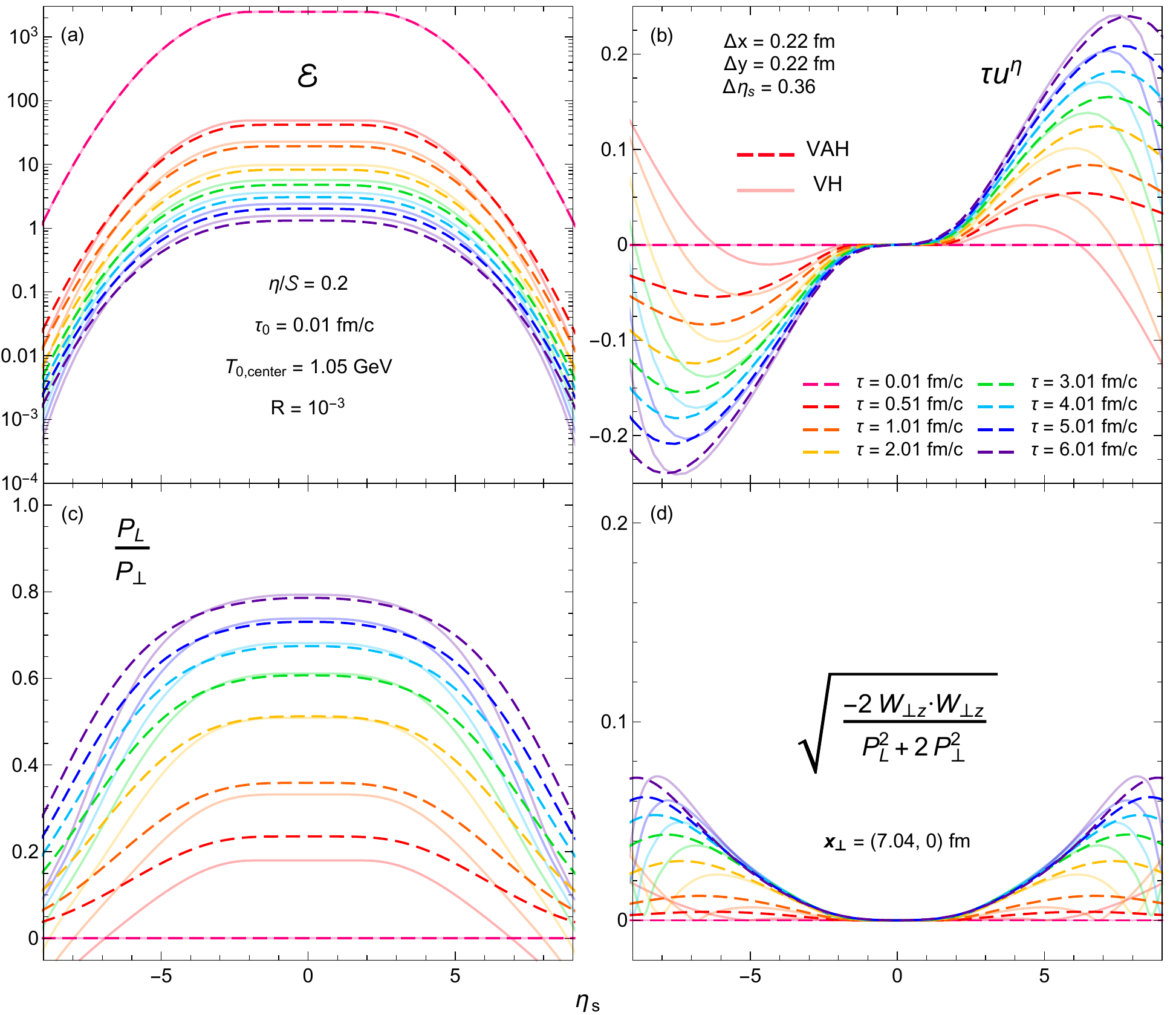}
\centering
\caption{(Color online)
\label{conformal_trento_z}
The evolution of (a) $\ene$ (GeV/fm$^3$), (b) $\tau u^\eta$, (c) $\PL/\Pperp$ and (d) $\sqrt{2 W_{\perp z} {\,\cdot\,} W_{\perp z}}$ /$\sqrt{\mathcal{P}_L^2 {+} 2\mathcal{P}_\perp^2}$ along the $\eta_s$--axis ($x{\,=\,}y{\,=\,}0$), computed with conformal anisotropic hydrodynamics (\cpuvah{}, dashed color) and second-order viscous hydrodynamics (\vh{}, transparent color), for the smooth (3+1)--d \trento{} initial condition described in the text. Note that in panel (d) the $\eta_s$--axis is shifted transversely to $(x,y)=(7.04,0)$\,fm.
}
\end{figure}

The longitudinal profile, however, evolves very differently in anisotropic hydrodynamics (\cpuvah{}) compared to second-order viscous hydrodynamics (\vh{}). This is shown in Fig.~\ref{conformal_trento_z} where we plot the evolution of the dimensionless longitudinal velocity $\tau u^\eta$ (as well as $\ene$ and $\PL/\Pperp$) along the $\eta_s$--axis ($x=y=0$). Similar to Fig.~\ref{conformal_trento_x}, the shear stress \eqref{eq:shear_vh} in viscous hydrodynamics quickly overpowers the equilibrium pressure, making $\PL$ negative initially. Along the longitudinal direction this causes a strong reversal of the longitudinal flow $\tau u^\eta$ at around $|\eta_s| \sim 7$. This unphysical feature (primarily driven by the strong longitudinal expansion rate at early times) results in a longitudinal contraction (in $\eta_s$) of the fireball near its forward and backward edges at the beginning of the simulation. In anisotropic hydrodynamics, the longitudinal pressure remains always positive, allowing for a stronger longitudinal flow profile whose signs are consistent with the gradients of the rapidity distribution of the energy density \eqref{eq:plateau} (and thus of the thermal pressure). This gives anisotropic hydrodynamics a clear advantage over second-order viscous hydrodynamics when simulating the longitudinal dynamics of a heavy-ion collision.

\begin{figure}[!t]
\includegraphics[width=0.6\linewidth]{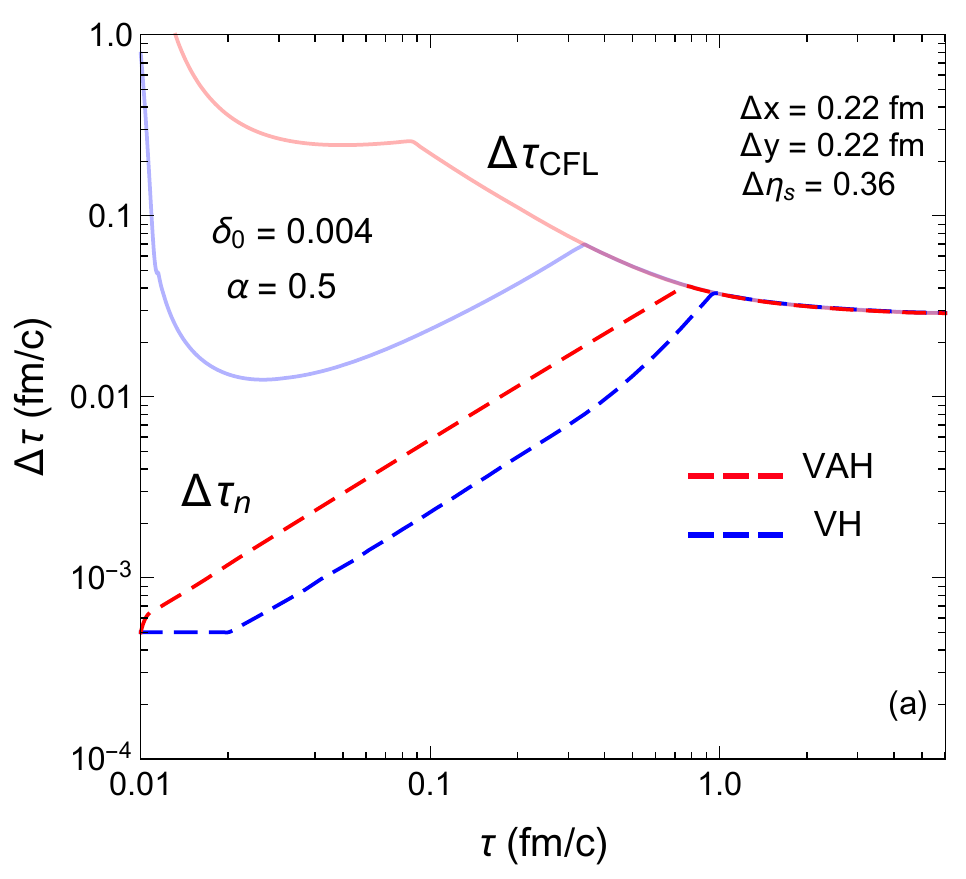}
\centering
\caption{(Color online)
\label{conformal_trento_timestep}
The evolution of the adaptive time step $\Delta \tau_n$ (dashed color) and CFL bound $\Delta \tau_\text{CFL}$ (transparent color) in conformal anisotropic hydrodynamics (red) and second-order viscous hydrodynamics (blue) for the smooth (3+1)--d \trento{} initial condition.
}
\end{figure}

In panel (d) of Fig.~\ref{conformal_trento_z} we also plot the spacetime rapidity dependence of the inverse Reynolds number $\sqrt{-2W_{\perp z}{\,\cdot\,} W_{\perp z} / (\mathcal{P}_L^2 + 2\mathcal{P}_\perp^2)}$. The longitudinal momentum diffusion current $\Wperp$ is nonzero only in regions that have both longitudinal and transverse gradients. For this reason we shift the $\eta_s$--axis transversely to $(x,y) = (7.04,0)$\,fm, which corresponds to the mid-right region of the grid. As expected, the momentum diffusion current is weakest around mid-rapidity, $\eta_s \sim 0$, where the fireball profile is approximately longitudinally boost-invariant, and strongest along the sloping edges of the rapidity plateau \eqref{eq:plateau}. However, this longitudinal edge region is also the place where its inverse Reynolds number differs most strongly between anisotropic and standard viscous hydrodynamics. Overall, $\Wperp$ only makes up a tiny fraction of the total shear stress \eqref{eq:shear_vh} in anisotropic hydrodynamics. The situation may change for initial conditions whose longitudinal and transverse gradients are larger than the ones used in this test comparison.

Finally, we plot in Fig.~\ref{conformal_trento_timestep} the adaptive time step for each of the two hydrodynamic simulations. One sees that in \vh{} $\Delta\tau_n$ stagnates until $\tau{\,\sim\,}0.02$\,fm/$c$ (see footnote \ref{adaptive_floor}) while the one in \cpuvah{} starts increasing immediately. This indicates that at early times viscous hydrodynamics generally has a faster evolution rate and therefore requires a smaller initial time step compared to anisotropic hydrodynamics. Nevertheless, both adaptive time steps remain below their CFL bound (initially dominated by the longitudinal velocity $u^\eta$) until $\tau{\,\sim\,}0.8{-}1$ fm/$c$. Notice that the CFL bounds for \vh{} and \cpuvah{} become virtually identical since they have almost the same transverse velocity profile (see Fig.~\ref{conformal_trento_x}b).

\subsection{Non-conformal Bjorken flow test}
\label{S4.4}

\begin{figure}[!t]
\includegraphics[width=\linewidth]{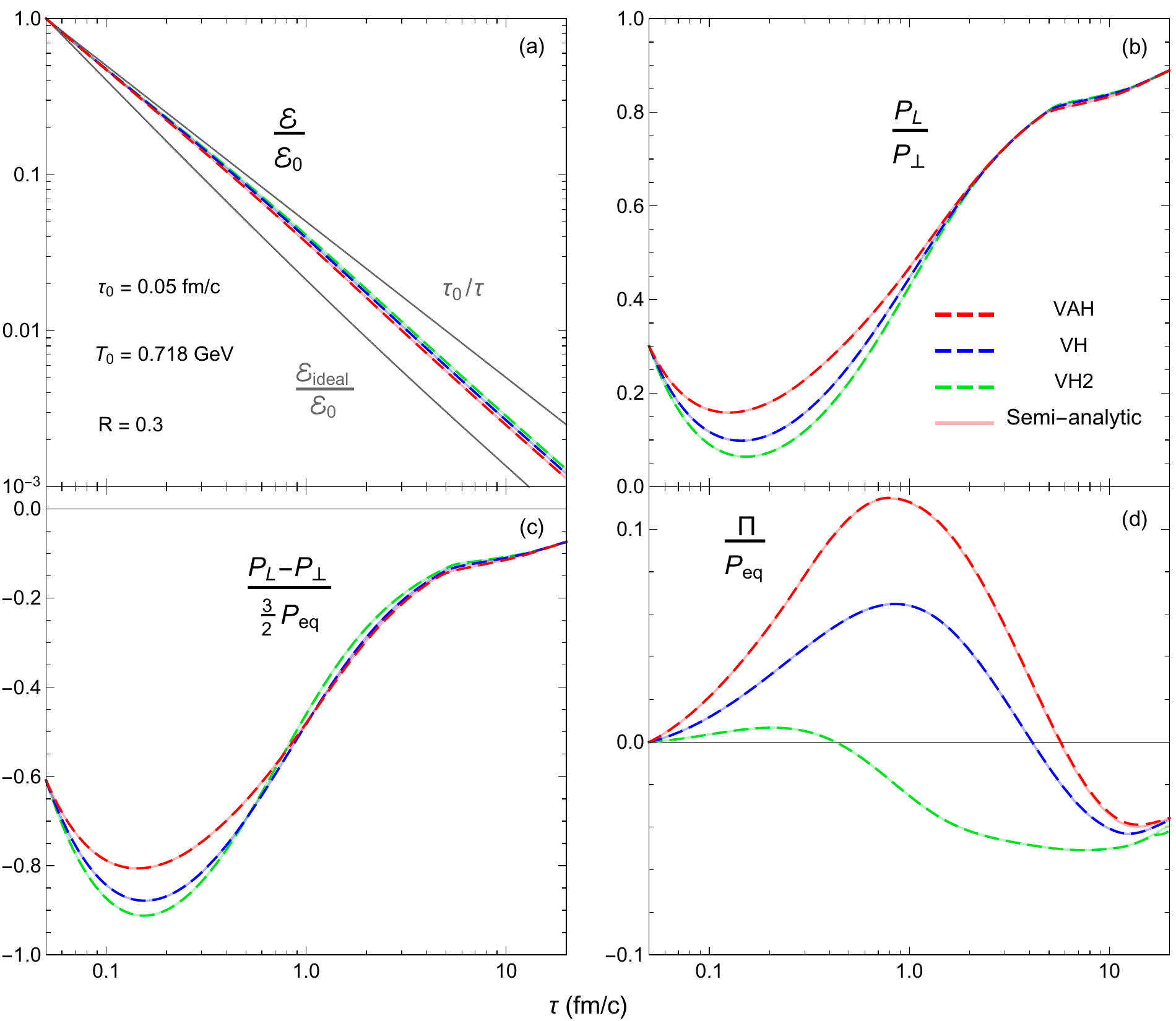}
\centering
\caption{(Color online)
\label{nonconformal_bjorken}
Non-conformal Bjorken evolution of (a) $\ene/\ene_0$, (b) the pressure ratio $\PL/\Pperp$, (c) the pressure anisotropy $\frac{2}{3}(\PL{-}\Pperp)$, and (d) the bulk viscous pressure $\Pi$ (the latter two normalized to the equilibrium pressure $\Peq$), given by anisotropic hydrodynamics (\cpuvah{}, dashed red), quasiparticle viscous hydrodynamics (\vh{}, dashed blue) and standard viscous hydrodynamics (\vh{}2, dashed green), along with their semi-analytic solutions (transparent color). The lower gray curve in panel (a) is from an ideal hydrodynamic calculation (i.e. $\PL = \Pperp = \Peq(\ene)$ and $\etas = \zetas = 0$).
}
\end{figure}

Now we turn to testing our non-conformal hydrodynamic simulation with the QCD equation of state from Fig.~\ref{eos}. First, we run (3+1)--d anisotropic hydrodynamics with Bjorken initial conditions and compare it to the semi-analytic solution of the equations \cite{McNelis:2018jho}
\bs
\label{eq:nonconformal_aniso_bjorken}
\beal
    \partial_\tau \ene &= - \frac{\ene + \PL}{\tau} \,,
\\
    \partial_\tau \PL &= \frac{\Peq {-} \bar{\mathcal{P}}}{\tau_\Pi} - \frac{\PL {-} \Pperp}{3\tau_\pi/2 } + \frac{\bar\zeta_z^L}{\tau}\,,
\\
    \partial_\tau \Pperp &= \frac{\Peq {-} \bar{\mathcal{P}}}{\tau_\Pi} + \frac{\PL {-} \Pperp}{3\tau_\pi} + \frac{\bar\zeta_z^\perp}{\tau}\,,
\\
    \partial_\tau B &= \frac{B_\text{eq} {-} B}{\tau_\Pi} + \frac{\ene{+}\PL}{m\tau} \frac{dm}{d\ene}(\ene{-}2\Pperp{-}\PL{-}4B)\,;
\end{align}
\es
here we use the shear and bulk relaxation times \eqref{eq:tau_r}--\eqref{eq:beta_r}, with the viscosities given by Eqs.~\eqref{eq:etas}--\eqref{eq:zetas} (see Fig.~\ref{viscosities}). The non-conformal anisotropic transport coefficients $\bar\zeta_z^L$ and $\bar\zeta_z^\perp$ are given by Eqs.~(\ref{eq:pl_coeff}a) and~(\ref{eq:pt_coeff}a), respectively. We start the simulation at $\tau_0 = 0.05$ fm/$c$ with initial temperature $T_0 = 0.718$ GeV and initial pressure ratio $R = 0.3$, and evolve the system until $\tau_f \sim 80$ fm/$c$ when the temperature falls below $T_\text{sw} = 0.136$\,GeV (we plot results only up to $\tau = 20$ fm/$c$). We also repeat this for quasiparticle viscous hydrodynamics (\vh{}) and standard viscous hydrodynamics (\vh{}2) (see footnote \ref{VH12}).\footnote{%
    \vh{} and \vh{}2 use the same equation of state $\Peq(\ene)$ as \cpuvah{} but different transport coefficients, see Appendix~\ref{appe} and Ref.~\cite{McNelis:2018jho}.}
Ideally, we would have preferred using smaller values for $\tau_0$ and $R$ to better match the longitudinally free-streaming initial condition used in the conformal hydrodynamic simulations (see Secs.~\ref{S4.1}--\ref{S4.3} and footnote \ref{free_stream}). However, we found that at earlier times the \cpuvah{} simulation has greater difficulty reconstructing the anisotropic variables ($\Lambda$, $\alpha_\perp$, $\alpha_L$), especially directly at initialization. This indicates that our anisotropic hydrodynamic model \cite{McNelis:2018jho} which integrates the QCD equation of state consistently with quasiparticle kinetic transport coefficients, has its limitations.

Figures~\ref{nonconformal_bjorken}a,b show the evolution of the normalized energy density $\ene/\ene_0$ and pressure ratio $\PL/\Pperp$ from the three hydrodynamic simulations. We also disentangle the longitudinal and transverse pressures' viscous components $\PL {-} \Peq = \frac{2}{3}\Delta \mathcal{P} {+} \Pi$ and $\Pperp {-} \Peq = -\frac{1}{3}\Delta \mathcal{P} {+} \Pi$ into the pressure anisotropy $\Delta \mathcal{P} = \PL{-}\Pperp$ and bulk viscous pressure $\Pi = \frac{1}{3}(\PL{+}2\Pperp)-\Peq$; their evolutions relative to $\Peq$ are shown in Figs.~\ref{nonconformal_bjorken}c,d. Although our initial condition for the pressure ratio is somewhat ad hoc, it quickly reaches its minimum value at $\tau \sim 0.1$ fm/c, allowing for the energy density to closely follow its free-streaming limit at the beginning of the simulation. The $\PL/\Pperp$ ratio is typically larger in anisotropic hydrodynamics compared to the two viscous hydrodynamic models (until $\tau \sim 1$ fm/$c$). This is mainly due to differences in the higher-order transport coefficients associated with the pressure anisotropy (i.e. beyond $\etas$ and $\tau_\pi$)~\cite{Bazow:2013ifa,Florkowski:2013lza,Bazow:2015cha,McNelis:2018jho}. Although in Bjorken flow the bulk viscous pressure $\Pi$ is much smaller than the pressure anisotropy $\Delta\mathcal{P}$, it evolves very differently in standard viscous hydrodynamics \vh{}2 compared to quasiparticle viscous hydrodynamics \vh{} and anisotropic hydrodynamics \cpuvah{}. Since the dimensionless bulk relaxation time $\tau_\Pi T < 1$ in standard viscous hydrodynamics \vh2 (see the dashed curve in Fig.~\ref{relaxation}b), the bulk viscous pressure quickly reaches its Navier-Stokes solution $\Pi_\text{NS} = -\zeta/\tau$ at $\tau \sim 2$\,fm/$c$. In contrast, the bulk relaxation time used in anisotropic hydrodynamics \cpuvah{} and quasiparticle viscous hydrodynamics \vh{} is significantly larger (solid curve in Fig.~\ref{relaxation}b). As a consequence, it takes a much longer time for $\Pi$ to relax to $\Pi_\text{NS}$ \cite{Tinti:2016bav, McNelis:2018jho}.

One also sees that the simulations agree very well with their corresponding semi-analytic solutions (continuous lines in transparent color). Although we do not display them explicitly here, the numerical errors are small enough that we can unambiguously distinguish the different dynamics of the three hydrodynamic models in comparison studies discussed in the next subsections.

\subsection{(3+1)--d non-conformal hydrodynamic models in central Pb+Pb collisions}
\label{S4.5}
\subsubsection{Smooth \trento{} initial conditions}
\label{S4.5.1}
Next we run non-conformal hydrodynamics for a central Pb+Pb collision with smooth, azimuthally symmetric \trento{} initial conditions; we set the initial time to $\tau_0 = 0.05$\,fm/$c$ and the initial pressure ratio to $R = 0.3$. The initial energy density profile is almost identical to the one in Sec.~\ref{S4.3}, except the normalization $\propto 1/\tau_0$ is five times smaller, making the initial temperature at the center of the fireball $T_{0,\text{center}} = 0.718$\,GeV. We evolve the system until, at $\tau_f{\,\sim\,}15{\,-\,}16$\,fm/$c$, all fluid cells are below $T_\text{sw} = 0.136$\,GeV.

\begin{figure}[thbp]
\includegraphics[width=0.94\linewidth]{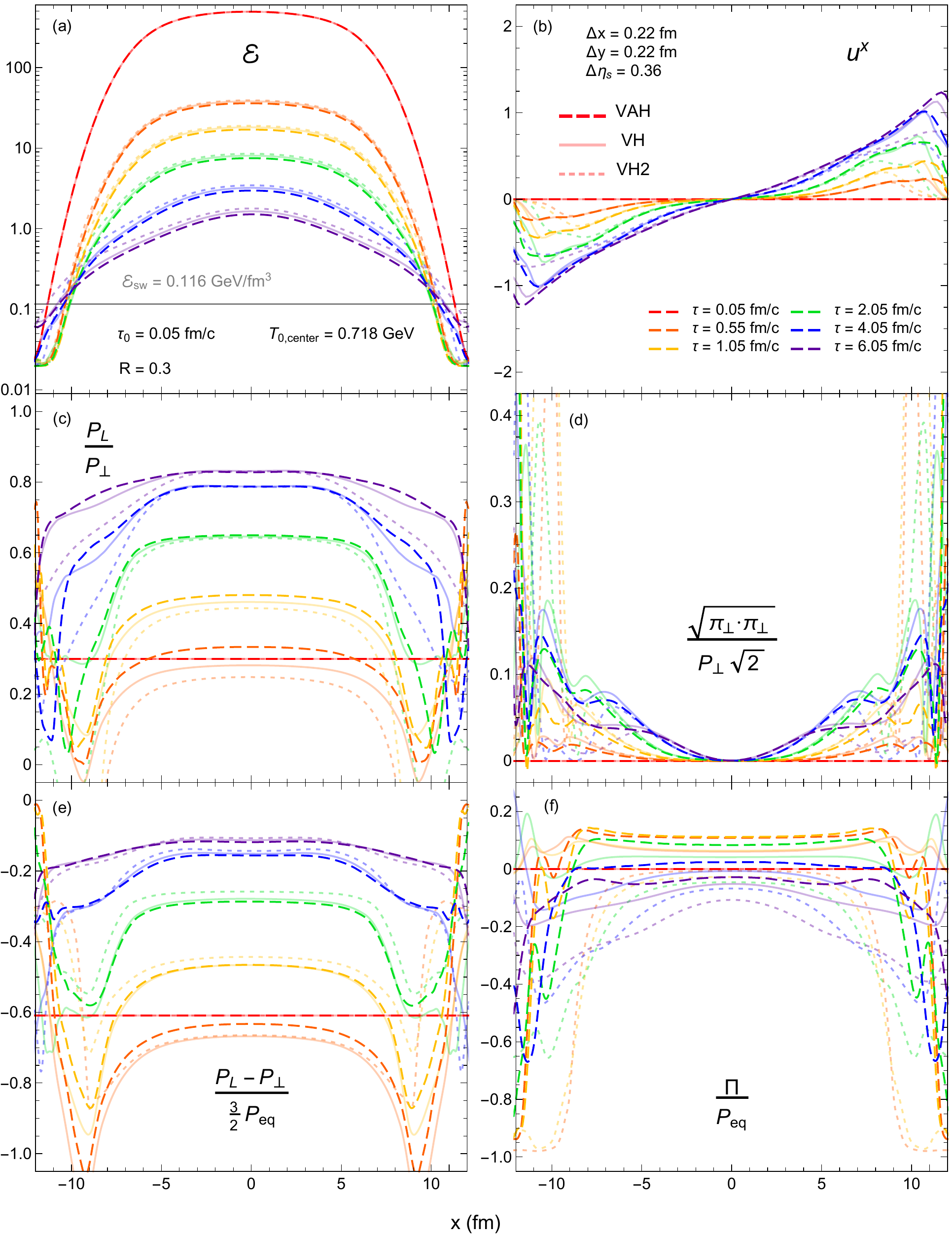}
\centering
\caption{(Color online)
\label{lattice_trento_x}
    The evolution of (a) $\ene$ (GeV/fm$^3$), (b) $u^x$, (c) $\PL/\Pperp$, (d) $\sqrt{\pi_\perp {\cdot\,} \pi_\perp }/(\Pperp\sqrt{2})$, (e) $\frac{2}{3}(\PL{-}\Pperp)/\Peq$ and (f) $\Pi/\Peq$ along the $x$--axis ($y{\,=\,}\eta_s{\,=\,}0$), given by non-conformal anisotropic hydrodynamics (\cpuvah{}, dashed color), quasiparticle viscous hydrodynamics (\vh{}, transparent color) and standard viscous hydrodynamics (\vh2, dotted transparent color), for the smooth (3+1)--d \trento{} initial condition. The minimum energy density parameter is set to $\ene_\text{min} = 0.02$\,GeV/fm$^3$.
}
\end{figure}

Figure~\ref{lattice_trento_x} shows, for the first $\Delta\tau = 6$ fm/$c$ of the simulation, the evolution of $\ene$, $u^x$, $\PL/\Pperp$, $\sqrt{\pi_\perp \cdot \pi_\perp}$/$(\Pperp\sqrt{2})$, $\frac{2}{3}(\PL-\Pperp)$/$\Peq$ and $\Pi/\Peq$ given by the three hydrodynamic models along the $x$--axis ($y=\eta_s = 0$). Here the fireball maintains higher temperatures for a longer duration than in conformal hydrodynamics since the QCD equilibrium pressure is much weaker than its Stefan--Boltzmann limit \eqref{eq:stefan} (see Fig.~\ref{eos}a). But similar to Fig.~\ref{conformal_trento_x}a, the energy density in anisotropic hydrodynamics cools down at a slightly faster rate compared to viscous hydrodynamics. This is initially due to the moderately larger $\PL/\Pperp$ ratio in the central fireball region, which closely follows the Bjorken evolution in Fig.~\ref{nonconformal_bjorken}b. The viscous corrections to the longitudinal pressure increase as we move towards the edges of the fireball, but $\PL$ still remains positive in anisotropic hydrodynamics. The longitudinal pressure in standard viscous hydrodynamics, however, is strongly negative at early times (and also, to a lesser extent, for the quasiparticle case), although it does not significantly alter the fluid's transverse dynamics. The $\PL/\Pperp$ ratios start to overlap at later times, mainly driven by the pressure anisotropies' convergence in Fig.~\ref{lattice_trento_x}e.

Compared to Fig.~\ref{conformal_trento_x}b, the transverse flow $u^x$ in non-conformal hydrodynamics is significantly weaker due to the softer QCD equation of state. The dip in the speed of sound at the quark-hadron phase transition (see Fig.~\ref{eos}b) also flattens the transverse velocity gradients near the edges of the fireball. This causes a significant decrease in the transverse shear stress relative to the conformal case in Fig.~\ref{conformal_trento_x}d. Among the hydrodynamic models, we see that \vh{}2 produces the smallest $u^x$ around the edges of the fireball since it has the largest negative bulk viscous pressure there. This is due to its ability to converge to the Navier-Stokes solution shortly after the simulation begins. In contrast, the bulk viscous pressure in \vh{} hovers slightly above zero across the grid before falling down toward negative values at $\tau \sim 3$ fm/$c$. Even then, it hardly catches up to the bulk viscous pressure curves in \vh{}2 since the differences between their bulk relaxation times grow as we move away from the center of the fireball. Overall, the bulk viscous pressure in \vh{} is much weaker compared to \vh{}2, which leads to a stronger transverse flow. Finally, the \cpuvah{} simulation has a bulk viscous pressure that is qualitatively similar to \vh{} (except at the edges of the grid) but slightly lags behind it. As a result, it generates the largest transverse flow out of the three simulations. The study here illustrates how the evolution of the bulk viscous pressure and transverse fluid velocity strongly depends on the choice of hydrodynamic model, specifically the temperature-dependent model for the bulk relaxation time $\tau_\Pi$. This is likely to have important implications for the phenomenological extraction of the bulk viscosity $\zetas$~\cite{Everett:2020yty,Everett:2020xug}.

\begin{figure}[thbp]
\includegraphics[width=0.94\linewidth]{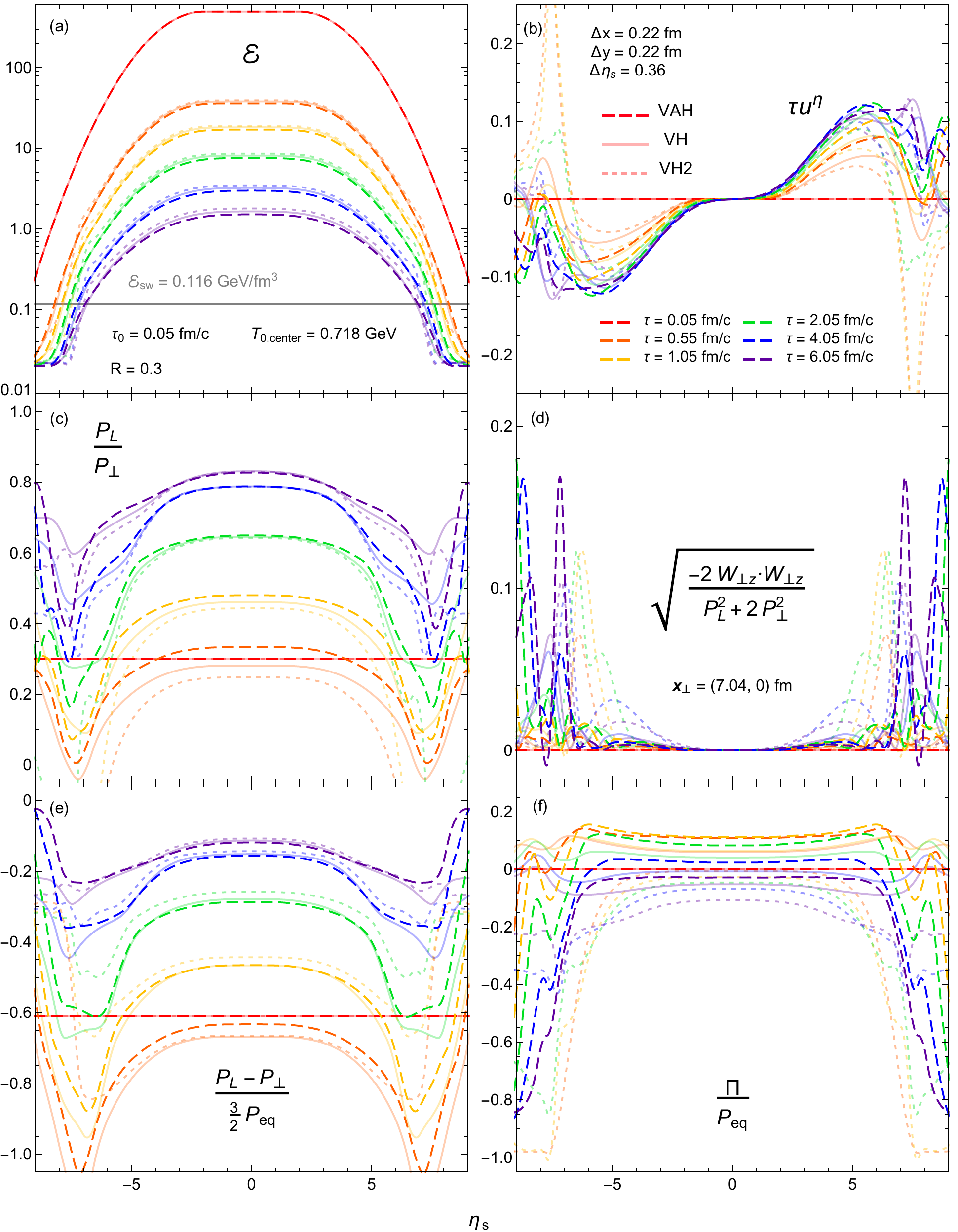}
\centering
\caption{(Color online)
\label{lattice_trento_z}
    The evolution of (a) $\ene$ (GeV/fm$^3$), (b) $\tau u^\eta$, (c) $\PL/\Pperp$, (d) $\sqrt{2 W_{\perp z} {\,\cdot\,} W_{\perp z}}$ /$\sqrt{\mathcal{P}_L^2 {+} 2\mathcal{P}_\perp^2}$, (e) $\frac{2}{3}(\PL{-}\Pperp) / \Peq$, and (f) $\Pi/\Peq$ along the $\eta_s$--axis ($x=y=0$), given by non-conformal anisotropic hydrodynamics (\cpuvah{}, dashed color), quasiparticle viscous hydrodynamics (\vh{}, transparent color) and standard viscous hydrodynamics (\vh2, dotted transparent color), for the smooth (3+1)--d \trento{} initial condition. Note that in panel (d) the $\eta_s$--axis is shifted transversely to $(x,y)=(7.04,0)$\,fm.
}
\end{figure}

Next, we compare the fireballs' longitudinal evolution along the $\eta_s$--axis ($x=y=0$) in Figure~\ref{lattice_trento_z}. When comparing the longitudinal evolution of the energy density and viscous pressures between the three hydrodynamic models, we note qualitatively similar features as already observed in Fig.~\ref{lattice_trento_x} for their transverse evolution. While the transverse velocities shown in Fig.~\ref{lattice_trento_x}b varied between the models as a result of differences in the transverse pressures (mainly the bulk viscous pressures), here differences in their longitudinal pressures result in very different longitudinal flow profiles (Figure~\ref{lattice_trento_z}b). Most notably, the large negative $\PL$ in standard viscous hydrodynamics (\vh{}2) initially causes $\tau u^\eta$ to reverse sign very sharply in the forward and backward rapidity regions. Only by imposing strong regulations on the \vh{}2 simulation is it able to recover the correct direction of longitudinal flow at later times as the gradients relax. The same breakdown at early times can also be seen in quasiparticle viscous hydrodynamics (\vh{}) although there the situation is not nearly as bad. With \cpuvah{} we can maintain positive longitudinal pressures so that the fireball can expand without imploding in regions where the longitudinal gradients are large. This greatly reduces the amount of regulation needed for the viscous pressures.

\begin{figure}[!t]
\includegraphics[width=0.6\linewidth]{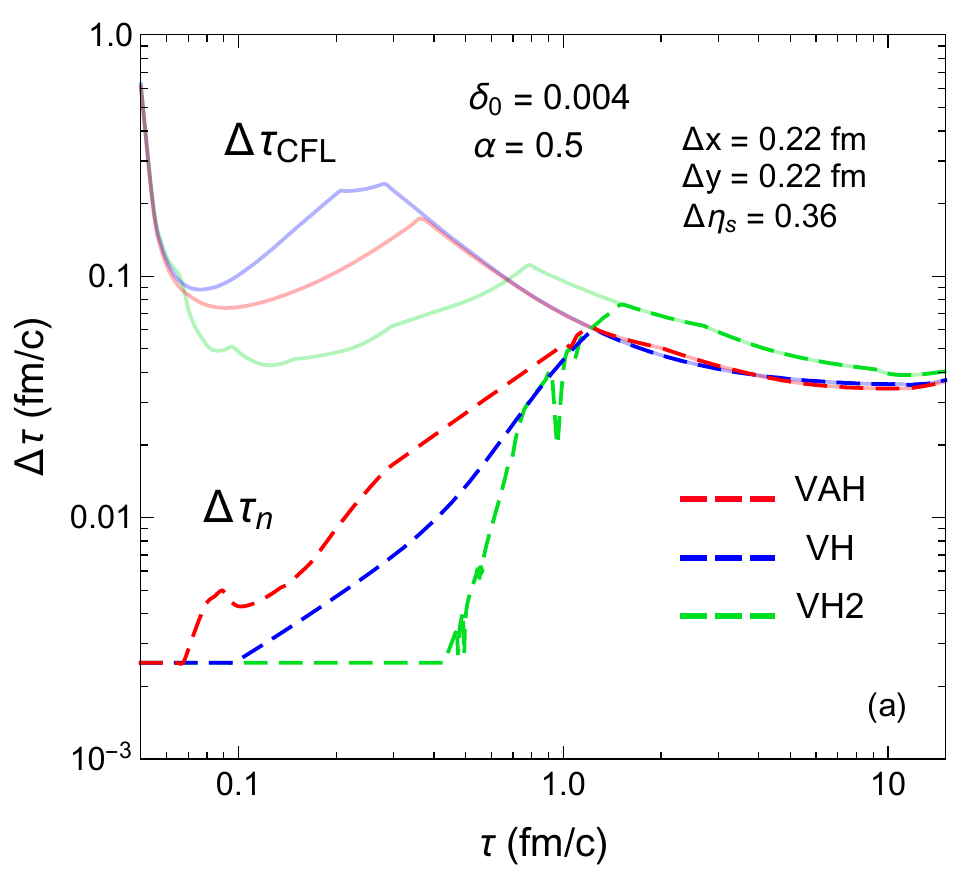}
\centering
\caption{(Color online)
\label{lattice_trento_timestep}
    The evolution of the adaptive time step $\Delta \tau_n$ (dashed color) and CFL bound $\Delta \tau_\text{CFL}$ (transparent color) in non-conformal anisotropic hydrodynamics (\cpuvah{}, red), quasiparticle viscous hydrodynamics (\vh{}, blue) and standard viscous hydrodynamics (\vh2, green), for the smooth (3+1)--d \trento{} initial condition.
}
\end{figure}

Finally, in Fig.~\ref{lattice_trento_timestep} we plot the adaptive time step for each of the three hydrodynamic simulations. One sees that in anisotropic hydrodynamics (\cpuvah{}) $\Delta\tau_n$ increases steadily until hitting the CFL bound at $\tau\sim 1$ fm/$c$. Its behavior is similar in quasiparticle viscous hydrodynamics (\vh{}). The two CFL bounds are slightly separated due to differences between their transverse velocity profiles (see Fig.~\ref{lattice_trento_x}b). On the other hand, the adaptive time step in standard viscous hydrodynamics (\vh2) does not pick up until $\tau \sim 0.4$ fm/$c$ (see footnote~\ref{adaptive_floor}). Even then, it fluctuates up and down before reaching its CFL bound, which is higher than the other two bounds since its transverse flow is suppressed by a large bulk viscous pressure. Although the code runs faster per time step (see Table~\ref{tab:runtime_table}), standard viscous hydrodynamics takes considerably more time steps to evolve the fluid than the other two hydrodynamic models, suggesting that it has a harder time resolving the fireball evolution in the presence of large gradients.

\begin{figure}[!t]
 \makebox[\textwidth][c]{\includegraphics[width=0.99\linewidth]{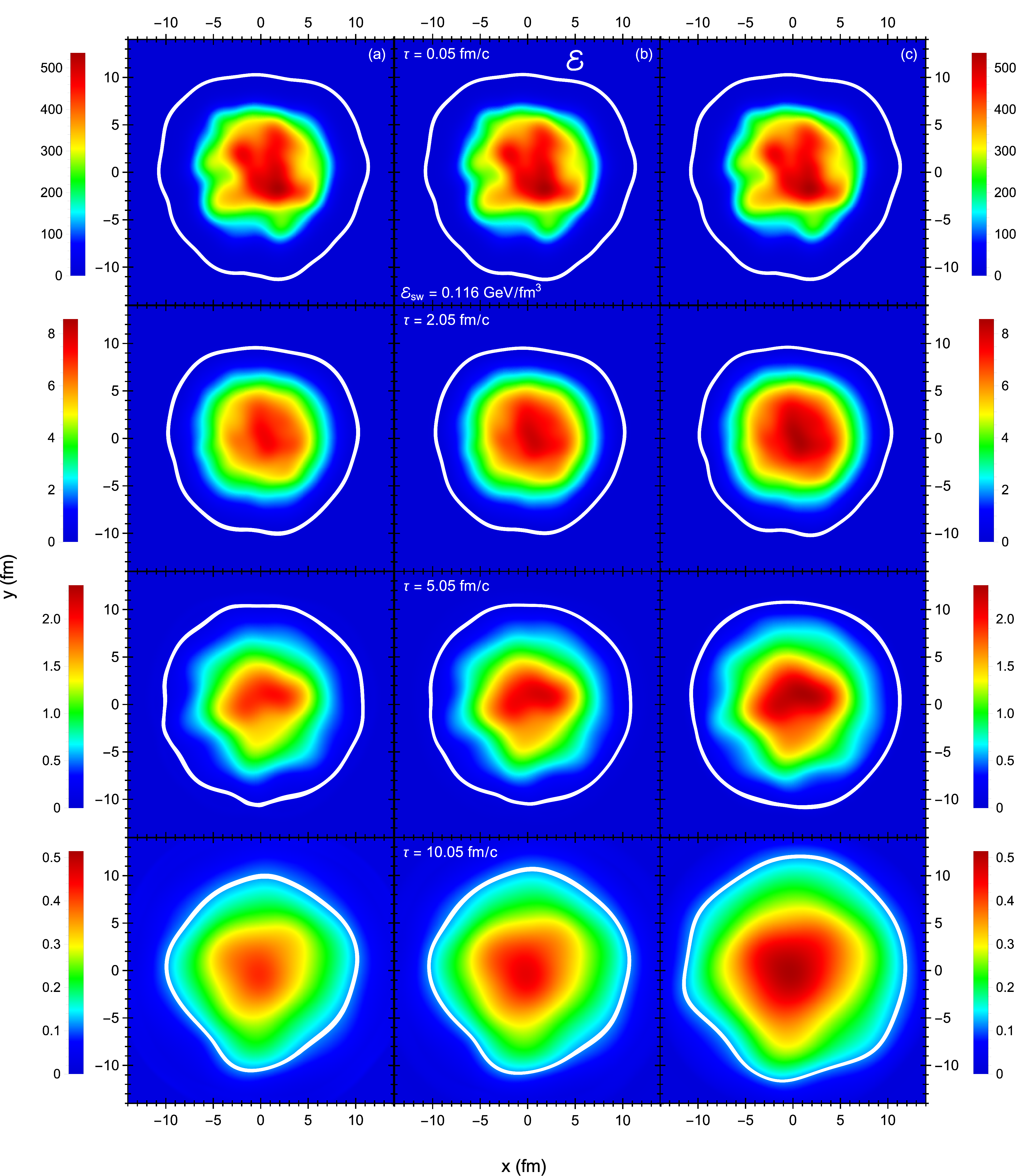}}
\centering
\caption{(Color online)
\label{fluctuating_profile}
    The evolution of the QCD energy density profile (GeV/fm$^3$) in the central transverse plane $(\eta_s{\,=\,}0)$, given by non-conformal anisotropic hydrodynamics (\cpuvah{}, left column), quasiparticle viscous hydrodynamics (\vh{}, middle column) and standard viscous hydrodynamics (\vh{}2, right column) for the fluctuating (3+1)--d \trento{} event described in the text. The white contour lines are $\eta_s{\,=\,}0$ slices at the shown time frames of a particlization hypersurface of constant energy density $\ene_\text{sw} = 0.116$\,GeV/fm$^3$.
}
\end{figure}

\subsubsection{Fluctuating \trento{} initial conditions}
\label{S4.5.2}

Finally we study the differences among the three non-conformal hydrodynamic models for a central Pb+Pb collision with a fluctuating initial energy density profile, using the same model parameters as in Sec.~\ref{S4.5.1}. In the \trento{} model used in this work, the energy density profile only has fluctuations in the transverse directions; the longitudinal profile is modeled with a finite rapidity plateau with smooth slopes at its longitudinal ends (see Appendix~\ref{appd}).

Figure~\ref{fluctuating_profile} shows the evolution of the fluctuating energy density profile in the transverse plane $(\eta_s = 0)$, along with the transverse slice of a particlization hypersurface of constant energy density $\ene_\text{sw} = 0.116$\,GeV/fm$^3$, for the three hydrodynamic simulations. Since anisotropic hydrodynamics generally has a smaller shear stress than viscous hydrodynamics (especially at early times), the transverse fluctuations across its fireball show the least dissipation or smearing. 
As a result, it can convert the initial-state eccentricities into anisotropic flow slightly more efficiently. The pressure anisotropy $\PL{-}\Pperp$ and bulk viscous pressure $\Pi$ mainly influence the overall size of the fireball on the particlization hypersurface, affecting the final-state particle yields. The initially strong longitudinal expansion rapidly cools down the system and temporarily shrinks the fireball size at early times; a more positive longitudinal pressure helps cool the fireball even further. Compared to the two viscous hydrodynamic models, anisotropic hydrodynamics has a larger longitudinal pressure, which means its hypersurface has a slightly narrower waist. At later times, the transverse expansion overtakes the longitudinal expansion, increasing the fireball size. Although anisotropic hydrodynamics has the largest transverse flow, it also has the smallest bulk viscous pressure, enabling the fireball to cool faster and evaporate more quickly. This ultimately results in a smaller maximum fireball size. In contrast, standard viscous hydrodynamics has a much larger bulk viscous pressure. This extends the fireball's lifetime by about $1$ fm$/c$ relative to the one in anisotropic hydrodynamics, allowing it to grow larger in size.
\begin{figure}[thbp]
\includegraphics[width=0.98\linewidth]{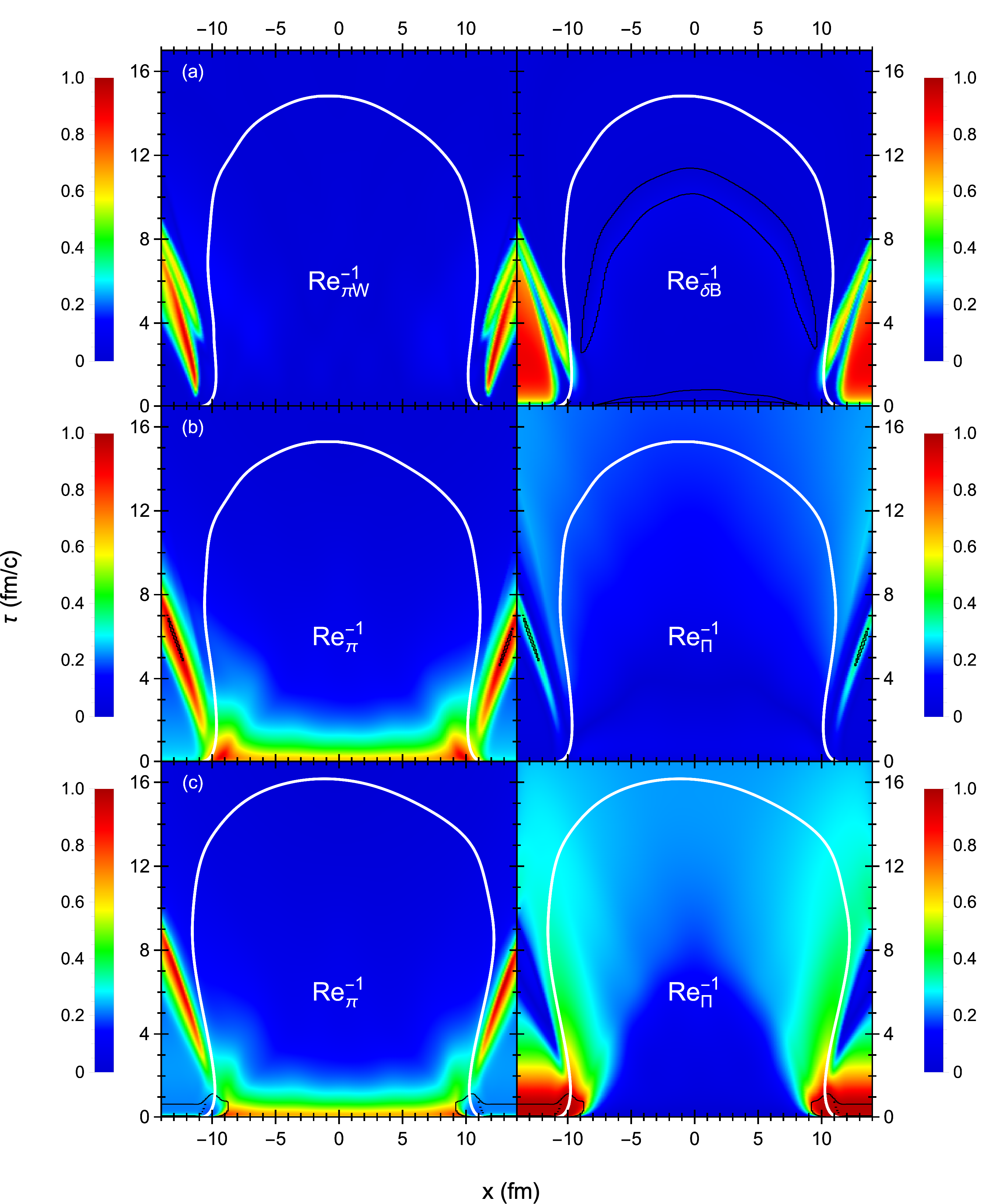}
\centering
\caption{(Color online)
\label{freezeout_x}
    The $\tau{-}x$ slice at $y{\,=\,}\eta_s{\,=\,}0$ of the residual shear and mean-field inverse Reynolds numbers in anisotropic hydrodynamics (\cpuvah{}, top row) and the shear and bulk inverse Reynolds numbers in quasiparticle viscous hydrodynamics (\vh{}, middle row) and standard viscous hydrodynamics (\vh{}2, bottom row), for the fluctuating (3+1)--d \trento{} event. The white contour lines are $\tau{-}x$ slices ($y=\eta_s=0$) of the same particlization hypersurface as Fig.~\ref{fluctuating_profile}. Spacetime regions that are regulated according to Sec.~\ref{S3.5} and App.~\ref{appe} are circled in black.
}
\end{figure}
\begin{figure}[thbp]
\includegraphics[width=0.98\linewidth]{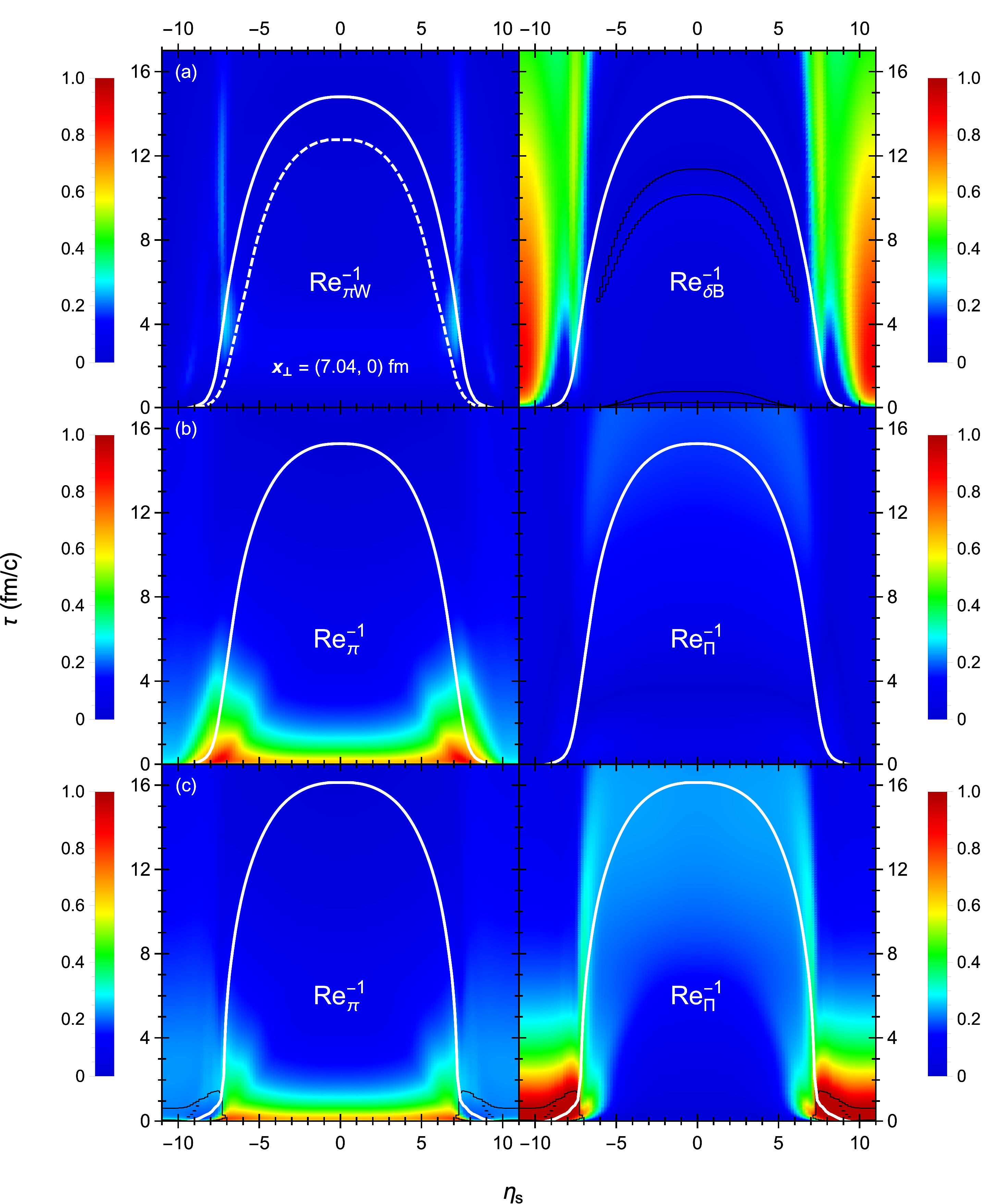}
\centering
\caption{(Color online)
\label{freezeout_z}
    The same as Fig.~\ref{freezeout_x} but for the $\tau{-}\eta_s$ plane at $x{\,=\,}y{\,=\,}0$. The $\eta_s$--axis in the upper left panel is shifted transversely to $(x,y) = (7.04, 0)$\,fm; the corresponding hypersurface slice (dashed white) is smaller than the one at $x{\,=\,}y{\,=\,}0$ (solid white).
}
\end{figure}

In Figs.~\ref{freezeout_x} and~\ref{freezeout_z} we plot the inverse Reynolds numbers from the three hydrodynamic simulations in the $\tau{-}x$ plane at $y{\,=\,}\eta_s{\,=\,}0$ and in the $\tau{-}\eta_s$ plane at $x{\,=\,}y{\,=\,}0$, respectively. Traditionally, the inverse Reynolds number measures the validity of the hydrodynamic expansion around local equilibrium \cite{Bazow:2016yra, Shen:2014vra}. Since anisotropic hydrodynamics expands the energy-momentum tensor~\eqref{eq:Tmunu} around an anisotropic background (i.e. $T^\munu_a = \ene u^\nu u^\nu + \PL z^\mu z^\nu - \frac{1}{2}\Pperp \Xi^\munu$), its validity is measured by the residual shear inverse Reynolds number~\cite{Bazow:2017ewq}
\bs
\allowdisplaybreaks
\beal
    \text{Re}^{-1}_{\pi W} =&\, \sqrt{\frac{\pi_\perp {\cdot\,} \pi_\perp - 2 W_{\perp z} {\,\cdot\,} W_{\perp z}}{\mathcal{P}_L^2 + 2\mathcal{P}_\perp^2}}\,,
\end{align}
\es
as opposed to the shear and bulk inverse Reynolds numbers in second-order viscous hydrodynamics \cite{Bazow:2016yra, Shen:2014vra}:
\be
    \text{Re}^{-1}_{\pi} =\, \frac{\sqrt{\pi \cdot \pi}}{\Peq\sqrt{3}}\,,\qquad
    \text{Re}^{-1}_{\Pi} =\, \frac{|\Pi|}{\Peq}\,,
\ee
with $\pi {\,\cdot\,} \pi = \pi_\munu \pi^\munu$. One sees that the residual shear inverse Reynolds numbers stay much smaller than unity inside the fireball, indicating that a first-order expansion in $O(\text{Re}^{-1}_{\pi W})$ is sufficient to capture the residual shear corrections to the anisotropic hydrodynamic equations. While there is no need to regulate their strength here, the regulation scheme might be needed for collision events with sharper initial-state fluctuations.

We also plot the contribution of the nonequilibrium mean-field component $\delta B$ to the bulk viscous pressure in anisotropic hydrodynamics:
\be
    \text{Re}_{\delta B}^{-1} = \frac{|\delta B|}{\Peq}\,.
\ee
We find that $\delta B$ is moderately large and positive near the edges of the fireball but small and negative inside the fireball. However, it is necessary to regulate the mean-field in the latter region via Eq.~\eqref{eq:dB_reg} to prevent unstable growth. Specifically, the instability seems to originate near the quark-hadron phase transition, where the driving force proportional to the kinetic trace anomaly $\ene^{(k)} - 2\mathcal{P}_\perp^{(k)} - \mathcal{P}_L^{(k)} = \ene - 2\Pperp - \PL - 4 B$ is at its strongest while the bulk relaxation rate $\tau_\Pi^{-1}$ is near a minimum.  After applying the regulation~\eqref{eq:dB_reg} for a brief period, $\delta B$ evolves freely from that point onward. While this initial instability is unfortunate, the overall impact of the non-equilibrium mean-field component on the longitudinal and transverse pressures inside the fireball region is limited.

In second-order viscous hydrodynamics, the shear inverse Reynolds number is large for a short period of time $\Delta\tau \sim 1$ fm/$c$ after the collision. The bulk inverse Reynolds number in standard viscous hydrodynamics is also very large near the peak of the bulk viscosity $\zetas$ during the same time frame. To prevent the code from crashing, our regulation scheme suppresses both the shear stress and bulk viscous pressure (see Appendix~\ref{appe}). Although it is still technically possible to run the heavy-ion simulation with standard viscous hydrodynamics, we cannot escape the effects of regulation around the base of the particlization hypersurface. In contrast, the viscous pressures in quasiparticle viscous hydrodynamics often do not require regulation\footnote{Even in the absence of regulation, the longitudinal pressure in quasiparticle viscous hydrodynamics can still be negative in some regions at early times ($\tau < 0.6$ fm/$c$) (e.g. see Figs.~\ref{lattice_trento_x} and~\ref{lattice_trento_z}).} since the bulk inverse Reynolds number is quite small.

We close this section with a comparison of the particlization hypersurfaces in the $\tau{-}x$ and $\tau{-}\eta_s$ planes. In Fig.~\ref{freezeout_x} one sees that standard viscous hydrodynamics has the largest hypersurface along the $x$--direction since its large bulk viscous pressure generates the most viscous heating. It also has the longest lifetime $\tau_f \sim 17$\,fm/$c$, for the same reason. In contrast, it has a pretty narrow waist along the $\eta_s$--direction because its longitudinal flow profile $u^\eta$ initially contracts the medium in the rapidity direction. In anisotropic hydrodynamics, the longitudinal flow begins transporting the medium away from the collision zone, following the direction of the longitudinal pressure gradients.
As a result, its hypersurface has a wider waist than the one in standard viscous hydrodynamics (about $\Delta\eta_s \sim 0.5$ larger for $\tau \sim 1-2$ fm/$c$) along the $\eta_s$--direction.

\section{Benchmarks}
\label{S5}

In this Section we benchmark the typical computational time needed to run (2+1)--d non-conformal hydrodynamic simulations of Pb+Pb collisions at LHC energies ($\sqrt{s_\text{NN}} = 2.76$\,TeV) with fluctuating initial conditions and different values for the impact parameter $b$ and model parameters. We also benchmark the OpenMP--accelerated runtime of (3+1)--d non-conformal hydrodynamics for a central Pb+Pb collision ($b = 0$\,fm) with the smooth \trento{} initial condition and model parameters from Sec.~\ref{S4.5.1}.

\begin{table}[t]
\centering
\footnotesize
\setlength{\tabcolsep}{0.875em} 
{\renewcommand{\arraystretch}{1.5}
 \begin{tabular}{|c|c|c|c|c|}
 \hline
 & Mean time $(s)$  & Max time $(s)$ & Mean time per step $(s)$ & Mean \# steps \\
 \hline
 VAH & 204 (136) & 1840 (1630) & 0.639 (0.410) & 272 \\ \hline
 VH  & 75.0 (49.9) & 561 (320) & 0.183 (0.119) & 350 \\ \hline
 VH2 & 89.7 (56.9) & 463 (358) & 0.175 (0.109) & 461 \\ \hline
 \end{tabular}}
 \caption{The mean runtime, maximum runtime, mean runtime per step and mean number of steps of the (2+1)--d non-conformal hydrodynamic simulations of Pb+Pb collisions on the fixed grid (auto--grid). (The number of time steps in the last column does not change for the auto--grid.) These statistics were generated by running a total of $10000$ simulations ($200$ test parameter samples times $50$ fluctuating \trento{} events) on single-core Intel Xeon E5-2680 v4 CPUs for each of the three hydrodynamic models.}
\label{tab:runtime_table}
\end{table}

\subsection{(2+1)--d non-conformal hydrodynamics on a fixed grid}
\label{S5.1}

In this test, we generate $200$ random samples for the impact parameter $b$ and Bayesian model parameters \cite{Everett:2020yty, Everett:2020xug}. The impact parameter is sampled from a piecewise linear distribution,
\be
P(b) = \Bigg\{
\begin{array}{ll}
      b / (2R_A^2) & (0 \leq b \leq 2R_{A}), \\
      0 & (b{\,>\,} 2R_{A}),
\end{array}
\ee
where we set $R_A = 7$ fm for the lead nuclear radius. The model parameters
\be
    P_B = \left[N, p, w, d_\text{min}, \sigma_k, T_\text{sw}, (\etas)_\text{kink}, T_\eta, a_\text{low}, a_\text{high}, (\zetas)_\text{max}, T_\zeta, w_\zeta, \lambda_\zeta\right]
\ee
are each sampled from a distribution that is uniform within the finite intervals used in the JETSCAPE SIMS Bayesian analysis of heavy-ion collisions \cite{Everett:2020yty,Everett:2020xug}.\footnote{%
    The \trento{} initial condition model parameters ($N$, $p$, $w$, $d_\text{min}$, $\sigma_k$), are defined in Appendix~\ref{appd}~\cite{Moreland:2014oya} and the viscosity model parameters ($(\etas)_\text{kink}$, $T_\eta$, $a_\text{low}$, $a_\text{high}$, $(\zetas)_\text{max}$, $T_\zeta$, $w_\zeta$, $\lambda_\zeta$) are discussed in Sec.~\ref{S2.6.1}. Finally, the switching temperature $T_\text{sw}$ determines the particlization hypersurface of constant energy density $\ene_\text{sw} = \ene(T_\text{sw})$.}%
$^,$\footnote{%
    We exclude the ratio between the dimensionless shear relaxation time and shear viscosity $b_\pi = \tau_\pi T / (\etas)$ as a continuous model parameter~\cite{Everett:2020yty,Everett:2020xug} and instead allow it to vary between discrete hydrodynamic models. The rapidity plateau model parameters $\eta_\text{flat}$ and $\sigma_\eta$ (see Appendix~\ref{appd}) are also not considered in this benchmark test.}
%
\begin{figure}[t]
\includegraphics[width=0.6\linewidth]{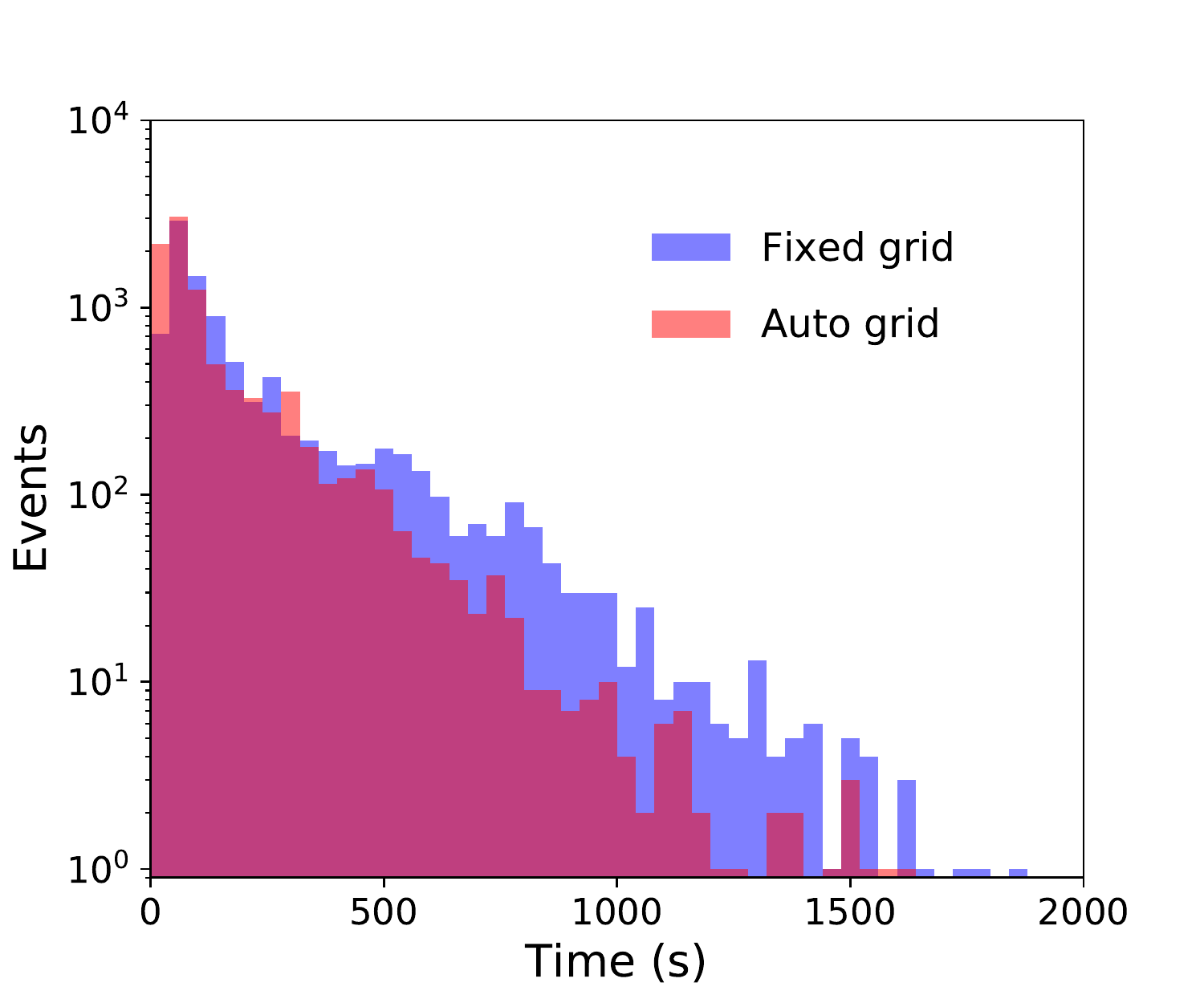}
\centering
\caption{(Color online)
\label{auto_grid_runtime}
    The runtime distribution of the (2+1)--d non-conformal anisotropic hydrodynamic simulations on the fixed grid (blue) and auto-grid (red).
}
\end{figure}

For each parameter sample $\{b,P_B\}$ we run $50$ (2+1)--d non-conformal hydrodynamic simulations with fluctuating \trento{} initial conditions on a $30$\,fm\,$\times$\,$30$\,fm transverse grid ($\eta_s = 0$).\footnote{%
    All other runtime parameters (e.g. the initial time $\tau_0$ and pressure ratio $R$) are the same as the ones used in Sec.~\ref{S4.5}.}%
$^,$\footnote{%
    We found that $93\%$ of the anisotropic hydrodynamic simulations finish successfully (the second-order viscous hydrodynamic simulations had a success rate of $99{+}\%$). For certain model parameter sets and/or fluctuating initial conditions, the code fails to reconstruct the anisotropic variables in the cold dilute regions, either because the maximum number of Newton iterations was exhausted or the kinetic longitudinal pressure $\mathcal{P}_{L}^{(k)}$ turned negative.} The resulting runtime statistics of the three hydrodynamic models are shown in Table~\ref{tab:runtime_table}. On average, it takes about $85s$ to run each (2+1)--d viscous hydrodynamic simulation on the fixed grid, compared to $200s$ for anisotropic hydrodynamics. The additional routine in Sec.~\ref{S3.4} makes the runtime per step in anisotropic hydrodynamics about $3.5\times$ slower than in viscous hydrodynamics, but this is partially compensated by the fewer time steps required to finish each simulation.

Figure~\ref{auto_grid_runtime} shows the runtime distribution of the anisotropic hydrodynamic simulations on the fixed grid. One sees that some runs take longer than others, depending on the parameter values used. For example, a smaller impact parameter $b$ increases the participant nucleon multiplicity, which extends the fireball's lifetime. The runtime is also very sensitive to the nucleon width parameter $w$: halving the value for $w$ doubles the spatial resolution required\footnote{%
    In this work, we set the transverse lattice spacing to $\Delta x = \Delta y = \frac{1}{5}w$.}
to capture the spatial variations of the fluctuating \trento{} profile (see Appendix~\ref{appd}) and therefore increases the runtime by roughly a factor of eight. As a result, the maximum runtime can be about an order of magnitude longer than the mean runtime (see Table~\ref{tab:runtime_table}).

\subsection{(2+1)--d non-conformal hydrodynamics on an automated grid}
\label{S5.2}

The previous benchmark test was performed on a fixed transverse grid of $30$\,fm\,$\times$\,$30$\,fm. Although this grid is large enough to fit all possible fireball sizes produced in Pb+Pb collisions, evolving peripheral collisions ($b \lesssim 2R_A$) on it is computationally inefficient. A large grid is also unnecessary for certain Bayesian model parameter combinations that further decrease the fireball size. In this section, we present a regression algorithm that predicts the fireball radius for a given set of parameters $\{b,P_B\}$. With this, we can automatically optimize the grid size to save computational time and memory.

\begin{figure}[!t]
\includegraphics[width=\linewidth]{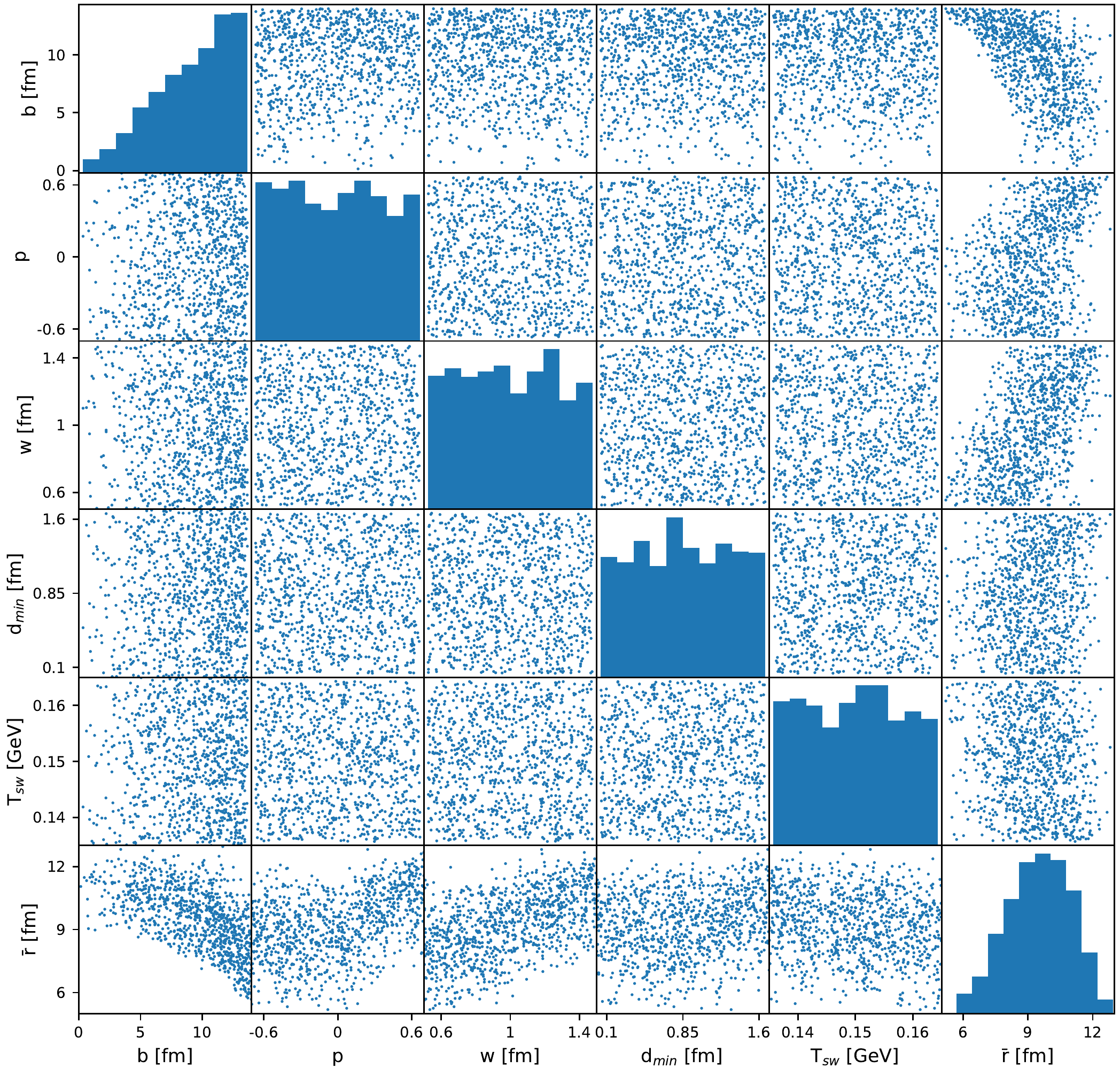}
\centering
\caption{(Color online)
\label{auto_training}
    A subset of the scattering matrix used to train the automated grid for (2+1)--d non-conformal anisotropic hydrodynamic simulations.
}
\end{figure}

To train the regression model, we generate $1000$ training parameter samples arranged in the column vectors
\be
\label{eq:feature}
    \boldsymbol{A} = \left[\boldsymbol{b}, \boldsymbol{P}_B \right],
\ee
to construct our feature matrix \cite{10.5555/3153997}. For each parameter sample, we run a single hydrodynamic simulation with a smooth, event-averaged \trento{} initial condition and compute the mean fireball radius $\bar{\boldsymbol{r}}$ (again arranged in vector); this will be our target variable\footnote{%
    We define the fireball radius $r$ as the maximum transverse radius of the particlization hypersurface in a given hydrodynamic simulation.}
\be
\label{eq:target}
    \boldsymbol{y} = \boldsymbol{\bar{r}} \,.
\ee
A sample subset of the scattering matrix containing the five most important features is shown in Fig.~\ref{auto_training}. As expected, the fireball size strongly decreases with the impact parameter. There are also some moderately positive correlations between the fireball radius and initial condition parameters
(e.g. increasing the nucleon width parameter $w$ increases the spread of the participant nucleons' thickness function, see Appendix~\ref{appd}, Eq.~\eqref{eq:Tpp}). While the viscosity parameters have a much smaller effect on the fireball radius individually, the model's fit to the train-validation data improves when we include all of them.

\begin{table}[t]
\centering
\small
\setlength{\tabcolsep}{0.875em} 
{\renewcommand{\arraystretch}{1.5}
 \begin{tabular}{|c|c|c|c|}
 \hline
 & VAH  & VH & VH2  \\
 \hline
 $\delta\bar{r}_\text{RMSE}$ & $0.29$ fm & $0.19$ fm & $0.22$ fm \\ \hline
 $\alpha$ & $0.013$ & $0.009$ & $0.007$ \\ \hline
 \end{tabular}}
 \caption{
    The cross-validated root-mean-square error $\delta\bar{r}_\text{RMSE}$ and regularization parameter $\alpha$ of the Lasso regression fit for each hydrodynamic model.}
\label{tab:lasso}
\end{table}

Next, we fit a cubic polynomial Lasso regression model with standardization feature scaling~\cite{10.5555/3153997} to the training data \eqref{eq:feature}--\eqref{eq:target}. The regularization parameter $\alpha$ is chosen to minimize the root-mean-square error $\delta\bar{r}_\text{RMSE}$, averaged over a five-fold cross-validation \cite{10.5555/3153997}.
The values for the cross-validated error and regularization parameter of the regression fit are listed in Table~\ref{tab:lasso}.

Finally, we rerun the simulations with fluctuating \trento{} initial conditions, using the 200 test parameter samples from earlier. For each parameter sample, we launch the regression model to predict the mean fireball radius $\bar{r}_\text{pred}$ and set the transverse grid lengths to
\be
L_x = L_y = 2\left(\bar{r}_\text{pred} + \delta\bar{r}_\text{RMSE} + \ell \right)\,,
\ee
where the margin parameter $\ell$ gives the fireball additional room to expand within the grid. Here we set $\ell = 2.5$\,fm not only for extra space but also to account for statistical fluctuations of the fireball radius in fluctuating \trento{} events, which were not considered in the training routine.\footnote{%
    With enough computing resources, the user has the option to run multiple fluctuating hydrodynamic simulations per parameter sample to produce a statistical distribution for the fireball radius $r$, which can be characterized with a mean radius and standard deviation $\boldsymbol{Y} = [\boldsymbol{\bar{r}}, \boldsymbol{\sigma_r}]$. For more details, we refer the reader to \url{https://github.com/mjmcnelis/fireball}.}
We find that the automated grid has about a $99.6\%$ success rate of enclosing the fireball without touching its edges. Fig.~\ref{auto_grid_runtime} shows the anisotropic hydrodynamic runtime distribution on the automated grid. On average, the automated grid algorithm reduces the grid area by $32 - 34\,\%$ relative to the fixed grid and provides the simulation with an additional $1.5 - 1.6\times$ speedup (see Table~\ref{tab:runtime_table}).

While (3+1)--d hydrodynamic simulations stand to benefit the most from an optimized grid volume, we would need to implement more realistic longitudinal initial conditions than the model used in this work before retraining the regression model to predict (in addition to its transverse radius) the fireball's average elongation along the $\eta_s$--direction. There may also be additional parameters that characterize such fluctuating (3+1)--d initial-state models, depending on the collision system of interest. Since it takes much more computational resources to produce such a training data set, we leave the (3+1)--d automated grid for future work.

\begin{figure}[!t]
\includegraphics[width=0.6\linewidth]{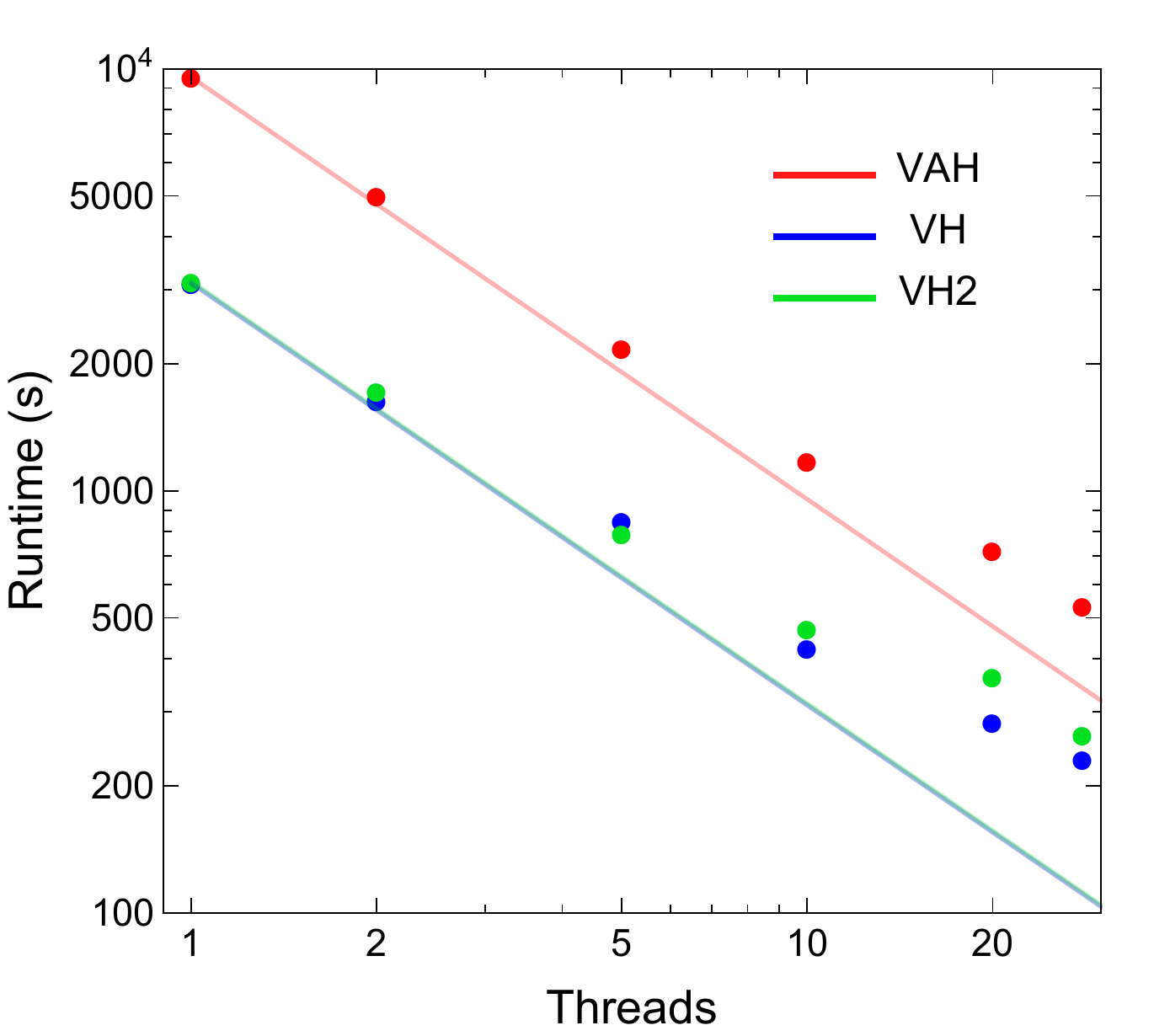}
\centering
\caption{(Color online)
\label{omp_test}
    The simulation runtime of (3+1)--d non-conformal anisotropic hydrodynamics (red dots), quasiparticle viscous hydrodynamics (blue dots) and standard viscous hydrodynamics (green dots) on an Intel Xeon E5-2680 v4 multi-core processor for the smooth \trento{} initial condition from Sec.~\ref{S4.5.1}. The ideal runtime is inversely proportional to the number of parallel CPU threads (same-colored continuous lines).
}
\end{figure}

\subsection{(3+1)--d non-conformal hydrodynamics with OpenMP acceleration}
\label{S5.3}

The code also includes the option to use OpenMP acceleration on a multi-core processor. This feature is especially useful for speeding up (3+1)--d hydrodynamic simulations, which run typically about two orders of magnitude longer than (2+1)--d simulations \cite{Bazow:2016yra}.\footnote{%
    For even faster runtimes, one can parallelize relativistic hydrodynamics on a graphics processing unit \cite{Bazow:2016yra, Pang:2018zzo}. The present version of \cpuvah{} does not yet offer this option.}
Fig.~\ref{omp_test} shows the runtime of the (3+1)--d non-conformal hydrodynamic simulations from Sec.~\ref{S4.5.1}. The anisotropic hydrodynamic simulation takes about two and a half hours to finish on a single-core CPU; the runtime of central Pb+Pb collisions can be about an order of magnitude longer or shorter than this depending on the values used for the nucleon-width parameter $w$ and rapidity plateau parameters $\eta_\text{flat}$ and $\sigma_\eta$ (see Appendix~\ref{appd}). On an Intel Xeon E5-2680 multi-core processor, we can significantly reduce the simulation time to about nine minutes with 28 CPU threads. However, we do not achieve a perfect speedup because some parts of the routine are not easily parallelizable (e.g. the construction of the particlization hypersurface).

\section{Summary and outlook}
\label{S6}

In this work we developed a (3+1)--dimensional anisotropic fluid dynamical simulation that evolves both the pre-equilibrium and viscous hydrodynamic stages of ultrarelativistic nuclear collisions. We validated the code's performance by reproducing the semi-analytic solutions of conformal and non-conformal Bjorken flow and conformal Gubser flow on a Cartesian grid. Thanks to the adaptive time step algorithm derived in Sec.~\ref{S3.6}, we can accurately capture the early-time dynamics of Bjorken and Gubser flow while finishing the simulation within a reasonable number of iterations. We also compared anisotropic hydrodynamics to two second-order viscous hydrodynamic models in central Pb+Pb collisions. Apart from the apparent sensitivity of the bulk viscous pressure evolution to the bulk relaxation time, the three hydrodynamic models have similar transverse dynamics, at least in the mid-rapidity region $\eta_s = 0$. However, the fluid's longitudinal evolution varies greatly near the longitudinal edges of the fireball. We showed that the longitudinal pressure in (3+1)--dimensional anisotropic hydrodynamics stays positive even in the presence of large gradients at very early times, unlike in viscous hydrodynamics. This causes the fluid to initially expand outward along the spacetime rapidity direction, as expected from the outward-pointing longitudinal gradients of the thermal pressure, reducing the risk of cavitation at the beginning of the simulation. With the new development presented here, we can for the first time model even the very early pre-equilibrium dynamics stage at $\tau_0{\,\ll\,}\tau_\text{hydro}$ with a QCD equation of state, as opposed to the conformal equation of state implicit in other pre-equilibrium models \cite{Chesler:2010bi, Chesler:2013lia, Liu:2015nwa, Chesler:2015bba, Attems:2016tby, Heller:2016rtz, Keegan:2016cpi, Kurkela:2018wud}.

In the near term, we plan to run \cpuvah{} in the JETSCAPE framework with the Maximum a Posteriori (MAP) model parameters extracted in Refs.~\cite{Everett:2020yty, Everett:2020xug}.\footnote{%
    Several model parameters associated with the conformal free-streaming module would not be used in our code, which replaces the free-streaming stage by anisotropic hydrodynamics.}
Since the JETSCAPE SIMS hybrid model \cite{Everett:2020yty, Everett:2020xug} uses conformal free-streaming and standard viscous hydrodynamics to evolve the pre-equilibrium and hydrodynamic stages, we will replace these two modules with our code and study the changes to the hadronic observables, such as the transverse momentum spectra and $p_T$--differential anisotropic flows. A full analysis of all available data will need to wait for a Bayesian recalibration of the full evolution model using \cpuvah{} as its hydrodynamic core.

Integration of \cpuvah{} with the JETSCAPE framework also requires an updated version of the particlization module \IS{} \cite{McNelis:2019auj} with a new option that allows choosing the leading-order anisotropic distribution $f_a$ (plus residual shear corrections $\delta\tilde{f}$) for the hadronic distribution in the Cooper--Frye formula \cite{PhysRevD.10.186, McNelis:2019auj}. Once completed, this will allow us to investigate how different selections among a discrete set of hydrodynamic models affect both the theoretical description of heavy-ion experimental observables and the shear and bulk viscosity constraints inferred from such data--theory comparisons. In particular, it would be interesting to test the predictions from anisotropic hydrodynamics against a variety of initializations of second-order viscous hydrodynamics using different pre-hydrodynamic evolution models\footnote{%
    Anisotropic hydrodynamics (\cpuvah{}) as presented here is not meant to be integrated with a pre-hydrodynamic module, especially one that uses a conformal approximation. This is mainly due to the technical difficulties in initializing the mean-field and anisotropic variables \cite{McNelis:2018jho}. Available second-order viscous hydrodynamic codes can read in the energy-momentum tensor from a pre-hydrodynamic module and thus initialize the dynamical and inferred variables more easily, but this has not yet been implemented in the \cpuvah{} code.}
(or no pre-hydrodynamic stage at all), such as the recent study done in Ref.~\cite{NunesdaSilva:2020bfs}.

The current code only evolves the fluid's energy-momentum tensor components and ignores the effects of net-baryon density and diffusion. Second-order viscous hydrodynamic codes with nonzero $n_B$ and $V_B^\mu$ have already been practically implemented \cite{Monnai:2012jc, Denicol:2018wdp}, and others that initialize viscous hydrodynamics with dynamical sources \cite{Hirano:2012kj, Shen:2017bsr, Shen:2018pty,Du:2018mpf, Du:2019obx} are currently under development. The \beshydro{} code \cite{Du:2018mpf, Du:2019obx}, in particular, shares a common ancestry \cite{Bazow:2017ewq} and therefore a number of similar features with \cpuvah{}. The integration of both codes into a single framework, in order to utilize the adaptive time step for capturing and resolving the dynamical production of energy and baryon sources at the onset of low-energy nuclear collisions, offers itself as an interesting project. On the theoretical side, the effects of non-zero chemical potentials for net charge, baryon number and strangeness have not yet been considered in our formulation of anisotropic hydrodynamics \cite{McNelis:2018jho}. An upgrade of our code package that will include them is planned for the future.
\section{Acknowledgements}
\label{S7}

M.M. would like to thank Seyed Sabok-Sayr from Rutgers University for assisting in the development of the automated grid algorithm during the Erd\H os Institute 2020 Data Science Boot Camp. Computational resources for the code validation, comparison and benchmark tests were provided by the Ohio Supercomputer Center under Project PAS0254 \cite{OhioSupercomputerCenter1987}. This work was supported by the National Science Foundation (NSF) within the framework of the JETSCAPE Collaboration under Award No. \rm{ACI-1550223}. Additional partial support by the U.S. Department of Energy (DOE), Office of Science, Office for Nuclear Physics under Award No. \rm{DE-SC0004286} and within the framework of the BEST and JET Collaborations is also acknowledged.

\begin{appendices}
\section{Gradient source terms}
\label{appa}
In this appendix, we list the gradients that appear in the source terms of the anisotropic hydrodynamic equations. The derivatives of the fluid velocity component $u^\tau$ are
\bs
\allowdisplaybreaks
\beal
\partial_\tau u^\tau &= v^x \partial_\tau u^x + v^y \partial_\tau u^y + \tau^2 v^\eta \partial_\tau u^\eta + \tau v^\eta u^\eta \,,\\
\partial_x u^\tau &= v^x \partial_x u^x + v^y \partial_x u^y + \tau^2 v^\eta \partial_x u^\eta \,,\\
\partial_y u^\tau &= v^x \partial_y u^x + v^y \partial_y u^y + \tau^2 v^\eta \partial_y u^\eta \,,\\
\partial_\eta u^\tau &= v^x \partial_\eta u^x + v^y \partial_\eta u^y + \tau^2 v^\eta \partial_\eta u^\eta \,.
\end{align}
\es
The derivatives of the longitudinal basis vector $z^\mu$ are
\bs
\allowdisplaybreaks
\beal
\partial_\tau z^\tau &= \frac{\tau}{\sqrt{1{+}u_\perp^2}}\left(\partial_\tau u^\eta - \frac{u^\eta\left(u^x \partial_\tau u^x {+} u^y \partial_\tau u^y \right)}{1{+}u_\perp^2}\right) + \frac{z^\tau}{\tau} \,,\\
\partial_x z^\tau &= \frac{\tau}{\sqrt{1{+}u_\perp^2}}\left(\partial_x u^\eta - \frac{u^\eta\left(u^x \partial_x u^x {+} u^y \partial_x u^y \right)}{1{+}u_\perp^2}\right) \,,\\
\partial_y z^\tau &= \frac{\tau}{\sqrt{1{+}u_\perp^2}}\left(\partial_y u^\eta - \frac{u^\eta\left(u^x \partial_y u^x {+} u^y \partial_y u^y \right)}{1{+}u_\perp^2}\right) \,,\\
\partial_\eta z^\tau &= \frac{\tau}{\sqrt{1{+}u_\perp^2}}\left(\partial_\eta u^\eta - \frac{u^\eta\left(u^x \partial_\eta u^x {+} u^y \partial_\eta u^y \right)}{1{+}u_\perp^2}\right) \,,\\
\partial_\tau z^\eta &= \frac{1}{\tau\sqrt{1{+}u_\perp^2}}\left(\partial_\tau u^\tau - \frac{u^\tau\left(u^x \partial_\tau u^x {+} u^y \partial_\tau u^y \right)}{1{+}u_\perp^2}\right) - \frac{z^\eta}{\tau} \,,\\
\partial_x z^\eta &= \frac{1}{\tau\sqrt{1{+}u_\perp^2}}\left(\partial_x u^\tau - \frac{u^\tau\left(u^x \partial_x u^x {+} u^y \partial_x u^y \right)}{1{+}u_\perp^2}\right) \,,\\
\partial_y z^\eta &= \frac{1}{\tau\sqrt{1{+}u_\perp^2}}\left(\partial_y u^\tau - \frac{u^\tau\left(u^x \partial_y u^x {+} u^y \partial_y u^y \right)}{1{+}u_\perp^2}\right) \,,\\
\partial_\eta z^\eta &= \frac{1}{\tau\sqrt{1{+}u_\perp^2}}\left(\partial_\eta u^\tau - \frac{u^\tau\left(u^x \partial_\eta u^x {+} u^y \partial_\eta u^y \right)}{1{+}u_\perp^2}\right) \,.
\end{align}
\es
The divergence of the spatial fluid velocity $v^i = u^i/u^\tau$ is
\be
\partial_i v^i = \frac{\partial_x u^x - v^x \partial_x u^\tau + \partial_y u^y - v^y \partial_y u^\tau + \partial_\eta u^\eta - v^\eta \partial_\eta u^\tau}{u^\tau} \,.
\ee
The longitudinal expansion rate is
\be
\begin{split}
\theta_L &=  z_\mu D_z u^\mu = - z_\mu z^\nu \partial_\nu u^\mu - z_\mu z^\nu \Gamma^\mu_{\nu\lambda} u^\lambda \\
&= - (z^\tau)^2 \partial_\tau u^\tau + z^\tau z^\eta (\tau^2 \partial_\tau u^\eta - \partial_\eta u^\tau) + (\tau z^\eta)^2 \partial_\eta u^\eta + \tau (z^\eta)^2 u^\tau\,.
\end{split}
\ee
The transverse expansion rate is
\be
\theta_\perp = \nabla_{\perp\mu}u^\mu = \theta - \theta_L\,,
\ee
where
\be
\label{eq:theta}
\begin{split}
\theta &= D_\mu u^\mu = \partial_\mu u^\mu + \Gamma^\mu_{\mu\nu} u^\nu \\
&= \partial_\tau u^\tau + \partial_x u^x + \partial_y u^y + \partial_\eta u^\eta + \frac{u^\tau}{\tau}
\end{split}
\ee
is the scalar expansion rate.

The components of the fluid acceleration
\be
a^\mu = Du^\mu = u^\nu\partial_\nu u^\mu + u^\nu \Gamma^\mu_{\nu\lambda} u^\lambda
\ee
are
\bs
\allowdisplaybreaks
\beal
a^\tau &= u^\tau \partial_\tau u^\tau + u^x \partial_x u^\tau + u^y \partial_y u^\tau + u^\eta \partial_\eta u^\tau + \tau(u^\eta)^2 \,,\\
a^x &= u^\tau \partial_\tau u^x + u^x \partial_x u^x + u^y \partial_y u^x + u^\eta \partial_\eta u^x \,,\\
a^y &= u^\tau \partial_\tau u^y + u^x \partial_x u^y + u^y \partial_y u^y + u^\eta \partial_\eta u^y \,,\\
a^\eta &= u^\tau \partial_\tau u^\eta + u^x \partial_x u^\eta + u^y \partial_y u^\eta + u^\eta \partial_\eta u^\eta + \frac{2u^\tau u^\eta}{\tau}\,.
\end{align}
\es
The components of the longitudinal vector's comoving time derivative
\be
\dot{z}^\mu = D z^\mu = u^\nu\partial_\nu z^\mu + u^\nu \Gamma^\mu_{\nu\lambda} z^\lambda
\ee
are
\bs
\allowdisplaybreaks
\beal
\dot{z}^\tau &= u^\tau \partial_\tau z^\tau + u^x \partial_x z^\tau + u^y \partial_y z^\tau + u^\eta \partial_\eta z^\tau + \tau u^\eta z^\eta \,,\\
\dot{z}^\eta &= u^\tau \partial_\tau z^\eta + u^x \partial_x z^\eta + u^y \partial_y z^\eta + u^\eta \partial_\eta z^\eta + \frac{u^\tau z^\eta {+} u^\eta z^\tau}{\tau}\,.
\end{align}
\es
The components of the fluid velocity's longitudinal derivative
\be
D_z u^\mu = - z^\nu\partial_\nu u^\mu - z^\nu \Gamma^\mu_{\nu\lambda} u^\lambda
\ee
are
\bs
\allowdisplaybreaks
\beal
D_z u^\tau &= -z^\tau \partial_\tau u^\tau - z^\eta \partial_\eta u^\tau - \tau u^\eta z^\eta \,,\\
D_z u^x &= -z^\tau \partial_\tau u^x - z^\eta \partial_\eta u^x \,,\\
D_z u^y &= -z^\tau \partial_\tau u^y - z^\eta \partial_\eta u^y \,,\\
D_z u^\eta &= -z^\tau \partial_\tau u^\eta - z^\eta \partial_\eta u^\eta - \frac{u^\tau z^\eta{+} u^\eta z^\tau}{\tau} \,.
\end{align}
\es
The transverse gradient of the fluid velocity projected along the longitudinal direction is
\be
z_\nu \nabla_\perp^\mu u^\nu = \Xi^\mu_\alpha z_\nu D^\alpha u^\nu \,,
\ee
where the components of
\be
z_\nu D^\alpha u^\nu = g^{\alpha\beta} z_\nu \partial_\beta u^\nu + g^{\alpha\beta}z_\nu \Gamma^\nu_{\beta\lambda} u^\lambda
\ee
are
\bs
\allowdisplaybreaks
\beal
z_\nu D^\tau u^\nu &= z^\tau \partial_\tau u^\tau - \tau^2 z^\eta \partial_\tau u^\eta - \tau u^\eta z^\eta \,,\\
z_\nu D^x u^\nu &= - z^\tau \partial_x u^\tau + \tau^2 z^\eta \partial_x u^\eta \,,\\
z_\nu D^y u^\nu &= - z^\tau \partial_y u^\tau + \tau^2 z^\eta \partial_y u^\eta \,,\\
z_\nu D^\eta u^\nu &= -\frac{1}{\tau^2}\left(z^\tau \partial_\eta u^\tau - \tau^2 z^\eta \partial_\eta u^\eta + \tau \left(u^\eta z^\tau {-} u^\tau z^\eta\right) \right)
\end{align}
\es
The transverse velocity-shear tensor is\footnote{%
    In the code we obtain $\sigma_\perp^\munu$ by applying the projector $\Xi^\munu_{\alpha\beta}$ onto $D^{(\alpha} u^{\beta)}$ directly, rather than simplifying the expression \eqref{eq:sigmaT}.}
\be
\label{eq:sigmaT}
\sigma_\perp^\munu = \Xi^\munu_{\alpha\beta} D^{(\alpha} u^{\beta)} \,,
\ee
where the components of
\be
D^{(\alpha} u^{\beta)} = \frac{1}{2}\left(g^{\alpha\rho}\partial_\rho u^\beta + g^{\beta\rho}\partial_\rho u^\alpha + g^{\alpha\rho} \Gamma^\beta_{\rho\lambda}u^\lambda + g^{\beta\rho} \Gamma^\alpha_{\rho\lambda}u^\lambda\right)
\ee
are
\bs
\allowdisplaybreaks
\beal
D^{(\tau} u^{\tau)} &= \partial_\tau u^\tau  \,,\\
D^{(\tau} u^{x)} &= \frac{1}{2}\left(\partial_\tau u^x - \partial_x u^\tau \right) \,,\\
D^{(\tau} u^{y)} &= \frac{1}{2}\left(\partial_\tau u^y - \partial_y u^\tau \right) \,,\\
D^{(\tau} u^{\eta)} &= \frac{1}{2}\left(\partial_\tau u^\eta - \frac{\partial_\eta u^\tau}{\tau^2} \right) \,,\\
D^{(x} u^{x)} &= - \partial_x u^x \,,\\
D^{(x} u^{y)} &= -\frac{1}{2}\left(\partial_x u^y + \partial_y u^x \right) \,,\\
D^{(x} u^{\eta)} &= -\frac{1}{2}\left(\partial_x u^\eta + \frac{\partial_\eta u^x}{\tau^2} \right) \,,\\
D^{(y} u^{y)} &= -\partial_y u^y \,,\\
D^{(y} u^{\eta)} &= -\frac{1}{2}\left(\partial_y u^\eta + \frac{\partial_\eta u^y}{\tau^2} \right) \,,\\
D^{(\eta} u^{\eta)} &= -\frac{1}{\tau^2}\left(\partial_\eta u^\eta + \frac{u^\tau}{\tau} \right) \,.
\end{align}
\es
The transverse vorticity tensor is
\be
\begin{split}
\omega_\perp^\munu &= \Xi^\mu_\alpha \Xi^\nu_\beta D^{[\alpha} u^{\beta]} \\
&= D^{[\mu} u^{\nu]} - \frac{u^\mu a^{\nu} {-} u^\nu a^{\mu} {+} z^\mu (D_zu^\nu {+} z_\alpha D^\nu u^\alpha) {-} z^\nu (D_z u^\mu {+} z_\alpha D^\mu  u^\alpha)}{2}
\end{split}
\ee
where the components of
\be
\label{eq:anti_Du}
D^{[\alpha} u^{\beta]} = \frac{1}{2}\left(g^{\alpha\rho}\partial_\rho u^\beta - g^{\beta\rho}\partial_\rho u^\alpha + g^{\alpha\rho} \Gamma^\beta_{\rho\lambda}u^\lambda - g^{\beta\rho} \Gamma^\alpha_{\rho\lambda}u^\lambda\right)
\ee
are
\bs
\allowdisplaybreaks
\beal
D^{[\tau} u^{\tau]} &= 0  \,,\\
D^{[\tau} u^{x]} &= \frac{1}{2}\left(\partial_\tau u^x + \partial_x u^\tau \right) \,,\\
D^{[\tau} u^{y]} &= \frac{1}{2}\left(\partial_\tau u^y + \partial_y u^\tau \right) \,,\\
D^{[\tau} u^{\eta]} &= \frac{1}{2}\left(\partial_\tau u^\eta + \frac{\partial_\eta u^\tau}{\tau^2} \right) + \frac{u^\eta}{\tau}\,,\\
D^{[x} u^{x]} &= 0 \,,\\
D^{[x} u^{y]} &= -\frac{1}{2}\left(\partial_x u^y - \partial_y u^x \right) \,,\\
D^{[x} u^{\eta]} &= -\frac{1}{2}\left(\partial_x u^\eta - \frac{\partial_\eta u^x}{\tau^2} \right) \,,\\
D^{[y} u^{y]} &= 0 \,,\\
D^{[y} u^{\eta]} &= -\frac{1}{2}\left(\partial_y u^\eta + \frac{\partial_\eta u^y}{\tau^2} \right) \,,\\
D^{[\eta} u^{\eta]} &= 0 \,.
\end{align}
\es
\section{Geometric source terms}
\label{appb}

Here we list the components of the geometric source terms $G_W^\mu$ and $G_\pi^\munu$ (Eq.~\eqref{eq:geometric_source}) that appear in the relaxation equations~\eqref{eq:relax_residual} for $\Wperp$ and $\piperp$, respectively:
\bs
\allowdisplaybreaks
\beal
G_W^\tau &= \tau u^\eta W_{\perp z}^\eta \,, \\
G_W^x &= 0 \,,\\
G_W^y &= 0 \,,\\
G_W^\eta &= \frac{u^\tau W_{\perp z}^\eta + u^\eta W_{\perp z}^\tau}{\tau} \,.
\end{align}
\es
\bs
\allowdisplaybreaks
\label{eq:geometric_piperp_comps}
\beal
G_\pi^{\tau\tau} &= 2 \tau u^\eta \pi_\perp^{\tau\eta} \,, \\
G_\pi^{\tau x} &= \tau u^\eta \pi_\perp^{x\eta} \,,\\
G_\pi^{\tau y} &= \tau u^\eta \pi_\perp^{y\eta} \,,\\
G_\pi^{\tau\eta} &= \tau u^\eta \pi_\perp^{\eta\eta}  +  \frac{u^\tau \pi_\perp^{\tau\eta} {+} u^\eta \pi_\perp^{\tau\tau}}{\tau} \,,\\
G_\pi^{xx} &= 0 \,,\\
G_\pi^{xy} &= 0 \,,\\
G_\pi^{x\eta} &= \frac{u^\tau \pi_\perp^{x\eta} + u^\eta \pi_\perp^{\tau x}}{\tau} \,, \\
G_\pi^{yy} &= 0 \,,\\
G_\pi^{y\eta} &= \frac{u^\tau \pi_\perp^{y\eta} + u^\eta \pi_\perp^{\tau y}}{\tau} \,, \\
G_\pi^{\eta\eta} &= \frac{2\left(u^\tau \pi_\perp^{\eta\eta} + u^\eta \pi_\perp^{\tau\eta}\right)}{\tau} \,.
\end{align}
\es
\section{Conformal anisotropic transport coefficients}
\label{appc}

In the conformal limit $m = B = 0$, the anisotropic transport coefficients~\eqref{eq:pl_coeff} -- \eqref{eqB4} only depend the functions $\I_{nrqs}$, which reduce to
\be
    \I_{nrqs} = \frac{g(n{+}s{+}1)!\,\alpha_L^{r+1}\Lambda^{n+s+2}\mathcal{R}_{nrq}}{4\pi^2(2q)!!}\,,
\ee
where $g$, $\alpha_L$ and $\Lambda$ are given by Eqs.~\eqref{eq:degeneracy},~\eqref{eq:aL_conformal} and~\eqref{eq:Lambda_conformal}, respectively. We list the functions $\mathcal{R}_{nrq}$ used in this work~\cite{McNelis:2018jho}:
\bs
\allowdisplaybreaks
\beal
\mathcal{R}_{200} & = \alpha_L \big(1+(1+\xi_L)t_L\big) \,,\\
\mathcal{R}_{220} & = \frac{-1+(1+\xi_L)t_L}{\xi_L \alpha_L} \,,\\
\mathcal{R}_{201} & = \frac{1+(\xi_L-1)t_L}{\xi_L \alpha_L} \,,\\
\mathcal{R}_{240} & = \frac{3+2\xi_L-3(1+\xi_L)t_L)}{\xi_L^2\alpha_L^3} \,,\\
\mathcal{R}_{202} & = \frac{3+\xi_L+(1+\xi_L)(\xi_L-3)t_L}{\xi_L^2(1+\xi_L)\alpha_L^3} \,,\\
\mathcal{R}_{221} & = \frac{-3+(3+\xi_L)t_L)}{\xi_L^2\alpha_L^3} \,,\\
\mathcal{R}_{441} & = \frac{-15+13\xi_L+3(1+\xi_L)(5+\xi_L)t_L}{4\xi_L^3\alpha_L^3} \,,\\
\mathcal{R}_{402} & = \frac{3(\xi_L-1)+(\xi_L(3\xi_L-2)+3)t_L}{4\xi_L^2\alpha_L} \,,\\
\mathcal{R}_{421} & = \frac{3+\xi_L+(1+\xi_L)(\xi_L-3)t_L}{4\xi_L^2\alpha_L} \,,\\
\mathcal{R}_{422} & = \frac{15+\xi_L+(\xi_L(\xi_L-6)-15)t_L}{4\xi_L^3\alpha_L^3} \,,\\
\mathcal{R}_{403} & = \frac{(\xi_L-3)(5+3\xi_L)+3(1+\xi_L)(\xi_L(\xi_L-2)+5)t_L}{4\xi_L^3(1+\xi_L)\alpha_L^3}\,,
\end{align}
\es
where $\xi_L = \alpha_L^{-2} - 1$ and $t_L = \mathrm{arctan}\sqrt{\xi_L}/\sqrt{\xi_L}$.

\section{\trento{} energy deposition model}
\label{appd}
In the \trento{} model, the transverse energy deposition (GeV/fm$^2$) of a single fluctuating nuclear collision event in the mid-rapidity region is~\cite{Moreland:2014oya}
\be
\label{eq:trento_perp}
\frac{dE_T}{dxdyd\eta_s}_{|{\eta_s=0}} = N \times T_R(x,y)\,,
\ee
where $N$ is the normalization parameter and
\be
T_R(x,y) = \left(\frac{T^p_A(x,y) + T^p_B(x,y)}{2}\right)^{1/p}
\ee
is the reduced nuclear thickness function, with $p$ being the geometric parameter. The nuclear thickness function of nucleus $A$ ($B$) is\footnote{In this work, we only consider Pb+Pb collisions ($A=B=208$) at LHC energies $\sqrt{s_\text{NN}} = 2.76$ TeV; the inelastic nucleon--nucleon cross section is set to $\sigma_\text{NN} = 6.4$ fm$^{-2}$.}
\be
T_{A,B}(x,y) = \sum_{n=1}^{N_{\text{part},A,B}} \gamma_n \,T_p(x-x_n,y-y_n)\,,
\ee
where $N_{\text{part},A,B}$ are the number of participant nucleons from nucleus $A$ ($B$); the nucleon positions are sampled from a Woods--Saxon distribution under the constraint that each nucleon--nucleon pair in nucleus $A$ ($B$) maintains a minimum separation $d_\text{min}$~\cite{Moreland:2014oya}. The participant nucleon's thickness function is centered around its sampled transverse position ($x_n$, $y_n$):
\be
\label{eq:Tpp}
    T_p(x-x_n, y - y_n) = \frac{1}{2\pi w^2} \times \exp\left[-\frac{(x{-}x_n)^2+(y{-}y_n)^2}{2w^2}\right]\,,
\ee
where $w$ is the nucleon width parameter.\footnote{This parameter does not control the nucleon's charge radius but rather the spread of thermal energy it deposits in the collision zone along the transverse directions.} Furthermore, the multiplicity factor $\gamma_n$ of each participant nucleon is sampled from the gamma distribution
\be
P(\gamma) = \frac{k^k \gamma^{k-1}\exp\left[-k\gamma\right]}{\Gamma[k]}\,,
\ee
where $k = \sigma_k^{-2}$ and $\sigma_k$ is the standard deviation.

For a very brief period $\tau_0$ after the collision, we assume the system is longitudinally free-streaming and static in Milne coordinates (i.e. $\PL/\Peq \sim 0$ and $\boldsymbol{u} \sim \boldsymbol{0}$) so that $\ene(x) \propto 1/\tau_0$. Therefore, we initialize the energy density profile of the hydrodynamic simulation at the starting time $\tau_0$ as
\be
\label{eq:trento_3d}
\ene(\tau_0,x,y,\eta_s) = \frac{1}{\tau_0} \times \frac{dE_T}{dxdyd\eta_s}_{|\eta_s=0} \times f_L(\eta_s)\,.
\ee
Here we also extend the transverse energy density profile along the spacetime rapidity direction with a smooth plateau distribution (unitless)~\cite{Pang:2018zzo}:
\be
\label{eq:plateau}
f_L(\eta_s) = \exp\left[-\frac{\big(|\eta_s| - \half\eta_\text{flat}\big)^2\, \Theta\big(|\eta_s| - \half\eta_\text{flat}\big)}{2\sigma_\eta^2}\right]\,,
\ee
where $\eta_\text{flat}$ is the plateau length, $\sigma_\eta$ is the standard deviation of the half-Gaussian tails and $\Theta$ is the Heaviside step function. The initial condition parameter values used in this work are $N = 14.19$ GeV, $p = 0.06$, $w = 1.11$ fm, $d_\text{min} = 1.45$ fm, $\sigma_k = 1.03$, $\eta_\text{flat} = 4.0$ and $\sigma_\eta = 1.8$~\cite{Pang:2018zzo,Everett:2020yty,Everett:2020xug}.

We set the lattice spacings to $\Delta x = \Delta y = \frac{1}{5}w$ and $\Delta \eta_s = \frac{1}{5}\sigma_\eta$ to resolve the fluctuating energy density profile~\eqref{eq:trento_3d} (or event-averaged profile). For the grid size, we set the longitudinal length to $L_\eta = \eta_\text{flat} + 10\sigma_\eta$ to fit the rapidity plateau~\eqref{eq:plateau}. The transverse lengths $L_x$ and $L_y$ are automatically configured by the algorithm described in Sec.~\ref{S5.2}.
\section{Numerical implementation of second-order viscous hydrodynamics}
\label{appe}

In this appendix, we summarize how second-order viscous hydrodynamics is implemented the code. We evolve viscous hydrodynamics with the same numerical algorithm discussed in Sec.~\ref{S3} except the energy-momentum tensor~\eqref{eq:Tmunu} is decomposed as
\be
\label{eq:Tmunu_viscous}
T^\munu = \ene u^\mu u^\nu - (\Peq {+} \Pi) \Delta^\munu + \pi^\munu \,,
\ee
where $\pi^\munu = \Delta^\munu_\ab T^{\alpha\beta}$ is the shear stress tensor and $\Pi = -\frac{1}{3}\Delta_\munu T^\munu - \Peq$ is the bulk viscous pressure. We also define the spatial projector $\Delta^\munu = g^\munu - u^\mu u^\nu$ and traceless double spatial projector $\Delta^\munu_\ab = \half(\Delta^\mu_\alpha\Delta^\nu_\beta +\Delta^\nu_\beta\Delta^\mu_\alpha - \frac{2}{3}\Delta^\munu\Delta_{\alpha\beta})$. The corresponding dynamical variables
\be
\label{eq:q_viscous}
    \boldsymbol{q} = (T^{\tau\mu}, \pi^\munu, \Pi)
\ee
are propagated along with $\ene$ and $\boldsymbol{u}$ (the mean-field $B$ and anisotropic variables ($\Lambda$, $\alpha_\perp$, $\alpha_L$) are not propagated). Although $\pi^\munu$ has only five independent components, we propagate all ten components in the simulation~\cite{Bazow:2016yra}.\footnote{%
    For longitudinally boost-invariant systems, we do not propagate the shear stress components $\pi^{\tau\eta}$, $\pi^{x \eta}$ and $\pi^{y \eta}$.}
%
%
\subsection{Hydrodynamic equations}

Here we list the evolution equations for the dynamical variables~\eqref{eq:q_viscous} (we refer the reader to Refs.~\cite{Bazow:2016yra, Denicol:2012cn} for details on their derivation):
\bs
\allowdisplaybreaks
\label{eq:viscous_hydro_equations}
\beal
\partial_\tau T^{\tau\tau} + \partial_i (v^i T^{\tau\tau}) =& - \frac{T^{\tau\tau} {+} \tau^2 T^{\eta\eta}}{\tau} + (\pi^{\tau\tau} {-} \Peq {-} \Pi) \partial_i v^i \\ \nonumber
& + v^i \partial_i (\pi^{\tau\tau} {-} \Peq {-} \Pi) - \partial_i \pi^{\tau i} \,, \\\nonumber\\
\partial_\tau T^{\tau x} + \partial_i (v^i T^{\tau x}) =& - \frac{T^{\tau x}}{\tau} - \partial_x (\Peq{+}\Pi) + \pi^{\tau x} \partial_i v^i + v^i \partial_i \pi^{\tau x} \\ \nonumber
&- \partial_i \pi^{xi} \,,\\\nonumber\\
\partial_\tau T^{\tau y} + \partial_i (v^i T^{\tau y}) =& - \frac{T^{\tau y}}{\tau} - \partial_y(\Peq{+}\Pi) + \pi^{\tau y} \partial_i v^i + v^i \partial_i \pi^{\tau y} \\ \nonumber
& - \partial_i \pi^{yi} \,,\\\nonumber\\
\partial_\tau T^{\tau\eta} + \partial_i (v^i T^{\tau\eta}) =& - \frac{3T^{\tau\eta}}{\tau} - \frac{\partial_\eta (\Peq{+}\Pi)}{\tau^2} + \pi^{\tau\eta} \partial_i v^i + v^i \partial_i \pi^{\tau\eta} \\\nonumber
&- \partial_i \pi^{\eta i}\,,\\\nonumber\\
\partial_\tau \pi^\munu + \partial_i (v^i \pi^\munu) =& \,\pi^\munu \partial_i v^i + \frac{1}{u^\tau}\left[-\frac{\pi^\munu}{\tau_\pi} + \I_{\pi^\prime}^\munu - \mathcal{P}_{\pi^\prime}^\munu - \mathcal{G}_{\pi^\prime}^\munu \right] \,,\\\nonumber\\
\partial_\tau \Pi + \partial_i (v^i \Pi) =& \,\Pi \partial_i v^i + \frac{1}{u^\tau}\left[-\frac{\Pi}{\tau_\Pi} + \I_\Pi\right]  \,,
\end{align}
\es
where $T^{\eta\eta} = (\ene{+}\Peq{+}\Pi)(u^\eta)^2 + (\Peq{+}\Pi)/\tau^2+\pi^{\eta\eta}$,
\bs
\beal
\mathcal{I}^\munu_{\pi^\prime} =&\, 2 \beta_\pi \sigma^\munu  +\Delta^\munu_{\alpha\beta}\big(2 \pi^{\lambda (\alpha} \omega^{\beta)}_{\,\,\,\lambda} - \bar{\tau}_{\pi\pi} \pi^{\lambda(\alpha} \sigma^{\beta)}_{\,\,\,\lambda}\big) - \bar{\delta}_{\pi\pi} \pi^\munu \theta \\ \nonumber
& + \bar{\lambda}_{\pi\Pi} \Pi \sigma^\munu \,,\\\nonumber\\
\mathcal{I}_\Pi =&\, -\beta_\Pi \theta - \bar{\delta}_{\Pi\Pi}\Pi\theta + \bar{\lambda}_{\Pi\pi} \pi^\munu \sigma_\munu\,,
\end{align}
\es
are the gradient source terms for $\pi^\munu$ and $\Pi$ and
\bs
\label{eq:geometric_shear_prime}
\beal
\mathcal{P}^\munu_{\pi^\prime} =&\, \left(\pi^{\mu\alpha}u^\nu + \pi^{\nu\alpha}u^\mu\right)a_\alpha \,,\\
\mathcal{G}^\munu_{\pi^\prime} =&\, u^\gamma \Gamma^\mu_{\gamma\lambda} \pi^{\nu\lambda}
+ u^\gamma \Gamma^\nu_{\gamma\lambda} \pi^{\mu\lambda}\,,
\end{align}
\es
are the spatial projection and geometric source terms for $\pi^\munu$.
We also define the velocity-shear tensor $\sigma^\munu = \Delta^\munu_\ab D^{(\alpha}u^{\beta)}$ and vorticity tensor $\omega^\munu = \Delta^\mu_\alpha \Delta^\nu_\beta D^{[\alpha}u^{\beta]}$.

The components of $\sigma^\munu$ and $\mathcal{G}^\munu_{\pi^\prime}$ are the same as $\sigma_\perp^\munu$ and $\mathcal{G}^\munu_{\pi}$ after replacing $\Xi^\munu_\ab \to \Delta^\munu_\ab$ and $\piperp \to \pi^\munu$ in Eqs.~\eqref{eq:sigmaT} and~\eqref{eq:geometric_piperp_comps}, respectively. The components of $\omega^\munu$ are
\be
\omega^\munu =  D^{[\mu}u^{\nu]} - \frac{u^\mu a^\nu {-} u^\nu a^\mu}{2}\,,
\ee
where $D^{[\mu}u^{\nu]}$ is given by Eq.~\eqref{eq:anti_Du}.
\subsection{Transport coefficients}
In quasiparticle viscous hydrodynamics~\cite{Tinti:2016bav} (i.e. $m = m(T)$ from Fig.~\ref{quasi}a), the relaxation times ($\tau_\pi$, $\tau_\Pi$) and first-order transport coefficients ($\beta_\pi$, $\beta_\Pi$) are given by Eqs.~\eqref{eq:tau_r} --~\eqref{eq:beta_r}. The second-order transport coefficients are~\cite{McNelis:2018jho}
\bs
\label{eq:second_coeff}
\allowdisplaybreaks
\beal
\tau_{\pi\pi} =&\, \frac{10 + 4\bar{c}_\pi m^2 \I_{22}}{7}  \,,\\
\delta_{\pi\pi} =&\, \frac{4}{3} + \bar{c}_\pi \I_{22} \Big(\frac{m^2}{3} - m\frac{dm}{d\ene}(\ene{+}\Peq)\Big) \,,\\
\lambda_{\pi\Pi} =&\, \frac{6}{5} - \frac{2m^4}{15}\big(\bar{c}_\ene \I_{00} + \bar{c}_\Pi \I_{01}\big)  \,,\\
\delta_{\Pi\Pi} =&\, 1 - c_s^2 - \frac{m^4}{9}\big(\bar{c}_\ene \I_{00} + \bar{c}_\Pi \I_{01}\big) \\\nonumber
&- m\frac{dm}{d\ene}(\ene{+}\Peq)\Big(\bar{c}_\ene\I_{21}+\frac{5}{3}\bar{c}_\Pi\I_{22}+\frac{3}{m^2}\Big)   \,,\\
\lambda_{\Pi\pi} =&\, \frac{1}{3} - c_s^2 + \frac{\bar{c}_\pi m^2 \I_{22}}{3}  \,,
\end{align}
\es
where
\bs
\allowdisplaybreaks
\beal
\bar{c}_{\pi} =&\, \frac{1}{2\I_{42}}  \,,\\
\bar{c}_{\ene} =&\, -\frac{\I_{41}}{\frac{5}{3}\I_{40}\I_{42} - \I_{41}^2}  \,,\\
\bar{c}_{\Pi} =&\, \frac{\I_{40}}{\frac{5}{3}\I_{40}\I_{42} - \I_{41}^2}  \,,
\end{align}
\es
and the function $\I_{nq}$ are given by Eq.~\eqref{eq:Inq_moments}.

In standard viscous hydrodynamics~\cite{Denicol:2014vaa,Bazow:2016yra} (i.e. $\bar{m} = m/T \ll 1$), the relaxation times are given by Eq.~\eqref{eq:tau_r_small} and
\bs
\allowdisplaybreaks
\beal
\beta_\pi \approx&\, \frac{\ene{+}\Peq}{5} + O(\bar{m}^2)  \,,\\
\beta_\Pi \approx&\, 15\Big(\frac{1}{3}-c_s^2\Big)^2(\ene{+}\Peq) + O(\bar{m}^5)  \,.
\end{align}
\es
The second-order transport coefficients~\eqref{eq:second_coeff} reduce to
\bs
\allowdisplaybreaks
\beal
\tau_{\pi\pi} \approx&\, \frac{10}{7} + O(\bar{m}^2)  \,,\\
\delta_{\pi\pi} \approx&\, \frac{4}{3} + O(\bar{m}^2) \,,\\
\lambda_{\pi\Pi} \approx&\, \frac{6}{5} + O(\bar{m}^2\ln\bar{m})  \,,\\
\delta_{\Pi\Pi} \approx&\, \frac{2}{3} + O(\bar{m}^2\ln\bar{m})  \,,\\
\lambda_{\Pi\pi} \approx&\, \frac{8}{5}\Big(\frac{1}{3}-c_s^2\Big)^2 + O(\bar{m}^4)  \,.
\end{align}
\es
Both models use the same shear and bulk viscosities as anisotropic hydrodynamics (e.g. Fig.~\ref{viscosities}).
\subsection{Reconstruction formulas for the energy density and fluid velocity}
We reconstruct the energy density by solving the following nonlinear equation via Newton's method~\cite{Shen:2014vra, Bazow:2016yra}:
\be
\label{eq:newton_energy}
f(\ene) = 0 \,,
\ee
where
\be
f(\ene) = \left(\bar{M}^\tau {-} \ene\right)\left(\bar{M}^\tau {+} \Peq {+} \Pi\right) - (\bar{M}^x)^2 - (\bar{M}^y)^2 - (\tau \bar{M}^\eta)^2\,,
\ee
with $\bar{M}^\mu = T^{\tau\mu} - \pi^{\tau\mu}$ and $\Peq = \Peq(\ene)$ being the QCD equation of state. Using the previous energy density for the initial guess, we iterate the solution to Eq.~\eqref{eq:newton_energy} as
\be
\label{eq:energy_iteration}
\ene \leftarrow \ene - \frac{f(\ene)}{df /d\ene}\,,
\ee
where
\be
\frac{df}{d\ene} = c_s^2 (\bar{M}^\tau {-} \ene) - \bar{M}^\tau - \Peq - \Pi\,,
\ee
with $c_s^2 = d\Peq/d\ene$ being the QCD speed of sound squared. We repeat the iteration~\eqref{eq:energy_iteration} until we achieve sufficient convergence or the energy density falls below $\ene_\text{min}$. If the bulk viscous pressure $\Pi < - \Peq$, we regulate it so that $\Peq + \Pi = 0$; this allows us to solve for $\ene$ explicitly~\cite{Shen:2014vra}:
\be
\ene = \bar{M}^\tau - \frac{(\bar{M}^x)^2 {+} (\bar{M}^y)^2 {+} (\tau \bar{M}^\eta)^2}{{\bar M}^\tau}\,.
\ee
Afterwards, we regulate the energy density via Eq.~\eqref{eq:energy_reg} and evaluate the fluid velocity components as
\bs
\allowdisplaybreaks
\beal
u^x &= \frac{\bar{M}^x}{\sqrt{\left(\ene{+}\Peq{+}\Pi\right)\left(\bar{M}^\tau{+}\Peq{+}\Pi\right)}} \,,\\
u^y &= \frac{\bar{M}^y}{\sqrt{\left(\ene{+}\Peq{+}\Pi\right)\left(\bar{M}^\tau{+}\Peq{+}\Pi\right)}} \,,\\
u^\eta &= \frac{\bar{M}^\eta}{\sqrt{\left(\ene{+}\Peq{+}\Pi\right)\left(\bar{M}^\tau{+}\Peq{+}\Pi\right)}}  \,.
\end{align}
\es
\subsection{Regulating the shear stress and bulk viscous pressure}
In this regulation scheme, we first adjust the shear stress components
\bs
\allowdisplaybreaks
\beal
\pi^{\eta\eta} \leftarrow&\, \frac{1}{\tau^2\left(1 {+} u_\perp^2\right)}\Big[\pi^{xx}\big((u^x)^2 {-} (u^\tau)^2\big) + \pi^{yy}\big((u^y)^2 {-} (u^\tau)^2\big) \nonumber\\
&+ 2\big(\pi^{xy} u^x u^y + \tau^2 (\pi^{x\eta}u^x {+} \pi^{y\eta}u^y)u^\eta\big)\Big] \,,\\
\pi^{\tau x} \leftarrow&\, \frac{\pi^{xx} u^x + \pi^{xy}u^y + \tau^2 \pi^{x\eta} u^\eta}{u^\tau}\,,\\
\pi^{\tau y} \leftarrow&\, \frac{\pi^{xy} u^x + \pi^{yy}u^y + \tau^2 \pi^{y\eta} u^\eta}{u^\tau}\,,\\
\pi^{\tau \eta} \leftarrow&\, \frac{\pi^{x\eta} u^x + \pi^{y\eta}u^y + \tau^2 \pi^{\eta\eta} u^\eta}{u^\tau}\,,\\
\pi^{\tau\tau} \leftarrow&\, \frac{\pi^{\tau x} u^x + \pi^{\tau y}u^y + \tau^2 \pi^{\tau\eta} u^\eta}{u^\tau} \,,
\end{align}
\es
so that $\pi^\munu$ satisfies the orthogonality and tracelessness conditions
\bs
\beal
\pi^\munu u_\nu &= 0 \,,\\
\pi^\mu_\mu &= 0 \,.
\end{align}
\es
Then we regulate the shear stress and bulk viscous pressure as
\bs
\beal
\pi^\munu &\leftarrow \gamma_\text{reg} \pi^\munu \,,\\
\Pi &\leftarrow \gamma_\text{reg} \Pi\,,
\end{align}
\es
where
\be
\label{eq:rescale_viscous}
\gamma_\text{reg} = \min\Bigg(1, \sqrt{\frac{3\mathcal{P}_\text{eq}^2}{\pi {\,\cdot\,} \pi + 3 \Pi^2}}\Bigg) \,,
\ee
with $\pi {\,\cdot\,} \pi = \pi_{\munu} \pi^\munu$. The regulation factor $\gamma_\text{reg}$ usually suppresses $\pi^\munu$ and $\Pi$ around the edges of the fireball at early times $\tau < 1$ fm/$c$, especially in standard viscous hydrodynamics (e.g. see Figs.~(\ref{freezeout_x}c) and~(\ref{freezeout_z}c)).
\end{appendices}
\bibliographystyle{elsarticle-num}
\bibliography{cpu_vah}
\end{document}